\documentclass[fleqn,usenatbib]{mnras}

\usepackage{newtxtext,newtxmath}

\usepackage[T1]{fontenc}

\DeclareRobustCommand{\VAN}[3]{#2}
\let\VANthebibliography\thebibliography
\def\thebibliography{\DeclareRobustCommand{\VAN}[3]{##3}\VANthebibliography}

\usepackage{graphicx}	
\graphicspath{{Figures/}}
\usepackage{amsmath}	
\usepackage{color}
\usepackage[dvipsnames]{xcolor}
\usepackage{comment}
\usepackage{threeparttable}
\usepackage{subfig}

\newcommand\dd{\mathrm{d}}
\newcommand\DD{\mathrm{D}}
\newcommand\ee{\mathrm{e}}
\newcommand\ii{\mathrm{i}}

\newcommand\f{\frac}
\newcommand\p{\partial}
\newcommand\cst{\mathrm{constant}}

\def\XXint#1#2#3{{\setbox0=\hbox{$#1{#2#3}{\int}$}
     \vcenter{\hbox{$#2#3$}}\kern-.5\wd0}}

\def\XXiint#1#2#3{{\setbox0=\hbox{$#1{#2#3}{\iint}$}
     \vcenter{\hbox{$#2#3$}}\kern-.5\wd0}}

\title[Disc breaking with differential precession]{Disc breaking and parametric instability in warped accretion discs driven by differential precession}

\author[L. E. Held and G. I. Ogilvie]{
Loren E. Held$^{1}$ and Gordon I. Ogilvie$^{1}$\thanks{E-mail: gio10@cam.ac.uk (GIO)}
\\
$^{1}$Department of Applied Mathematics and Theoretical Physics, University of Cambridge, Centre for Mathematical Sciences, Wilberforce Road,\\ 
Cambridge CB3 0WA, United Kingdom\\}

\date{Accepted XXX. Received YYY; in original form ZZZ}

\pubyear{2026}

\begin{document}
\label{firstpage}
\pagerange{\pageref{firstpage}--\pageref{lastpage}}
\maketitle

\begin{abstract}
We present the first local simulations of accretion disc breaking or tearing with differential precession. Precession can occur due to a non-spherical potential (e.g.\ tilted circumbinary discs), or relativistic effects (e.g.\ tilted discs around spinning black holes). This tends to induce a twist in the disc, potentially causing it to break into rings. We extend earlier work on free (unforced) warps \citep{held2026disc} to include differential precession in the shearing box, and carry out quasi-2D, isothermal, high-resolution ($N = 32$ cells per scale-height $H$) numerical simulations in which an imposed precession drives a warp in a tilted disc in either the wavelike ($\alpha < H/r$) or diffusive ($\alpha > H/r$) regime. Our key finding is that, at sufficiently large tilts and precession amplitudes, the disc breaks into multiple (4--5) mutually tilted rings exhibiting a rich morphology: from wide ($\sim 10H$), flat rings to compact tori of width $\sim H$. Interactions between rings occur as a result of oblique shocks that form in the gaps between them, leading to instances of disc rejoining. Smaller tilts and precession amplitudes lead instead to parametric instability of inertial waves in the disc. We develop a linear viscous theory of bending waves under the influence of an imposed precession. Finally, we present a systematic study of the effects of grid-based dissipation on tilted discs, and find that numerical dissipation damps the tilt and induces precession. This effect decreases with resolution as $\propto (1/N)^\beta$, $2.3 \lesssim \beta \lesssim 3.9$.
\end{abstract}


\section{Introduction}
\label{INTRO}

\subsection{Background}
\label{INTRO_Background}
Distorted accretion discs have now been observed across all types of disc system. Precessing warped discs have been inferred from modulations in the light curves of X-ray binaries, e.g.\ Her X-1, SS~433, LMC X-4, SMC X-1 and SMC MAXI J1820+070 \citep{gerend1976optical, katz1982nodding, brumback2020modeling, thomas2022large}, while superhumps in the light curves of SU UMa-type dwarf novae, e.g.~Z Cha, OY Car, have been interpreted as evidence of a precessing eccentric disc \citep{vogt1982z, han2026comprehensive}. Direct images have also revealed warps in several active galactic nuclei, e.g.\ NGC 4258 and NGC 1068 \citep{miyoshi1995evidence, gallimore2004parsec}, and in protoplanetary discs, e.g.~KH 15D, IRAS 04368+2557, L1489 IRS, and MWC 758 \citep{chiang2004circumbinary, lamitina2025continued, sakai2019warped, sai2020disk, boehler2018complex}.

Distorted discs exhibit a myriad of processes which are absent in circular and planar discs. In addition to undergoing various types of \textit{precession}, distorted discs are not in vertical hydrostatic equilibrium and thus vertically oscillate (or \textit{bounce}) in the plane perpendicular to the local disc mid-plane \citep{lubow1981vertically,ogilvie2013local,fairbairn2021non,held2024instability}. They are also susceptible to a purely hydrodynamic instability known as the \textit{parametric instability} (PI) \citep{gammie2000linear, paardekooper2019local, held2026disc}. Warped discs give rise to oscillatory radial flows known as \textit{horizontal sloshing motions} \citep{papaloizou1983time, dullemond2022equations}. If the disc is sufficiently warped (or twisted) it can even \textit{break} (or \textit{tear}) into rings, which has been observed both directly and indirectly (e.g.~through shadows) in protoplanetary discs, e.g.~GW Ori, HD14257, HD139614, HD143006, and TW Hydrae \citep{mathieu1995submillimeter, fang2017millimeter, kraus2020triple, marino2015shadows, muro2020shadowing,perez2018disk, debes2023surprising}. Disc breaking has been the subject of intense study in global numerical simulations \citep{larwood1996tidally, fragner2010evolution, nixon2013tearing, musoke2023disc, rabago2024warps, smallwood2025shedding, young2025high}, and, more recently, local ones \citep{held2026disc}.

A great deal of attention has been focused on precessing discs over the years, both because differential precession can drive a warp (sometimes also called a twist) in a tilted disc, and due to their link to observational signatures, mainly quasi-periodic oscillations (QPOs); see  \cite{fragile2026tilted} for a review. There are two types of disc precession. \textit{Apsidal} precession occurs in the plane of the disc, and involves rotation of the line of apsides of an eccentric orbit, while \textit{nodal} precession occurs only in tilted discs and involves rotation of the disc's orbital plane about an axis perpendicular to the orbital plane of a binary or the equatorial plane of a spinning central accretor. In general, both effects can occur together provided the disc is tilted, and their orientation and strength depend on whether relativistic effects are important or not. If the disc is tilted with respect to the symmetry plane of a non-spherical Newtonian potential (e.g.~a tilted circumbinary protoplanetary disc), apsidal and nodal precession will be equal and opposite \citep{papaloizou1995dynamics,nealon2020rocking, kimmig2024warped}. If, on the other hand, precession is driven by relativistic effects (such as in a disc around a tilted black hole) apsidal precession is faster than nodal precession and they are both in the same direction \citep{bardeen1975lense, nealon2015bardeen, nealon2015apsidal}.\footnote{Note that in a non-precessing Keplerian disc there is a resonance in the sense that the epicyclic frequency is equal to the orbital frequency. Precession results in a detuning of this resonance.}

In a recent companion work we showed how the standard shearing box formalism for modelling a local patch of a disc can be adapted in order to study tilted and warped discs \citep{held2026disc}. This enabled us to carry out the first simulations of disc breaking in the local model. The high resolution  in the breaking region afforded by our local approach revealed a new component to the breaking mechanism involving shocks formed from converging horizontal sloshing motions in the warp. In that paper we initialized the warp \textit{ab initio} and did not include precession. We refer to this as a \textit{free warp}, which has also been modelled in both global SPH and grid-based simulations \citep[e.g.][]{lodato2010diffusive, fragner2010evolution}. We now show how our previous work can be extended to drive the warp in a tilted disc by including differential precession in the shearing box. 

\subsection{Including precession in the local model}
\label{INTRO_IncludingPrecessionInTheLocalModel}
\begin{figure}
    \centering
    \includegraphics[width=1\linewidth]{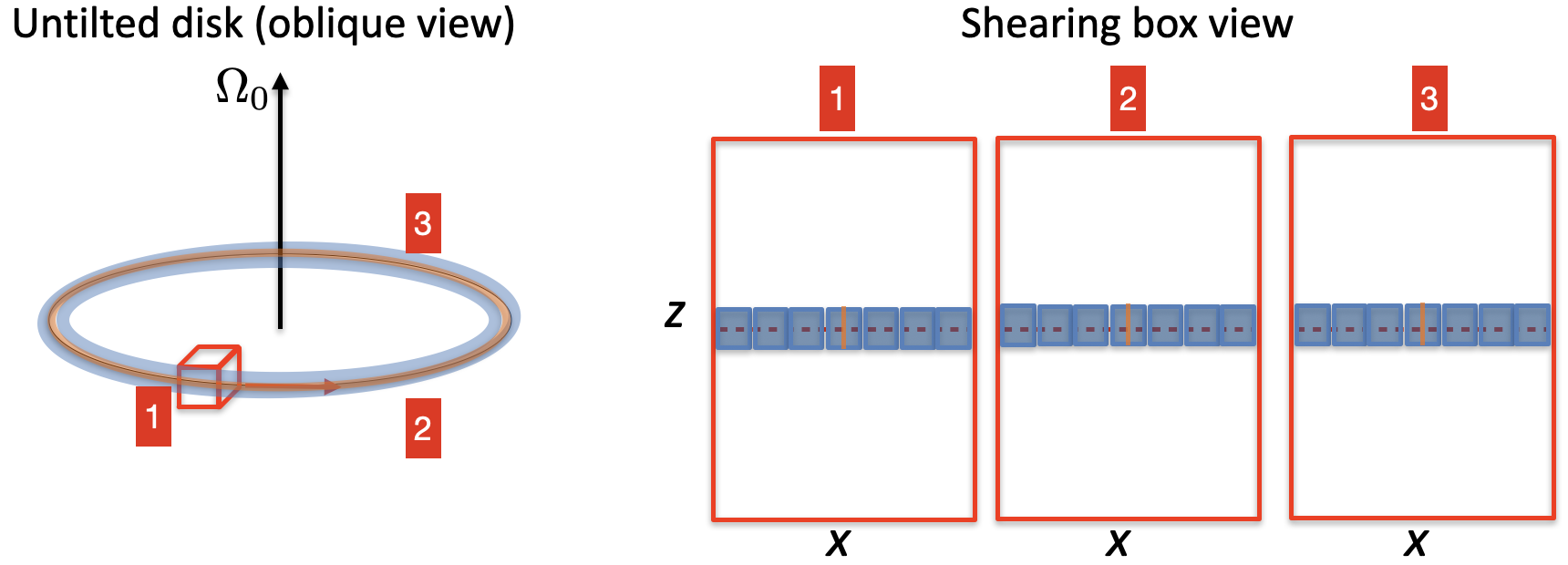}
    \includegraphics[width=1\linewidth]{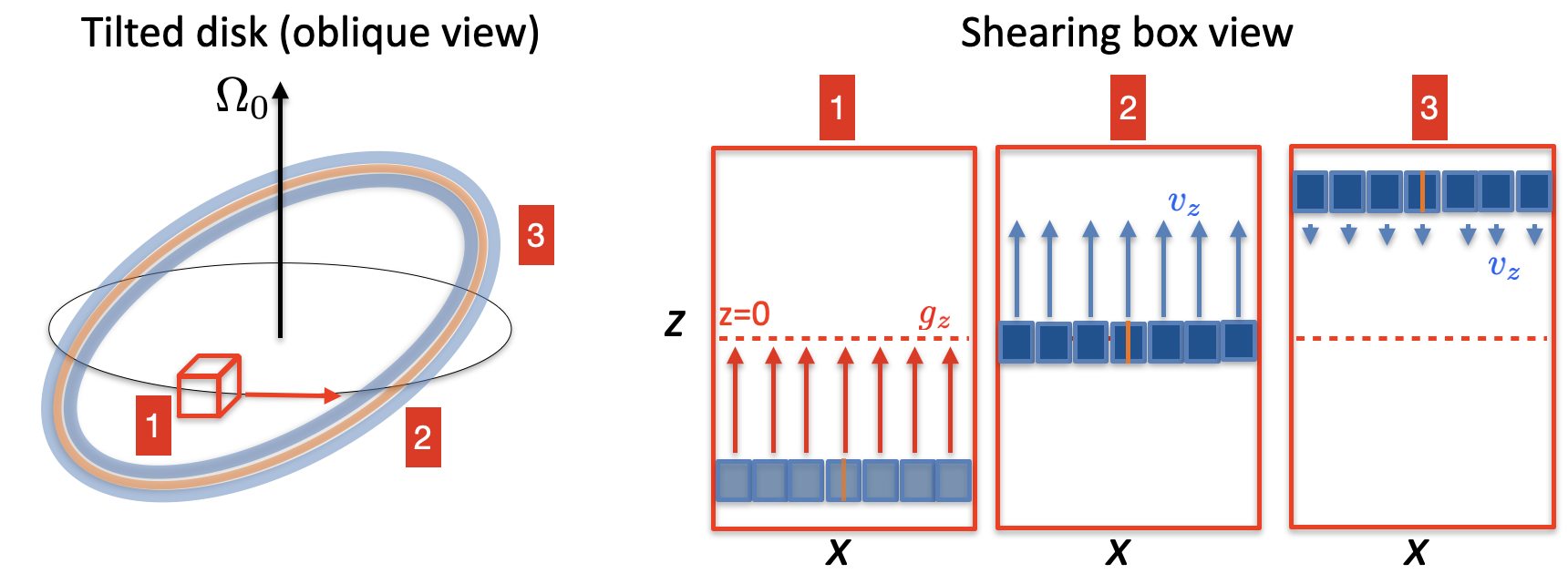}
    \includegraphics[width=1\linewidth]{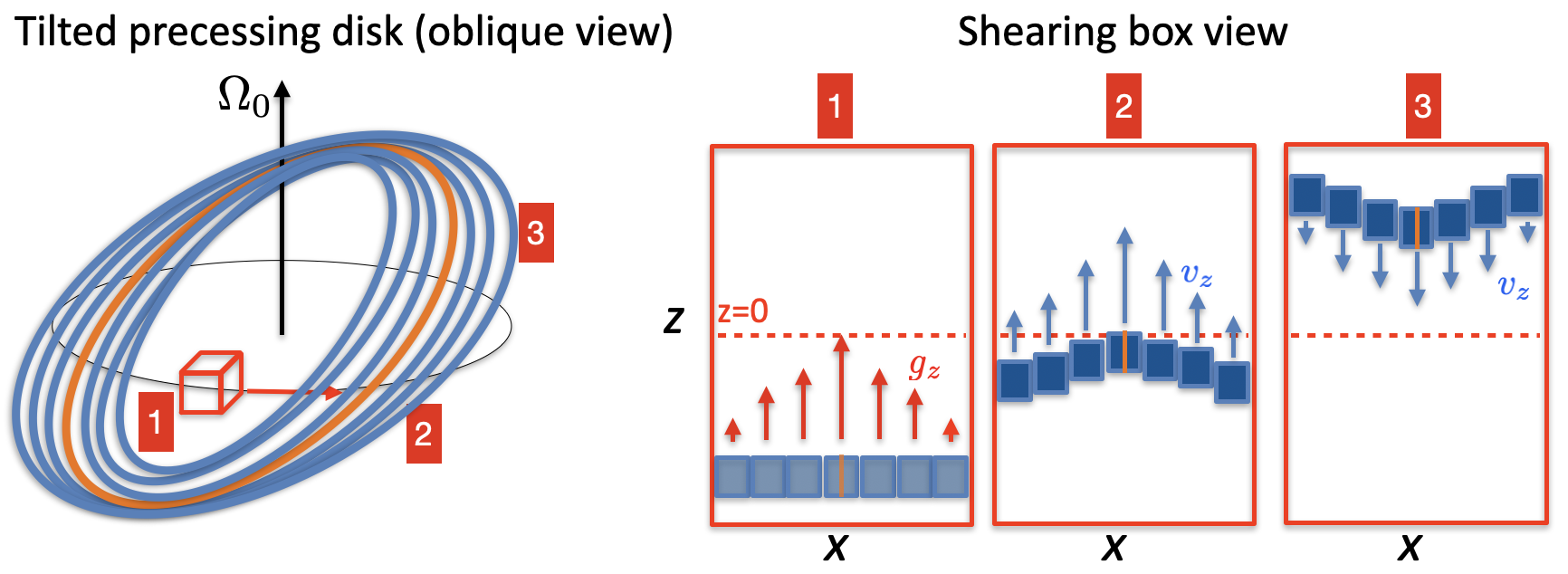}
    \caption{Schematic showing how to model tilt and precession (both rigid and differential) in the shearing box. \textit{Left}: disc annuli viewed from a global (inertial) reference frame. \textit{Right}: the corresponding local view relative to an untilted shearing box orbit. The box is anchored about some reference radius in the disc (orange annulus), and orbits with the angular frequency $\Omega_0$ of the gas that radius. The numbers 1, 2, and 3 show three different phases around an orbit. \textit{Top panel}: an untilted circular disc. \textit{Middle panel}: a tilted (but not precessing) disc (left) appears as a series of fluid columns that oscillate about the shearing box orbital plane ($z=0$) \textit{at} the local orbital frequency $\Omega_0$ (right). Red (blue) arrows illustrate the vertical gravitational acceleration $g_z$ (velocity $v_z$) felt by each fluid column. (The box should be tall enough to intersect the tilted disc, but we have reduced its size for visual clarity.)  If the tilted disc is instead \textit{rigidly precessing}, the columns will all oscillate with some frequency different from the local orbital frequency. \textit{Bottom panel}: a tilted \textit{differentially precessing} disc which develops a twist. This appears in the shearing box as a set of fluid columns which each oscillate independently (if we ignore pressure) about $z = 0$ at some frequency different from the local orbital frequency $\Omega_0$. Thus, differential precession can be related locally to a radial variation in the vertical gravitational acceleration.}
    \label{FIGURE_TiltedDiskShearingBoxView}
\end{figure}

Consider a disc that is tilted with respect to the equator of a spinning black hole, or the orbital plane of a binary. The key conceptual aspect of our model is that disc tilt can be related to a vertical oscillation of fluid columns in the shearing box \citep{ogilvie2022hydrodynamics, held2026disc}. In Fig.~\ref{FIGURE_TiltedDiskShearingBoxView} we illustrate what a tilted and precessing disc looks like relative to an observer (i.e.\ a shearing box) moving along an untilted circular orbit about some cylindrical reference radius $r_0$ (orange line) with angular frequency $\Omega_0(r_0)$. The top panel shows an untilted disc: in this case the disc simply straddles the shearing box orbital plane ($z=0$). The middle panel shows a tilted disc: if the disc is not precessing, it will appear to oscillate uniformly about the the shearing box orbital plane $z = 0$ at the orbital frequency $\Omega_0$; if, instead, the disc is \textit{rigidly} precessing, it will oscillate about $z=0$ at some frequency \textit{different} from the orbital frequency. Finally, if the disc is \textit{differentially precessing} (bottom panel), each fluid column in the disc will appear to oscillate about the orbital plane with a \textit{different} vertical frequency. This radial variation in the vertical oscillation frequency can be induced by introducing a radial variation in the vertical gravitational acceleration, i.e.~$g_z(z) \rightarrow g_z(x,z)$.

There are several limitations to this approach, some of which have already been discussed in our companion work \citep{held2026disc}: mainly we require both very wide ($L_x = 90H_0$) and tall ($L_z \gg H_0$) boxes in order to accommodate a long-wavelength warp and a tilted disc, respectively. Furthermore, in order to be consistent with the shear-periodic boundary conditions of the shearing box, the vertical gravitational acceleration has to be periodic in the radial direction as well, i.e.~$g_z \propto \cos(k x)$. In reality apsidal precession is usually of one sign and depends monotonically on radius.\footnote{For example, in a disc around a black hole, the radial epicyclic frequency $\kappa(r)$ is systematically less than the angular frequency $\Omega(r)$.} Nevertheless, our model includes all the salient physics needed to study the impact of differential precession on a tilted disc, including inducing a warp (or twist) in the disc, disc bouncing, parametric instability, and even disc breaking.  

\subsection{Motivation and aims}
\label{INTRO_MotivationAndAims}
In this work we aim to study how differential precession in an accretion disc leads to disc breaking and parametric instability. We do so through hydrodynamic, quasi-2D (in the $xz$-plane), vertically stratified, isothermal, high-resolution shearing box simulations. The key novelty is the inclusion of precession by means of a radially varying vertical gravitational acceleration $g_z = -\Omega_0^2 z[1+f(x)]$. This allows us to systematically determine which precession amplitudes ($a$) and initial tilts ($z_0$) cause the disc to break, to become parametrically unstable, or both. 

The paper is organized as follows: in Section \ref{METHODS} we summarize the governing equations, numerical setup, and key diagnostics. In Section \ref{THEORY} we present a linear theory of bending  waves with precession and viscosity. In Section \ref{RESULTS_FIDUCIALSIMULATION} we describe our fiducial simulation of disc breaking with precession. To our knowledge this is the first shearing box simulation of disc breaking or tearing as a result of an imposed precession. In Section \ref{RESULTS_PARAMETERSTUDY} we present a parameter study, where we investigate the effects of varying the initial disc tilt and the precession amplitude. We also discuss select runs carried out in the \textit{diffusive regime} of warp propagation ($\alpha > H/r$), where $\alpha$ is the dimensionless effective viscosity. In Section \ref{RESULTS_GAUSSIANRINGSTUDY} we present a study in which we model interactions between broken disc segments. We discuss the astrophysical implications of our results in Section \ref{RESULTS_DISCUSSION}, and conclude in Section \ref{CONCLUSIONS}. Finally, in Appendix \ref{APPENDIX_TiltDissipationStudy} we include a study of tilted discs (without precession or warping) to determine whether grid-based numerical dissipation can artificially damp the tilt and cause the disc to precess.

\section{Methods}
\label{METHODS}

\subsection{Governing equations}
\label{METHODS_GoverningEquations}
We employ the standard shearing-box formalism in all our simulations. The computational domain consists of a Cartesian box located at a fixed cylindrical radius $r_0$ in the disc. The box orbits with the angular frequency $\Omega_0$ of the gas at that radius. The governing equations are given by 

\begin{align}
&\partial_t \rho + \nabla \cdot (\rho \mathbf{u}) = 0, \label{EQUATION_SB1}\\
&\partial_t \mathbf{u} + \mathbf{u}\cdot\nabla \mathbf{u} = -\frac{1}{\rho} \nabla P - 2\Omega_0 \mathbf{e}_z \times \mathbf{u} + \mathbf{g_\text{eff}}+\frac{1}{\rho}\nabla \cdot \mathbf{T}, \label{EQUATION_SB2}
\end{align}
where $\rho$ is the gas density, $P$ is the gas pressure, and $\mathbf{u}$ is the total velocity. The viscous stress tensor (present only in our diffusive regime runs where we include an explicit viscosity) is given by $\mathbf{T} \equiv 2\rho\nu \mathbf{S}$, where $\nu$ is the kinematic viscosity and $\mathbf{S} \equiv (1/2)[\nabla \mathbf{u}+(\nabla\mathbf{u})^\text{T}] - (1/3)(\nabla\cdot\mathbf{u})\mathbf{I}$ is the traceless shear tensor. We use an isothermal equation of state $P = c^2_{s0} \rho$ in all our simulations, where $c_{\text{s}0}$ is the constant sound speed.

All our simulations are stratified in the vertical direction. In a standard shearing box, the gravitational acceleration is given by the tidal expansion of the effective gravitational potential $\mathbf{g}_{\text{eff}} = 2q\Omega_0^2 x \mathbf{e}_x -\Omega_0^2 z \mathbf{e}_z$, where $q \equiv -\dd \ln\Omega/\dd \ln{r}|_{r=r_0}$ is the dimensionless shear parameter. In Section \ref{METHODS_PrecessionFormalismInTheShearingBox} we explain how we can modify this potential in order to include precession effects. In this paper we will assume the disc is Keplerian, for which $q = 3/2$.\footnote{In a differentially precessing disc there is, in fact, a deviation from Keplerian motion. However, the manner in which we implement precession (Section \ref{METHODS_PrecessionFormalismInTheShearingBox}), results in a spatially oscillatory apsidal precession on top of this Keplerian flow. In this model, there are as many radial locations in the disc where apsidal precession is prograde as where it is retrograde (see top panel of Fig.~\ref{FIGURE_ComplexTiltAndWarpAnalytical}). Thus in our simulations, the disc is Keplerian in the mean and we set $q = 3/2$ in all our simulations.}

\subsubsection{Including precession through the gravitational acceleration}
\label{METHODS_PrecessionFormalismInTheShearingBox}
As explained in the introduction, relative to an observer moving along an untilted shearing box orbit, a tilted disc appears as a series of fluid columns oscillating about the orbital reference plane $z = 0$ at the orbital frequency $\Omega_0$. If the disc is rigidly \textit{precessing} the fluid columns will oscillate in tandem about $z=0$ with some frequency different from the orbital one. If the disc is \textit{differentially precessing} each fluid column will oscillate about $z = 0$ at a frequency different both from the orbital frequency and from its neighbours. We can therefore mimic the influence of differential precession in the standard shearing box by introducing a sinusoidal spatial dependence to the vertical gravitational acceleration $g_z$.\footnote{As mentioned in the introduction, this sinusoidal dependence is required in order to be compatible with the shear periodic radial boundaries of the domain. In reality, the precession frequency is not periodic \citep{fragile2026tilted}.}

To include apsidal and nodal precession, we modify the standard effective gravitational potential in the shearing box, given by $\Phi_\text{eff} = -q\Omega_{0}^2x^2+\frac{1}{2}\Omega_{0}^2z^2$. We decompose the potential as $\Phi_\text{eff}(x,z) = \Phi_1(x) + \Phi_2(x,z)$. The effects of spatially sinusoidal precession are embodied in these component parts as

\begin{align}
 \Phi_1 &= -q\Omega_0^2x^2 +\frac{a}{k^2}\Omega_0^2 \cos{kx},
 \label{EQUATION_Phiprecession1}\\
   \Phi_2 &= \frac{1}{2}\Omega_0^2z^2(1 + b\cos{kx}),
   \label{EQUATION_Phiprecession2}
\end{align}
where $a$ is a dimensionless parameter corresponding to the amplitude of \textit{apsidal} precession, while $b$ corresponds to the amplitude of \textit{nodal} procession, and $k$ is the radial wavenumber with which the potential varies. We take $k=2\pi/L_x$ in all our simulations, where $L_x$ is the radial size of the domain. The first two terms in each expression correspond to the usual gravitational potential in the shearing box (i.e.~without precession effects). The second terms in each equation correspond to apsidal and nodal precession, respectively. 

The inclusion of apsidal precession in this manner leads to a radial force which gives rise to a zonal flow, and results in a departure from Keplerian rotation. A complementary view is that the inclusion of periodic apsidal precession changes the epicyclic frequency in a spatially oscillatory manner. Indeed, the radial and vertical epicyclic frequencies are then given by $\kappa^2 = \Omega_0^2(1-a\cos{kx})$ and $\Omega_z^2 = \Omega_0^2(1+b\cos{kx})$, respectively.\footnote{Suppose that we enhance vertical gravity (by taking $b>0$). The vertical oscillation will occur slightly faster than the orbital frequency $\Omega_0$, so a particle returns to the mid-plane $z=0$ slightly early. This causes the orbit to precess backwards, hence the change in sign between $\omega_\text{a}$ and $\omega_\text{n}$. In the inner parts of a disc surrounding a black hole, on the other hand, precession is relativistic in origin, apsidal precession is faster than nodal, and they are both in the same direction. To model this we would take $a < 0$, $b > 0$, and $|a|> |b|$.}  The apsidal and nodal precession amplitudes $a$ and $b$, can be related to the more familiar apsidal and nodal precession \textit{frequencies} (in units of the disc angular frequency $\Omega_0$) through $\omega_\text{a} / \Omega_0 = (a/2)\cos{kx}$ and $\omega_\text{n} / \Omega_0 = -(b/2)\cos{kx}$, respectively. 

We can then obtain the components of the effective gravitational acceleration by taking the gradient of the potential, i.e.~$g_x = -\partial_x \Phi_\text{eff}$ and $g_z = -\partial_z \Phi_\text{eff}$, to obtain

\begin{align}
    g_x &= 2q\Omega_0^2x + \frac{\Omega_0^2 a}{k}\sin{k x} + \frac{1}{2}\Omega_0^2 z^2 b k \sin{k x},
    \label{EQUATION_gxprecession}\\
    g_z &= -\Omega_0^2 z(1+b\cos{k x}).
    \label{EQUATION_gzprecession}
\end{align}
Note that the gravitational force is conservative, as can be verified by checking that $\nabla \times \mathbf{g} = \mathbf{0}$. From the term $b\cos{k x}$ in $g_z$, we see that precession is strongest at $x/H_0 = -L_x/2, 0, L_x/2$, and zero at $x/H_0 = -L_x/4, L_x/4$. We plot a radial profile of $g_z$ and of the corresponding nodal precession frequency in the top panel of Fig.~\ref{FIGURE_ComplexTiltAndWarpAnalytical}.
 
Finally, in this paper we assume that nodal and apsidal precession are equal and opposite, i.e.~we set the precession amplitudes to be $b = a$, and take $a > 0$. This assumption holds for a non-self-gravitating disc tilted with respect to the symmetry plane of a non-spherical Newtonian potential, e.g.~in the inner parts of a circumbinary disc, or the outer parts of a circumstellar disc surrounded by a binary companion. Thus in our simulations we will control the amplitude of precession through a single parameter, the dimensionless precession amplitude $a$, which is related to the (nodal) precession frequency by $\omega_\text{n}(x)/\Omega_0 = -(a/2)\cos{kx}$.

\subsection{Numerical set-up}
\label{METHODS_NumericalSetup}

\subsubsection{Code}
\label{METHODS_Code}
All our simulations were run using the finite-volume code \textsc{PLUTO} (version 4.4, patch 2) \citep{mignone2007}. The solution is advanced in time using the 2nd-order-in-time Runge--Kutta algorithm. In addition, we employ the HLLC Riemann solver and 3rd-order-in-space parabolic reconstruction.\footnote{We tested other solvers and interpolation methods in our previous paper: see Appendix A2 of \cite{held2026disc}.} Reconstruction is carried out using characteristic variables, and we employ the shock flattening algorithm, in which the code reverts to the minmod limiter and HLL Riemann solver whenever a shock is detected \citep{plutocode_userguide}. We implement boundary conditions using ghost cells which border the active domain. When we include anexplicit viscosity $\nu$, we also employ the Super-Time-Stepping (STS) scheme \citep{alexiades1996super}, which reduces the computational time. We employ the built-in shearing box module in PLUTO \citep{mignone2012}, which evolves the primitive-variable equations (equations \ref{EQUATION_SB1}--\ref{EQUATION_SB2}) in conservative form.

\subsubsection{Units}
\label{METHODS_Units}
The time unit is given by the inverse angular frequency $\Omega_0$, speeds are measured in units of the isothermal sound speed $c_{\text{s}0}$, and the mass unit is given by the initial mid-plane density $\rho_0$. All other units are formed from combinations of these base units, e.g.~lengths are measured in units of the initial scale-height $H_0 = c_{\text{s}0}/\Omega_0$.

\subsubsection{Box size, density floor, and resolution}
\label{METHODS_BoxSizeDensityFloorResolution}
To remain as faithful as possible to the free (unforced) warp simulations in \cite{held2026disc}, we employ the same box size for most of our simulations, mainly a 2D (in  $x$- and $z$-) box of size $[L_x, L_y,L_z]=[90, 0.04, 68]H_0$ in the radial, azimuthal, and vertical directions, respectively. For this domain size we use a resolution of $[N_x,N_y,N_z]=[2880, 1, 2176]$, which corresponds to $32/H_0$ cells-per-scale-height $H_0$ in both the $x-$ and $z$-directions. Other resolutions and box sizes can be found in Appendix \ref{APPENDIX_TablesOfSimulations}. We employ a density floor of $\rho_{\text{floor}} = 1.3\times 10^{-14} \rho_0$ in all our simulations.

\subsubsection{Boundary conditions}
\label{METHODS_BoundaryConditions}
We employ standard shear-periodic boundary conditions in the radial direction, and periodic boundary conditions in the $y$-direction. In the vertical direction (zBCs) we use \textit{reflective} boundary conditions, $\rho \rightarrow \rho, u_z \rightarrow -u_z, u_t \rightarrow u_t$, where $t$ denotes the transverse (i.e.~the $x$- or $y$-) components of the velocity. (In \cite{held2026disc} we also tried outflow zBCs (see Appendix A3), but the results were very similar to those obtained using reflective zBCs.)

\subsubsection{Initial conditions for tilted disc simulations}
\label{METHODS_InitialConditionTiledDiskSims}
In Appendix \ref{APPENDIX_TiltDissipationStudy} we carry out a study of how numerical dissipation in our grid-based code damps the tilt in the disc, and whether it can induce an artificial precession (or even a twist) in the disc. We initialize these tilted-disc simulations using a Gaussian density profile of the form

\begin{equation}
    \rho = \rho_0\exp\left(-\frac{(z-z_0)^2}{2H_0^2}\right),
    \label{EQUATION_InitialDensityProfileTiltedDiskSims}
\end{equation}
where $z_0$ is a parameter that we can use to set the initial tilt of the disc (e.g.~$z_0 = 10H_0$ means the disc is initially displaced by $10H_0$ from the vertical centre of the domain $z = 0$, which corresponds to the orbital plane of the shearing box).

The background azimuthal velocity is given by its usual value of $v_{y0} = -(3/2)\Omega_0 x$. We also perturb each velocity component with random noise at initialization to seed any instabilities that might be present. Note that we do not include any imposed precession in these simulations.

\subsubsection{Initial conditions for precession simulations}
\label{METHODS_InitialConditionPrecessionSims}
In our previous work on disc breaking without precession, we studied the evolution of a \text{free} (unforced) warp, which was initialized using a sinusoidal density profile (see equation 3 of \cite{held2026disc}). Here, we instead start with a \textit{flat}, tilted disc, which will become warped (and twisted) due to the imposed precession. The extra precession terms in the gravitational potential (see equations \ref{EQUATION_Phiprecession1}-\ref{EQUATION_Phiprecession2}) give rise to the following density profile

\begin{equation}
    \rho = \rho_0\exp\left(-\frac{\Omega_z^2(x) (z-z_0)^2}{2H_0^2}\right),
    \label{EQUATION_InitialDensityProfilePrecessionSims}
\end{equation}
where the effects of precession are included in the dimensionless parameter $\Omega_z^2(x) = 1 + a\cos kx$, and the initial tilt of the disc is given by $z_0$ (in units of the initial mid-plane scale-height $H_0$).  

An untilted disc corresponds to $z_0/H_0 = 0$. To satisfy hydrostatic equilibrium we must also modify the initial background azimuthal velocity to

\begin{equation}
    u_{y0} = -\frac{3}{2}\Omega_0 x-\frac{a\Omega_0}{2k_x}\sin kx
    \label{EQUATION_InitialBackgroundAzVelProfilePrecessionSims}.
\end{equation}
To seed any instabilities that may be present, we perturb all three velocity components at initialization with random noise (we use different random noise for each component). The maximum amplitude of the perturbations $\delta \mathbf{u}$ is around $0.05 c_{\text{s}0}$. 

When $z_0 / H_0 = 0$ the disc is untilted and equations \ref{EQUATION_InitialDensityProfilePrecessionSims}--\ref{EQUATION_InitialBackgroundAzVelProfilePrecessionSims} correspond to an equilibrium. We have verified this by running a simulation (Precession10 in Table \ref{TABLE_PrecessionStudy}) for 100 orbits without any vertical displacement. The disc remained at the mid-plane in this case, and did not develop a warp or a twist.

There are several important differences between the set-up we use in this paper to model forced or driven warps under the action of differential precession, and the one we used in \cite{held2026disc} to model the evolution of a free warp. In the latter case we initialized a warp, but not horizontal sloshing motions, nor did we include any mechanism for sustaining the warp. The disc was \textit{not} in hydrostatic equilibrium by design, and so the warp could evolve freely. This led to the immediate growth of sloshing motions (due to the initial radial pressure imbalance), and ultimately to damping of the warp due to shocks or hydrodynamic instabilities. In our precession simulations, on the other hand, we instead start with a tilted (but \textit{flat}) disc, which would be in hydrostatic equilibrium if the tilt were zero. If it is not, the imposed precession will drive a warp in the disc, which will in turn force horizontal sloshing motions.

\subsubsection{Initial conditions for study of interactions between broken disc segments}
\label{METHODS_InitialConditionGaussianRingSims}
In both the free warp simulations described in \cite{held2026disc}, as well as the precession simulations described later on in this paper, we observe considerable interaction between broken disc segments or rings. To investigate the interaction between these rings in more detail, we carry out simulations in which we initialize only two mutually tilted rings. Motivated by the structure of some of the broken disc segments that develop in our simulations with precession (see the bottom-right-hand panel of Fig.~\ref{FIGURE_FiducialSimulationFlowField}), we initialize the rings with a Gaussian density profile in both the vertical and radial directions, given by $\rho = \rho_\text{L} + \rho_\text{R}$, where

\begin{align}
   &\rho_\text{L} = \rho_0\exp\left(-\frac{(x+(1/2)L)^2}{2w^2}-\frac{z^2}{2H_0^2}\right),
\label{EQUATION_GaussianRingsDensityProfile1}\\
   &\rho_\text{R} =\rho_0\exp\left(-\frac{(x-(1/2)L)^2}{2w^2}-\frac{(z-z_0)^2}{2H_0^2}\right).
\label{EQUATION_GaussianRingsDensityProfile2}
\end{align}
Here $L$ is the radial separation between the two rings and $z_0$ is the initial tilt (relative to $z = 0$) of the the right-hand ring. (Note that we always initialize the left-hand ring with zero tilt.) The initial radial thickness (radial scaleheight) of the rings is given by $w$, which we fix at $w = 3H_0$ in all our simulations.

The radial variation in density implies that there is a radial pressure gradient as well. This and the associated zonal flow result in a deviation from Keplerian flow. To maintain hydrostatic balance in the radial direction, we need to modify the background azimuthal velocity to

\begin{align}
    &u_{y0} = -\frac{3}{2}\Omega_0 x - \frac{1}{2}\left(\frac{H_0}{w}\right)^2 \Omega_0 (x+(1/2)L)\,\,\,\,\,\,(x < 0),\\
    &u_{y0} = -\frac{3}{2}\Omega_0 x - \frac{1}{2}\left(\frac{H_0}{w}\right)^2 \Omega_0 (x-(1/2)L)\,\,\,\,\,\,(x > 0).
\end{align}
This results in a discontinuity of the $y$-velocity at $x=0$, where the solutions do not smoothly overlap. The discontinuity is subsonic for $L=10H_0$ and $L=15H_0$, and transonic for $L=20H_0$. We also perturb each velocity component with random noise at initialization to seed any instabilities that might be present. 

Finally, we do \textit{not} include an imposed precession in any of these broken disc segment simulations: the gravitational acceleration is given by the usual form: $\mathbf{g}_{\text{eff}} = 3\Omega_0^2 x \mathbf{e}_x -\Omega_0^2 z \mathbf{e}_z$.

\subsection{Diagnostics}
\label{METHODS_Diagnostics}
In \cite{held2026disc} we presented many novel diagnostics for analysing the evolution of a warped disc in the shearing box. We reproduce these here for the reader's convenience.

\subsubsection{Energetics}
\label{METHODS_Energetics}
The total energy density is given by

\begin{equation}
    E_{\text{total}} = \frac{1}{2}\rho u^2 + \rho \Phi + c_{\text{s}0}^2\rho\ln{\rho}. 
    \label{EQUATION_TotalEnergy}
\end{equation}
Here $ \frac{1}{2}\rho u^2$ is the total kinetic energy density $E_{\text{kin}}$, $\Phi$ is the effective gravitational potential in the shearing-box approximation for a Keplerian disc, given either by its standard form of $\Phi=-\frac{3}{2}\Omega_{0}^2x^2+\frac{1}{2}\Omega_{0}^2z^2$ in the absence of precession, or by equations (\ref{EQUATION_Phiprecession1})--(\ref{EQUATION_Phiprecession2}) if we include precession. The term $c_{\text{s}0}^2 \rho \ln{\rho}$ embodies the variable part of the Helmholtz free
energy, which replaces the internal energy in an isothermal gas (see footnote 5 in \cite{held2024instability} for a brief derivation).

\subsubsection{Averaged quantities}
\label{METHODS_Averages}

The volume-average of a quantity $X$ is denoted $\langle X \rangle$ and is defined as 
\begin{equation}
\langle X \rangle(t) \equiv \frac{1}{V} \int_V X(x, y, z, t) \, \dd V,
\label{EQUATION_VolumeAverageDefinition}
\end{equation}
where $V=L_x L_y L_z$ is the volume of the box. 

The orbit-average of a quantity $X$ is denoted as $\langle X \rangle_{\tau}$ and is defined as

\begin{equation}
\langle X \rangle_{\tau} \equiv \frac{1}{\tau} \int X(x,y,z,t)\,\dd t,
\label{EQUATION_OrbitAverageDefinition}
\end{equation}
where $\tau = 2\pi/\Omega_0$ is an orbital period. Note that averaging quantities over one orbit in our local simulations is equivalent to carrying out an azimuthal average in a global simulation. \footnote{For various diagnostics defined below, the notation $\langle X \rangle_{\tau}$ will also imply an additional mass-weighted average $(1/\Sigma)\int \rho X\,\,\dd z$ or a vertical average $ \langle X \rangle_z \equiv (1/L_z)\int X\,\,\dd z$, and possibly multiplication by $\ee^{\ii\Omega_0 t}$.} 

\subsubsection{Surface density, centre of mass, and disc thickness}
\label{METHODS_SurfaceDensityCOMDiskThickness}
The profile of the surface density at time $t$ (in units of $\rho_0 H_0$) is defined as:

\begin{equation}
    \Sigma(x,t) \equiv \int_{-L_z/2}^{L_z/2}  \rho\,\,\dd z.
    \label{EQUATION_SurfaceDensity}
\end{equation}
The orbit-averaged radial profile of surface density is $\langle \Sigma \rangle_{\tau}(x) \in \mathbb{R}$.

The instantaneous radial profile of the vertical centre-of-mass coordinate (in units of $H_0$) is given by

\begin{equation}
    \bar{z}(x,t) \equiv \frac{1}{\Sigma(x,t)}\int_{-L_z/2}^{L_z/2}  \rho z\,\,\dd z.
    \label{EQUATION_COM}
\end{equation}

Warped discs are expected to bounce twice per orbit \citep{ogilvie2013local, fairbairn2021non}, so we expect the thickness of the disc to change in time. An intrinsic measure of disc thickness is the dynamical scale-height $H(x,t)$ (in units of $H_0$) defined as:
\begin{equation}
    H^2(x,t) = \frac{1}{\Sigma(x,t)}\int_{-L_z/2}^{L_z/2}  \rho (z-\bar{z})^2 \dd z.
     \label{EQUATION_DiskThickness}
\end{equation}
This definition differs slightly from that used in \cite{held2024instability}. There $H^2(t) \equiv (1/\Sigma)\int \langle \rho \rangle_{x} z^2 \dd z$, where $\langle \rho \rangle_{x}$ denotes a radial average, which made sense for a purely bouncing disc for which we did not expect $\bar z\ne0$. For our warped disc set-up here, it is more prudent to use $z-\bar z$ (in order to pick up the location of the local disc mid-plane), and to calculate a radial profile (because the thickness varies with radius as well as time).

\subsubsection{Reynolds stress}
\label{METHODS_ReynoldsStress}
In purely hydrodynamic discs, the radial transport of angular momentum is related to the $xy$-component of the Reynolds stress tensor $R_{xy} \equiv \rho u_x \delta u_y$, where $\delta u_y \equiv u_y + q\Omega_0 x$ is the perturbation of the $y$-component of the total velocity $u_y$ about the background Keplerian flow $u_{y0} = -(3/2) \Omega_0 x$. The Reynolds stress is related to the classical dimensionless angular momentum transport parameter $\alpha$ by $\alpha \equiv \langle R_{xy} \rangle / \langle P \rangle$.

The orbital- and vertically-averaged radial profile of the $xy$-component of Reynolds stress is defined as

\begin{equation}
    \langle R_{xy} \rangle_{\tau}(x) \equiv \left\langle \frac{1}{L_z}\int \rho u_x \delta u_y\,\, \dd  z\right\rangle_{\tau},
    \label{EQUATION_ReyonldsStress}
\end{equation}
where $\langle R_{xy} \rangle_{\tau} \in \mathbb{R}$ and has units of $\rho_0 c_{\text{s}0}^2$.

\subsubsection{Complex tilt}
\label{METHODS_ComplexTilt}
Following \cite{ogilvie2022hydrodynamics}, we can define the local equivalent of a complex tilt, $Z = (z+\ii u_z/\Omega_0)\ee^{\ii\Omega_0t}$.

This quantity can be thought of as the complex amplitude of a harmonic vertical oscillation and uses information from the vertical position and velocity. The mass weighted and orbit-averaged oscillation amplitude is then

\begin{equation}
   \langle  Z \rangle_{\tau} = \left\langle \frac{1}{\Sigma} \int \rho Z\,\,\dd z \right\rangle_{\tau},
   \label{EQUATION_ComplexTilt}
\end{equation}
where $\langle Z \rangle_{\tau} \in \mathbb{C}$ and has units of $H_0$. Note that in the literature,  the tilt is often defined in terms of a tilt vector $\mathbf{l} = [l_x, l_y, l_z] = [\cos{\gamma}\sin{\beta}, \sin{\gamma}\sin{\beta}, \cos{\beta}]$, where $\beta(r,t)$ and $\gamma(r,t)$ are the tilt and twist angles, respectively \citep{nixon2016warp}. The horizontal (misaligned) components of the tilt vector can then be combined into a complex tilt vector $l_x+\ii l_y$, and there is a correspondence between this quantity and its local equivalent via $-Z/r_0 \leftrightarrow l_x + \ii l_y = \sin{\beta} \, \ee^{\ii\gamma}$, where $r_0$ is the radius at which the local model is constructed.

\begin{figure}
    \centering
    \includegraphics[width=1.0\linewidth]{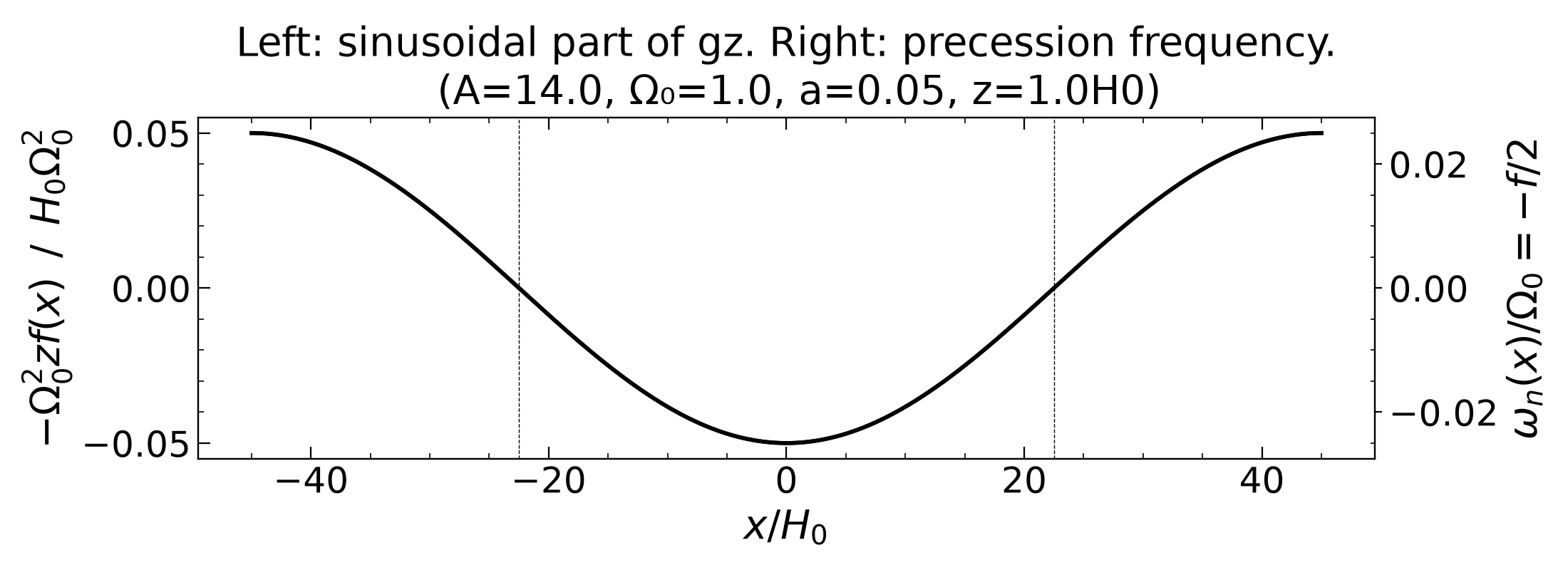}
    \includegraphics[width=1.0\linewidth]{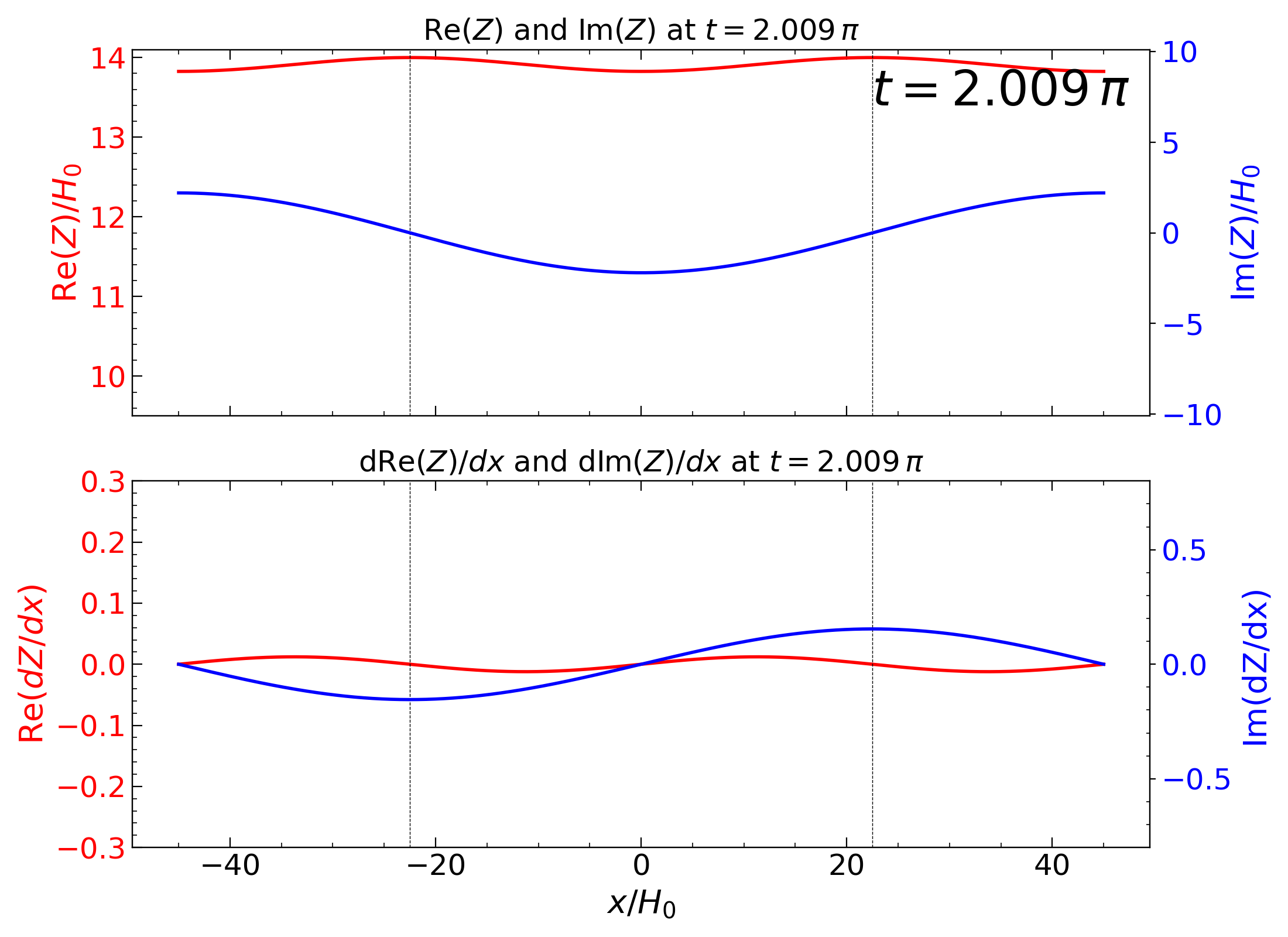}\\
    \includegraphics[width=1.0\linewidth]{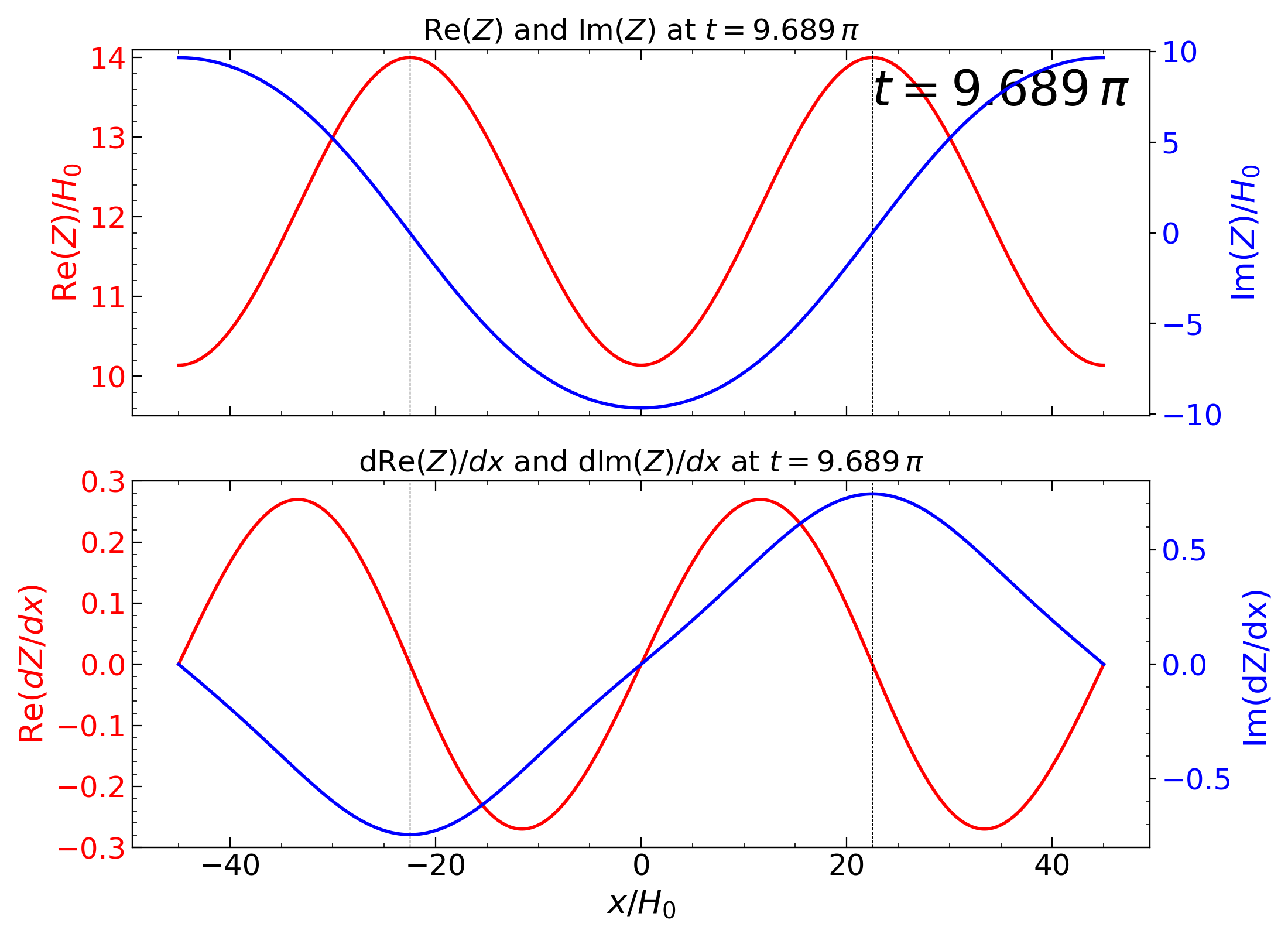}\\
    \caption{Top panel: radial profile of the oscillatory part of the vertical gravitational acceleration $g_z = -\Omega^2_0z[1+f(x)]$ (in units of $H_0\Omega_0^2)$, where $f(x) = b\cos{k x}$ (see equation \ref{EQUATION_gzprecession}) at $z = H_0$ for a precession amplitude of $a = 0.05$. (We have taken nodal and apsidal precession to be equal and opposite such that $b=a$.) The right-hand axis shows the corresponding radial profile of the nodal precession frequency in units of $\Omega_0$, $\omega_\text{n}(x)/\Omega_0 = -(f/2)$. The remaining panels show radial profiles of the theoretical value of the real (in red) and imaginary (in blue) components of the complex tilt $Z = A \exp[\ii\omega_\text{n}(x) t]$ that would develop in the absence of gas pressure, and of their radial derivatives, at two different times: $t = 2.009\pi\Omega_0^{-1}$ (orbit 1) and $t = 9.689\pi \Omega_0^{-1}$ (orbit 4.8). The initial amplitude (tilt) was taken to be $A = 14H_0$. The dashed vertical lines mark the locations where the imposed precession is zero.}
    \label{FIGURE_ComplexTiltAndWarpAnalytical}
\end{figure}

If we ignore pressure effects, a column of fluid at each $x$ just oscillates at its local vertical frequency about the shearing box orbital plane $z/H_0 = 0$. This will give rise to something equivalent to differential precession, which then appears as an evolving phase of the complex tilt variable $Z$ and an increasing warp. In Fig.~\ref{FIGURE_ComplexTiltAndWarpAnalytical} we show analytical profiles of $Z = A\exp{[\ii \omega_\text{n}(x) t]}$ and its radial derivative (which is related to the warp) at two separate times $t = 2\pi\,\Omega_0^{-1}$ (two  middle panels) and $t = 10\pi\,\Omega_0^{-1}$ (two bottom panels). The sinusoidal part of the vertical component of the gravitational acceleration $g_z$ (equation \ref{EQUATION_gzprecession}) is shown in the top panel for reference. 

\subsubsection{Complex warp amplitude}
\label{METHODS_ComplexWarpAmplitude}
The orbit-averaged complex warp amplitude can be defined in terms of the complex tilt $Z$ as

\begin{equation}
    \langle \psi \rangle_{\tau}(x) = \left\langle -\frac{\partial}{\partial x}\left(\frac{1}{\Sigma} \,\int \rho Z \,\,\dd z \right) \right\rangle_{\tau},
    \label{EQUATION_ComplexWarpAmplitude}
\end{equation}
where $\langle \psi \rangle_{\tau} \in \mathbb{C}$ and is dimensionless. The correspondence between this quantity and the warp amplitude in cylindrical coordinates is $(\psi/r_0)\,\ee^{\ii\gamma} \leftrightarrow (l_x + \ii l_y)' = (\beta'\cos{\beta} + \ii\gamma'\sin{\beta})\,\ee^{\ii\gamma}$, where here (and only here) the prime denotes a radial derivative.

\subsubsection{Complex torque}
\label{METHODS_ComplexTorque}
The orbit-averaged complex `torque' is defined as

\begin{equation}
 \langle G \rangle_{\tau}(x) =  \left\langle \int  -\rho u_x Z\,\, \dd z \right\rangle_{\tau},
 \label{EQUATION_ComplexTorque}
 \end{equation}
 where $\langle G \rangle_{\tau} \in \mathbb{C}$ and has units of $\rho_0 H_0^{3} \Omega_0$. We associate this quantity with the radial flux of the misaligned ($x$ and $y$) components of angular momentum. Note that $\rho u_x Z$ itself has units of $\rho_0 H_0^2 \Omega_0$, while an angular momentum flux (or torque) has units of $\rho_0 H_0^5 \Omega_0^2$. However, the quantity $\int -\rho u_x Z \dd z \cdot 2\pi r_0 \cdot r_0 \Omega_0$ does have the same dimensions as an angular momentum flux or torque (units of $\rho_0 H_0^5 \Omega_0^2$), so, for simplicity, we simply refer to equation~(\ref{EQUATION_ComplexTorque}) as a complex torque.

\subsubsection{Amplitude of horizontal sloshing motions}
The warp gives rise to a vertical shear $u_x(z)$ and $\delta u_y(z)$, also referred to as horizontal `sloshing' motions \citep{dullemond2022equations}. These sloshing motions play an important role in mediating the torque. Since the sloshing motions consist of a horizontal velocity proportional to the height above the warped mid-plane, the orbit-averaged amplitudes of these motions can be defined as

\begin{equation}
    \langle U_x \rangle_{\tau}(x) \equiv \left\langle \left(\frac{1}{\Sigma H^2} \int \rho u_x (z-\bar{z})\,\,\dd z\right)\ee^{\ii\Omega_0 t}  \right\rangle_{\tau},
    \label{EQUATION_ComplexShearAmplitudeXcomponent}
\end{equation}
\begin{equation}
    \langle U_y \rangle_{\tau}(x) \equiv \left\langle \left(\frac{1}{\Sigma H^2} \int \rho \, \delta u_y (z-\bar{z})\,\,\dd z \right)\ee^{\ii\Omega_0 t} \right\rangle_{\tau},
     \label{EQUATION_ComplexShearAmplitudeYcomponent}
\end{equation}
where $U_x,U_y \in \mathbb{C}$ and have units of $\Omega_0$.\footnote{There is a typo in the equations $14$-$15$ in \cite{held2026disc} for $U_x$ and $U_y$ : the integrand should contain a factor of $(z-\bar{z})$, not $(z-\bar{z})^2$.}

\section{Theory}
\label{THEORY}

\subsection{Governing equations}
\label{THEORY_GoverningEquations}
Consider axisymmetric motions of an inviscid, isothermal gas in a Keplerian shearing box. The governing equations, equivalent to equations~(\ref{EQUATION_SB1}) and~(\ref{EQUATION_SB2}), are
\begin{align}
  &\DD u_x-2\Omega_0u_y=-\p_x(\Phi_\text{eff}+h),\label{theory1}\\
  &\DD u_y+2\Omega_0 u_x=0,\label{theory2}\\
  &\DD u_z=-\p_z(\Phi_\text{eff}+h),\label{theory3}\\
  &\DD h=-c_{\text{s}0}^2(\p_xu_x+\p_zu_z),\label{theory4}
\end{align}
where $\Phi_\text{eff}(x,z)=\Phi_1(x) + \Phi_2(x, z) = \Phi(x)+\f{1}{2}z^2\Psi(x)$ is the effective potential described in Section \ref{METHODS_PrecessionFormalismInTheShearingBox} (for convenience we separate $\Phi_2$ into a part that is $\propto z^2$ and a part $\Psi(x) = \Omega_0^2(1+b\cos{kx})$ that depends only on $x$), the enthalpy per unit mass of an isothermal gas is given by $h=c_{\text{s}0}^2\ln\rho$ \citep{held2024instability}, and $\DD=\p_t+u_x\p_x+u_z\p_z$ is the Lagrangian time-derivative for axisymmetric motions. The basic state we consider (which has density and pressure independent of~$x$ in the midplane $z=0$) is
\begin{equation}
  h_0=\cst-\f{1}{2}z^2\Psi(x),\qquad u_{y0}=\f{1}{2\Omega_0}\f{\dd\Phi}{\dd x},
  \label{EQUATION_TheoryNoDissipationBackgroundState}
\end{equation}
as well as $u_{x0}=u_{z0}=0$,
corresponding to an $x$-dependent hydrostatic equilibrium with a Gaussian vertical distribution of density and pressure with scale-height $c_{\text{s}0}/\Omega_z(x)$, where $\Omega_z^2=\Psi$.

\subsection{Linearized equations}
\label{THEORY_LinearizedEquations}
Small perturbations (denoted by a prime) from the basic state satisfy the linearized equations
\begin{align}
  &\p_tu'_x-2\Omega_0u'_y=-\p_xh',\label{linearized1}\\
  &\p_tu'_y+\f{\kappa^2}{2\Omega_0} u'_x=0,\label{linearized2}\\
  &\p_tu'_z=-\p_zh',\label{linearized3}\\
  &\p_th'-\f{1}{2}z^2\f{\dd\Omega_z^2}{\dd x}u_x'-\Omega_z^2zu'_z=-c_{\text{s}0}^2(\p_xu'_x+\p_zu'_z),\label{linearized4}
\end{align}
where $h'=p'/\rho=c_{\text{s}0}^2\rho'/\rho$ is the Eulerian enthalpy perturbation and $\kappa^2=4\Omega^2+\f{\dd^2\Phi}{\dd x^2}$. These are similar to equations 21--24 of \citet{held2026disc} but include the effects of the modified potential.

\subsection{Vertical structure}
\label{METHODS_VerticalStructure}
The second term on the left-hand side of equation~(\ref{linearized4}) involves the radial enthalpy gradient (which vanishes in the midplane) and the radial velocity perturbation. If this term can be neglected, then equations (\ref{linearized1})--(\ref{linearized4}) admit solutions of the form
\begin{align}
  &u_x'=\hat u_x(x,t)z,\\
  &u_y'=\hat u_y(x,t)z,\\
  &u_z'=\hat u_z(x,t),\\
  &h'=\hat h(x,t)z,
\end{align}
appropriate for describing tilts, warps, bending waves and sloshing motions, provided that
\begin{align}
  &\p_t\hat u_x-2\Omega_0\hat u_y=-\p_x\hat h,\label{linearized1z}\\
  &\p_t\hat u_y+\f{\kappa^2}{2\Omega_0}\hat u_x=0,\label{linearized2z}\\
  &\p_t\hat u_z=-\hat h,\label{linearized3z}\\
  &\p_t\hat h-\Omega_z^2\hat u_z=-c_{\text{s}0}^2\p_x\hat u_x.\label{linearized4z}
\end{align}
The neglected term is then $\propto z^3$ rather than $\propto z$ and so is incompatible with this separation of variables. We will check \textit{a posteriori} that it is relatively small.

\subsection{Modified tilt mode}
\label{METHODS_ModifiedTiltMode}
In a standard shearing box without the modified potential, these equations admit a solution independent of $x$ and involving only $\hat u_z$ and $\hat h$, representing a uniform vertical translation of the disc that oscillates at the orbital frequency. This solution corresponds to an equilibrium disc viewed from a slightly misaligned reference orbit, or equivalently to a disc with a slight uniform tilt.

When the potential is modified and the precessional effects are present, this tilt mode is also modified and becomes non-uniform. In our set-up, which has equal amounts of positive and negative precession, the modified tilt mode still oscillates at the orbital frequency. We therefore seek a solution of the form
\begin{align}
  &\hat u_x=\check u_x(x)\sin\Omega_0t,\\
  &\hat u_y=\check u_y(x)\cos\Omega_0t,\\
  &\hat u_z=\check u_z(x)\sin\Omega_0t,\\
  &\hat h=\check h(x)\cos\Omega_0t,
\end{align}
which means that
\begin{align}
  &\Omega_0\check u_x-2\Omega_0\check u_y=-\p_x\check h,\label{linearized1zt}\\
  &-\Omega_0\check u_y+\f{\kappa^2}{2\Omega_0}\check u_x=0,\label{linearized2zt}\\
  &\Omega_0\check u_z=-\check h,\label{linearized3zt}\\
  &-\Omega_0\check h-\Omega_z^2\check u_z=-c_{\text{s}0}^2\p_x\check u_x.\label{linearized4zt}
\end{align}
Eliminating $\check u_y$ and $\check u_z$ leads to
\begin{align}
  &(\Omega_0^2-\kappa^2)\check u_x=-\Omega_0\p_x\check h,\\
  &(\Omega_z^2-\Omega_0^2)\check h=-c_{\text{s}0}^2\Omega_0\p_x\check u_x
\end{align}
and so to
\begin{align}
  &\Omega_0a\cos kx\,\check u_x=-\p_x\check h,\\
  &\Omega_0b\cos kx\,\check h=-c_{\text{s}0}^2\p_x\check u_x.
\end{align}
The term that was dropped in the previous subsection can now be compared in magnitude to the retained term $\propto c_{\text{s}0}^2\p_x\check u_x$. Assuming that $\check u_x$ and $\Omega_z^2$ vary on a similar lengthscale, the neglected term is smaller by a factor of order $b$.

In terms of the variable
\begin{equation}
  s=\sqrt{ab}\f{\Omega_0}{c_{\text{s}0}k}\sin kx,
\end{equation}
we find
\begin{equation}
  \f{\dd^2\check h}{\dd s^2}=\check h
\end{equation}
and a similar equation\footnote{If $a$ and $b$ are taken to have opposite signs, then $s$ becomes imaginary and the hyperbolic functions becomes trigonometric ones.} for~$\check u_x$. The modified tilt mode corresponds to the solution
\begin{equation}
  \check h=A\Omega_0^2\cosh s,\qquad
  \check u_x=-\sqrt{\f{b}{a}}\f{A\Omega_0^2}{c_{\text{s}0}}\sinh s,
\end{equation}
where~$A$ is a tilt amplitude (in units of $H_0$).

In the absence of precession, $s=0$ and $\check h$ is independent of $x$, corresponding to a uniform tilt of amplitude~$A$, while the sloshing amplitude $\check u_x$ vanishes. 

If $\sqrt{ab}\,\Omega_0\lesssim c_{\text{s}0}k$,
which means that precession is slow compared to bending-wave propagation,
then the range of~$s$ is small and the $\cosh$ function does not differ greatly from unity, meaning that the motion involves a uniform tilt plus a small warp. In this regime the solution can be approximated using 
\begin{align}
  &\cosh s\approx1+\f{1}{2}s^2=1+\f{ab\Omega_0^2}{2c_{{\text s}0}^2k^2}\sin^2kx,\\
  &\sinh s\approx s=\sqrt{ab}\f{\Omega_0}{c_{\text{s}0}k}\sin kx,
\end{align}
meaning that the sloshing amplitude is 
\begin{equation}
  \check u_x\approx-\f{bA\Omega_0^3}{c_{\text{s}0}^2k}\sin kx.
\end{equation}
Note that this has dimensions of frequency, i.e.~it is a sloshing rate or shear rate. This approximate solution can be interpreted as follows. The leading approximation to~$\check h$ is a constant, corresponding to a uniform tilt. The tilted disc is subject to differential nodal precession ($b$). In order for the tilted disc \emph{not} to precess nodally (i.e.\ for the mode to have a frequency of exactly $\Omega_0$), an internal torque distribution is required to compensate for the differential nodal precession. The sloshing motion ($\check u_x\propto b$) provides this internal torque. In turn, the sloshing motion must be driven by the pressure gradients associated with a small warp. This is described by the non-uniform part of $\check h$, which is $\propto a$ as well as $\propto b$ because the sloshing motion is detuned from resonance by apsidal precession ($a$).

\subsection{Slow warp dynamics}
\label{METHODS_SlowWarpDynamics}

More generally, evolving solutions of the warp dynamics in a linear regime can be described using the complex tilt and torque variables $Z(x,t)$ and $G(x,t)$. The relevant evolutionary equations for our system are [cf.\ equations 46--47 of \citet{held2026disc}]
\begin{align}
  &\frac{\partial Z}{\partial t}=\ii\omega_\text{n}Z+\f{1}{\Sigma}\frac{\partial G}{\partial x},\\
  &\frac{\partial G}{\partial t}=\ii\omega_\text{a}G-\alpha\Omega_0G+\frac{1}{4}\Sigma c_{\text{s}0}^2\frac{\partial Z}{\partial x},
\end{align}
where $\omega_\text{a}=\f{1}{2}\Omega_0a\cos kx$ and $\omega_\text{n}=-\f{1}{2}\Omega_0b\cos kx$ are the apsidal and nodal precession rates and we have also included viscosity. (Here we regard $\Sigma$ as being independent of $x$, even though it is not exactly so in the actual system.) The modified tilt mode described above corresponds to the stationary solution $Z\propto\cosh s$ in the inviscid case.

\begin{figure*}
\centering
\includegraphics[scale=0.36]{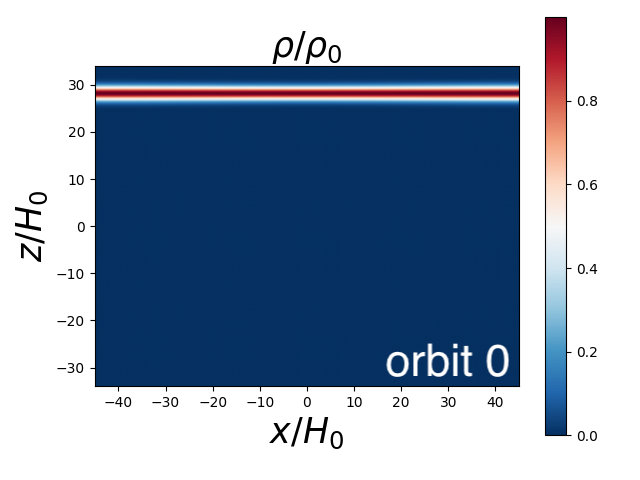}
\includegraphics[scale=0.36]{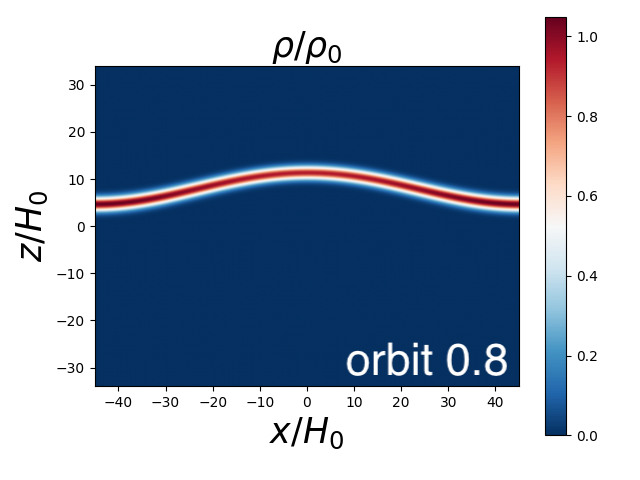}
\includegraphics[scale=0.36]{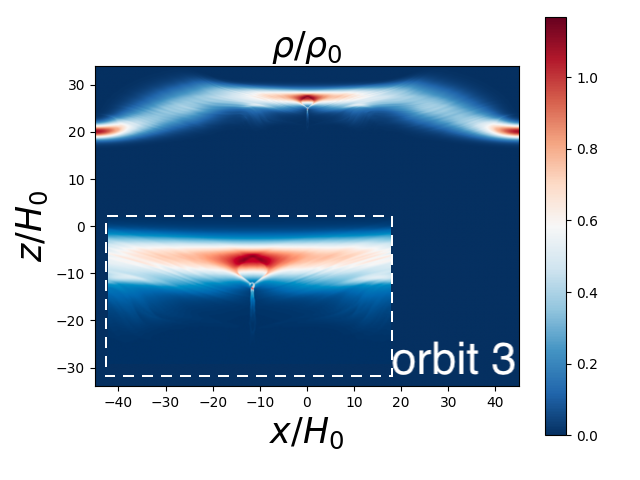}
\includegraphics[scale=0.36]{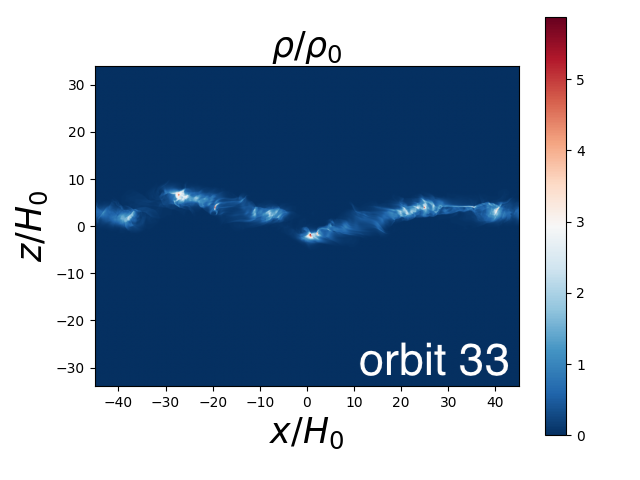}
\includegraphics[scale=0.36]{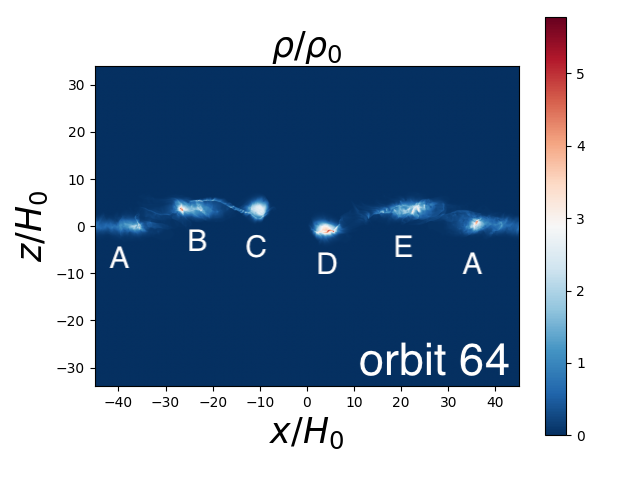}
\includegraphics[scale=0.36]{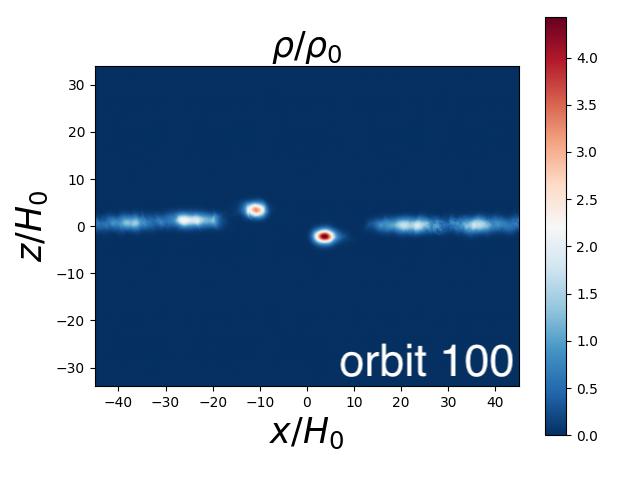}
\caption{Evolution of a tilted disc (in the wavelike regime) under the influence of an imposed differential precession. The figures show density snapshots in the $xz$-plane from our fiducial simulation (Precession02) with a large initial tilt ($z_0=28H_0$) and precession amplitude ($a = 0.05$). Top-left: the initial condition (orbit 0) is a flat, tilted disc. Top-middle (orbit 0.8): the imposed precession induces a warp in the disc. Top-right (orbit 3): horizontal sloshing motions in the disc converge into shocks at certain locations (magnified in the inset). Bottom-left: by orbit 33 the disc has broken into three rings. Bottom-middle (orbit 64): the disc begins to break into two more rings. We have labelled the different rings as A, B, C, D, and E, respectively. Bottom-right (orbit 100): the disc has broken into five rings (the two rings at the radial boundaries are actually the same ring). Three of the rings have lost most of their tilt and appear to be `rejoining' again (at $x \sim \pm35H_0$). There are two highly compact, tilted, and well-separated disc segments resembling tori at $x \sim -10H_0$ and $x \sim 5H_0$.}
\label{FIGURE_FiducialSimulationFlowField}
\end{figure*}

Assuming that the precession is slow compared to bending-wave propagation, the solution can plausibly be separated into a slowly evolving uniform tilt plus a small warp.

In this regime we can find an approximate solution in the form of a rapidly converging Fourier expansion in~$x$:
\begin{equation}
  Z=\sum_{j=0}^\infty Z_j(t)\cos(jkx),\qquad G=\sum_{j=1}^\infty G_j(t)\sin(jkx). 
\end{equation}
Retaining only the most important terms, this leads to
\begin{align}
  &\dot Z_0=-\f{1}{4}\ii\Omega_0bZ_1,\label{EQUATION_ODEZ0dot}\\
  &\dot Z_1\approx-\f{1}{2}\ii\Omega_0bZ_0+\f{kG_1}{\Sigma},\label{EQUATION_ODEZ1dot}\\
  &\dot G_1\approx-\alpha\Omega_0G_1-\f{1}{4}\Sigma c_{\text{s}0}^2kZ_1.
  \label{EQUATION_ODEG1dot}
\end{align}
These coefficients can be interpreted as follows: $Z_0(t)$ describes the uniform tilting of the disc; if the initial vertical displacement is $z_0$ and there is no initial vertical velocity (as in Section~\ref{METHODS_InitialConditionPrecessionSims}), then the initial value of $Z_0$ is $z_0$.
Then $Z_1(t)$ describes the longest-wavelength bending wave or warp that fits into the periodic box; $G_1(t)$ is the component of the internal torque with the same wavelength.

If we initially neglect the time-dependence of~$Z_0$, then the solution for $Z_1$ and $G_1$ consists of a transient solution (decaying if $\alpha>0$) plus a steady solution with $G_1/\Sigma\approx\f{1}{2}\ii\Omega_0bZ_0/k$ and $Z_1\approx-2\ii\alpha\Omega_0^2bZ_0/c_{{\text s}0}^2k^2$, implying a slow decay of the mean tilt at the rate
\begin{equation}
  \f{\dot Z_0}{Z_0}\approx-\f{\alpha\Omega_0}{2}\left(\f{b\Omega_0}{c_{{\text s}0}k}\right)^2.
\label{tilt_decay}
\end{equation}
The approximation is self-consistent if the bracketed quantity is $\lesssim1$.

Indeed, if $Z_0$ is treated as a constant, then the solution of the remaining initial-value problem in the wavelike regime starting with the initial condition $Z_1=G_1=0$ is
\begin{align}
  \f{G_1}{\Sigma}&=\f{\ii\Omega_0bZ_0}{2k}\left[1-\left(\cos\omega t+\f{\alpha\Omega_0}{2\omega}\sin\omega t\right)\ee^{-\alpha\Omega_0t/2}\right],\\
  Z_1&=-\f{2\ii\alpha\Omega_0^2bZ_0}{c_{\text{s}0}^2k^2}\left[1-\left(\cos\omega t-\f{4\omega^2-\alpha^2\Omega_0^2}{4\alpha\omega\Omega_0}\sin\omega t\right)\ee^{-\alpha\Omega_0t/2}\right],
\label{z1_transient}
\end{align}
where
\begin{equation}
  \omega=\f{1}{2}\left(c_{\text{s}0}^2k^2-\alpha^2\Omega_0^2\right)^{1/2}
  \label{EQUATION_BendingWaveFrequency}
\end{equation}
is the angular frequency of the bending wave (reduced by viscous damping).\footnote{Note that precession does not appear in this expression. This is because the imposed precession in our simulations is spatially oscillatory, and so the mean precession rate is zero.} According to this solution, the decay of the mean tilt is given by equation~(\ref{tilt_decay}) but multiplied by the factor in square brackets in equation~(\ref{z1_transient}). If we are comfortably in the wavelike regime ($\alpha\Omega_0\ll c_{\text{s}0}k$), then the first peak of this factor occurs close to $\omega t=\pi/2$ and is large, approximately $\omega/\alpha\Omega_0$, corresponding to $Z_1/Z_0\approx-\ii(b\Omega_0/c_{\text{s}0}k)$. This means that the transient bending wave is much larger than the steady-state one and causes a more rapid damping of the tilt than estimated above.

\subsection{Discussion of nonlinear effects}
\label{METHODS_DiscussOfNonlinearEffects}
One way of assessing the validity of the linear solution follows \citet{held2026disc} (end of section~3.6) by estimating the product of the warp $\p\xi_z/\p x$ and the shear $\p\xi_x/\p z$, where $\boldsymbol{\xi}$ is the displacement, and requiring this product to be small compared to unity. In the present notation this condition corresponds to $|Z_1G_1|\lesssim\Sigma H_0^2\Omega_0k$, which in turn means (for the steady solution mentioned above) $\alpha b^2|Z_0/H_0|^2\lesssim (kH_0)^2$ or (for the transient evolution in the wavelike regime) $b^2|Z_0/H_0|^2\lesssim2kH_0$.

\section{Fiducial simulation}
\label{RESULTS_FIDUCIALSIMULATION}
In the previous section we presented a linear theory governing the evolution of bending waves under the influence of an imposed precession. This theory omitted nonlinear effects such as vertical bouncing of the disc or shocks. To explore this nonlinear regime we now present a simulation with an imposed precession in which the disc is initialized with a large initial tilt of $z_0 = 28H_0$. The disc is initially flat (see equations \ref{EQUATION_InitialDensityProfilePrecessionSims}--\ref{EQUATION_InitialBackgroundAzVelProfilePrecessionSims}). The warp is driven by a large precession amplitude amplitude of $a = 0.05$. Taking $b=a$, $z_0 = Z_0$, and $kH_0=2\pi/L_x$ (with $L_x = 90H_0)$, and substituting into the criterion for linearity given at the end of the last section, we find $a^2|z_0/H_0|^2 \gg 2kH_0$, so we expect the transient evolution to be highly nonlinear. We use a resolution of $32/H_0$, and the runtime is 125 orbits. The simulation is inviscid and the disc is therefore in the \textit{wavelike} regime of warp propagation (we will consider the diffusive regime in Section \ref{RESULTS_ParameterStudy_ViscosityStudy}). This fiducial simulation is called Precession02 and is listed in Table~\ref{TABLE_PrecessionStudy}.

\subsection{Flow field evolution}
\label{RESULTS_FiducialSimulation_FlowFieldEvolution}
We start by describing the evolution of the flow field, mainly the density in the $xz$-plane (see Fig.~\ref{FIGURE_FiducialSimulationFlowField}). Already during the first orbit, the imposed precession induces a warp in the disc (top-middle-hand panel of Fig.~\ref{FIGURE_FiducialSimulationFlowField}). Sloshing motions in $u_x$ and $\delta u_y$ (not shown) begin to grow in the disc due to the radial pressure imbalance associated with the warp. Around orbit 3, the sloshing motions are large enough in amplitude that they converge into shocks at $x = 0$ (see top-right-hand panel of Fig.~\ref{FIGURE_FiducialSimulationFlowField}, particularly the inset). These shocks are similar those in the free warp simulations of \cite{held2026disc} (compare to their Fig. 3). Once the disc become warped it also bounces twice orbit (this effect is largest where the warp is most pronounced).

By orbit 10 the disc has broken around $x/H_0 \sim \pm35$. There is a large and flat central ring of width $\sim 70H_0$. By orbit 20 the central ring has developed a noticeable warp, symmetric about $x = 0$ (not shown), and by orbit 30, the central ring has broken again, this time at $x/H_0 \sim 0$, as shown in the bottom-left-hand panel of Fig.~\ref{FIGURE_FiducialSimulationFlowField}). We now have three broken disc segments: two relatively flat rings near the left- and right-hand part of the domain (these are actually the same ring, owing to the radially periodic boundary conditions), and two warped rings, one to the left of $x = 0$, and the other to the right of $x = 0$.

Around orbit 50 the ring immediately to the left of $x = 0$ starts to break once more into two pieces. A few orbits later, the same happens to the ring immediately to the right of $x = 0$. By orbit 60 the disc has broken into a total of \textit{five} rings (see bottom-middle panel of Fig.~\ref{FIGURE_FiducialSimulationFlowField}): three flat, tilted rings of radial size $\sim10H_0$, and two more compact rings that more closely resemble tori with a Gaussian structure in both the radial and vertical directions. [We will sometimes refer to these as \textit{Gaussian rings} (or tori).] The Gaussian rings are centred around $\sim -11H_0$ and $\sim 4H_0$, respectively. For convenience, we have labelled the individual rings A, B, C, D, E, A, moving from left to right. There is also a noticeable symmetry breaking about $x = 0$.

We observe considerable interactions between broken disc segments, especially between the rings A and B, and rings E and A. The relative tilt between the rings results in a varying radial pressure gradient which induces horizontal sloshing motions. The combination of these and the vertical motion causes an oblique shock to form in the gap between the rings. Starting around orbit 80, rings A and B, and rings E and A, begin to rejoin. (We will refer to these rejoined rings as AB and EA, respectively.) In contrast, the gaps between these rejoined rings and the central broken disc segments C and D appear to have \textit{deepened}, forming a nearly clean break (see the bottom-right-hand panel of Fig.~\ref{FIGURE_FiducialSimulationFlowField}). Throughout this entire process the tilt has decreased significantly: each fluid column in the disc started with an initial tilt of $z_0 = 28H_0$, but by orbit 100 the AB and EA rings have a tilt of around $\sim H_0$, while the central rings have larger tilts of around $\sim 5H_0$.

In conclusion, we started with a smooth and flat disc with a large initial tilt of $z_0 = 28H_0$ (top-right-hand panel of Fig.~\ref{FIGURE_FiducialSimulationFlowField}). The imposed differential precession caused the disc to develop a warp or twist within the first orbit of initialization (top middle-panel of Fig.~\ref{FIGURE_FiducialSimulationFlowField}), which in turn induced horizontal sloshing motions that converged into shocks at the location where the warp amplitude was zero ($x/H_0 = 0$). Over the next 50 orbits the disc broke at five separate locations: $x/H_0 \sim -35, -15, -5, 0, 10, 30$. We find considerable variability in the size and shape of the resultant broken disc segments. Some are flat rings of width around $10H_0$, similar to the type of rings we observed in the free warp simulations in \cite{held2026disc}. However, here we also find the emergence of a new type of ring, which has a more compact structure and is more similar to a torus. These rings have a width and thickness around $\sim H_0$, and have a Gaussian density profile in both the radial and vertical directions. Some of the rings are separated by deep gaps of width $\sim 10$--$12H_0$ and do not appear to interact by the end of the simulations, while other rings are close enough to their neighbours that we observe significant interactions between them in the form of oblique shocks. Repeated interactions between these rings result in a synchronization of their tilts and, ultimately, even in \textit{disc rejoining}. We will investigate disc interactions in more detail in Section \ref{RESULTS_GAUSSIANRINGSTUDY}.

\begin{figure}
    \centering
    \includegraphics[scale=0.24]{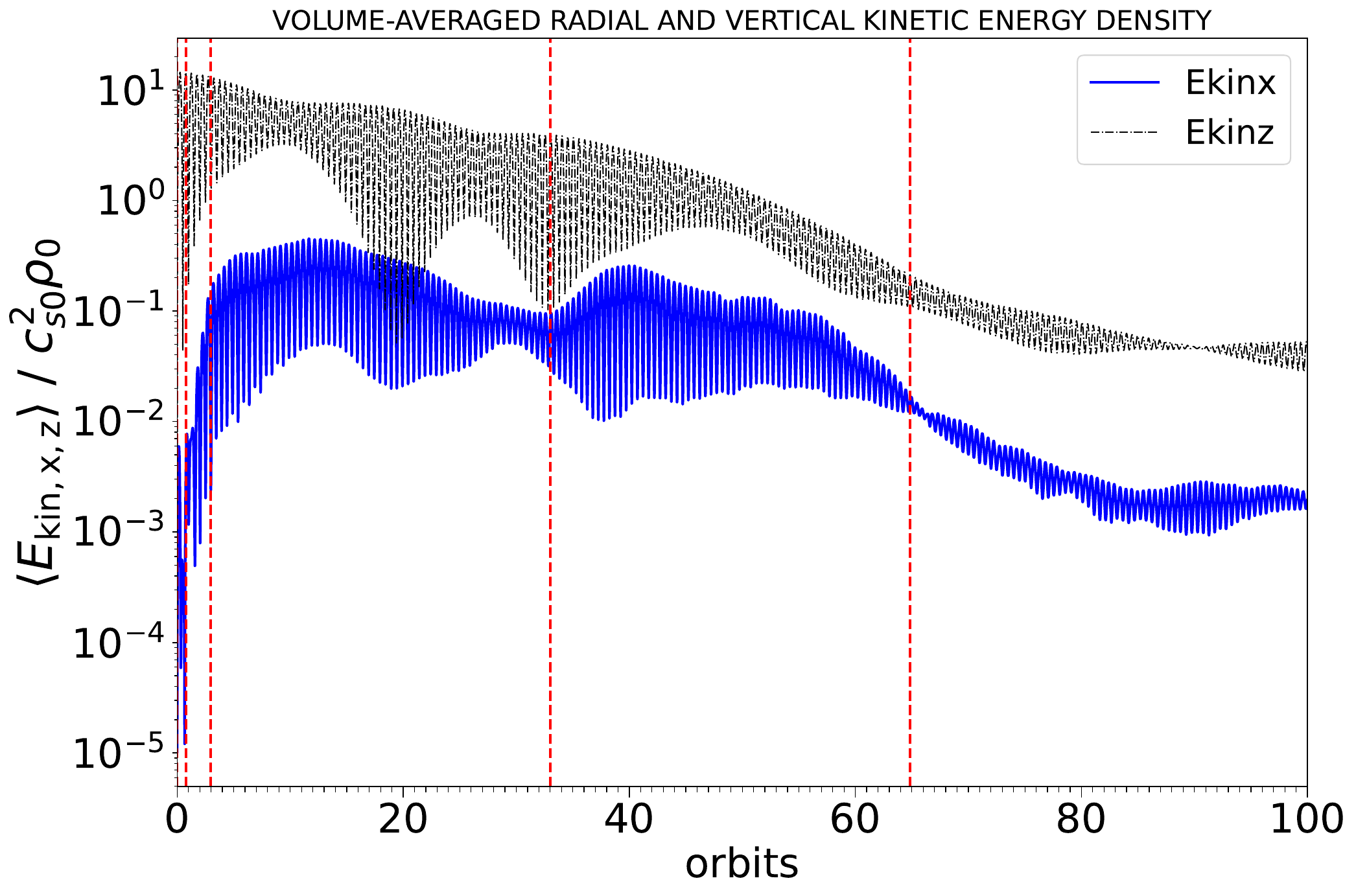}
    \caption{Time-evolution of volume-averaged radial (blue solid curve) and vertical kinetic energy density (black dot-dashed curve) taken from our fiducial simulation run with an initial disc tilt of $z_0 = 28H_0$ and a precession amplitude of $a = 0.05$. Vertical dashed lines mark the times of the snapshots shown in Fig.~\ref{FIGURE_FiducialSimulationFlowField}.}
\label{FIGURE_FiducialSimulationTimeSeriesEkinxEkinz}
\end{figure}

\subsection{Other diagnostics}
\label{RESULTS_FiducialSimulation_OtherDiagnostics}
\subsubsection{Time-series of volume-averaged quantities}
\label{RESULTS_FiducialSimulation_TimeSeries}
In Fig.~\ref{FIGURE_FiducialSimulationTimeSeriesEkinxEkinz} we show the time-series of the volume-averaged vertical component of the kinetic energy (KE) density (blue solid line) and the corresponding radial component (black dot-dashed line). As in our free warp simulations (compare to Fig. 7 in \citealt{held2026disc}), both the radial and kinetic energies exhibit behaviour on three different time-scales: short oscillations on orbital time-scales, a beating on a time-scale of around 14 orbits, and a decay on a time-scale of around 100 orbits.

The vertical kinetic energy oscillates twice per orbit and dominates over the radial KE at nearly all times. These orbital time-scale oscillations come from the disc tilt, and, to a certain degree, also from the bouncing of the disc. The radial kinetic energy energy density also oscillates twice per orbit. Its envelope initially increases as the disc becomes warped and horizontal sloshing motions grow. The vertical KE also exhibits a mild beating on a time-scale of around 14 orbits, which is likely related to the period of a bending wave that is trying to propagate through the disc. From our linear theory of bending waves we predicted that the bending wave frequency is given by equation~\ref{EQUATION_BendingWaveFrequency}. In the inviscid limit, and for a warp of wavelength $k = 2\pi/L_x$ ($L_x = 90H_0$), the bending wave period is $1/(\frac{1}{2}c_{\text{s}0}k) \approx 28.6$ orbits. The \textit{beat} in the kinetic energy should be \textit{half} this, i.e.~14.3 orbits, which is exactly what we measure in the simulation.

\begin{figure*}
\centering
\includegraphics[scale=0.43]{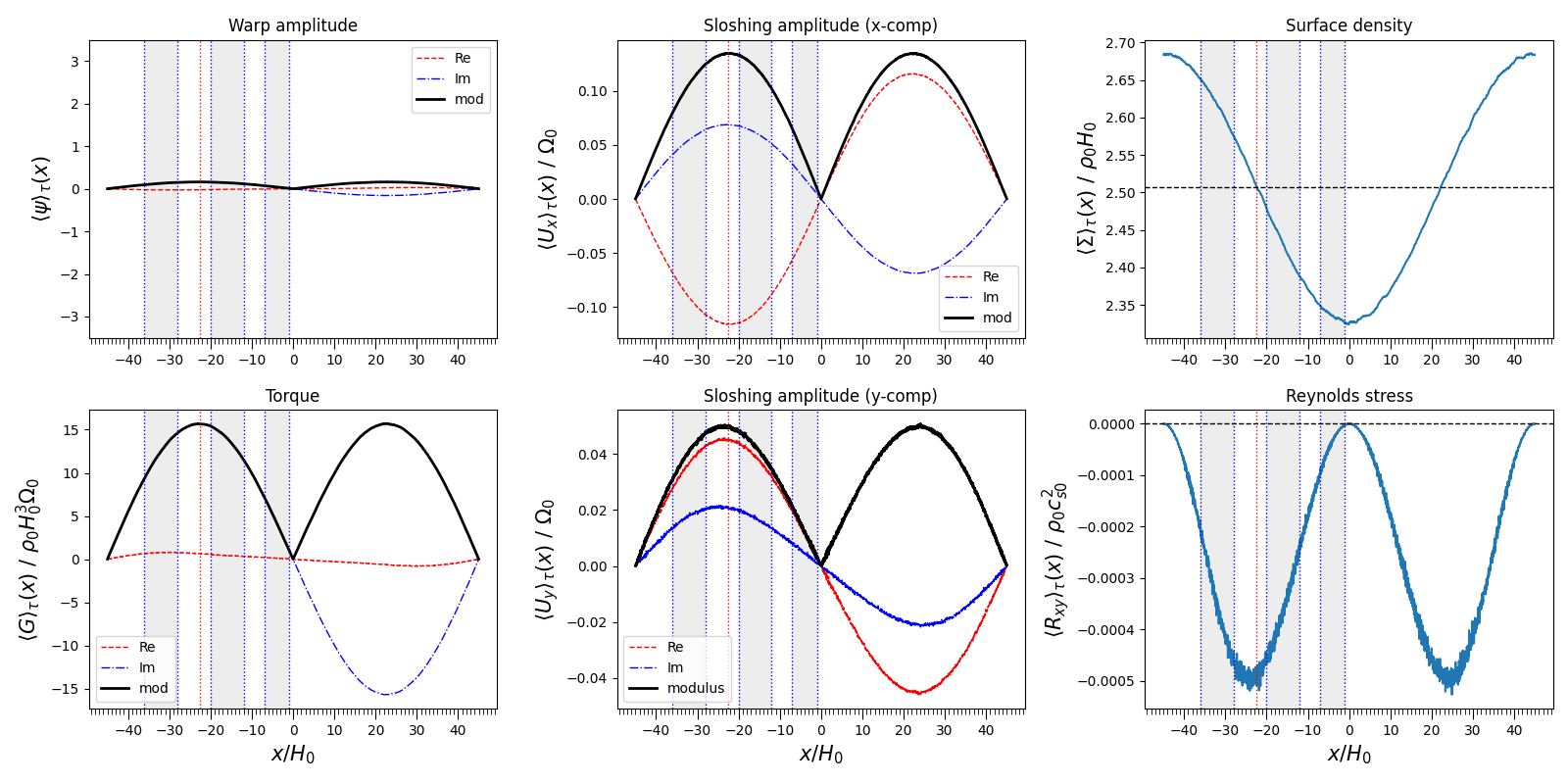}
\caption{Orbit-averaged radial profiles from our fiducial simulation (Precession02) with a large initial tilt ($z_0 = 28H_0$) and a large precession amplitude ($a = 0.05$). See Section \ref{METHODS_Diagnostics} for definitions. The orbit-average was taken over the first orbit (orbit 0-1). \textit{Top-left}: warp amplitude $\langle \psi \rangle_{\tau}$. \textit{Top-middle}: shear amplitude ($x$-component) $\langle U_x \rangle_{\tau}$. \textit{Top-right}: surface density $\langle \Sigma \rangle_{\tau}$. \textit{Bottom-left}: torque $\langle G \rangle_{\tau}$. \textit{Bottom-middle}: shear amplitude ($y$-component) $\langle U_y \rangle_{\tau}$. \textit{Bottom-right}: $xy$-component of the Reynolds stress $\langle R_{xy} \rangle_{\tau}$. The shaded areas show the approximate regions (in the left-hand part of the disc, only) at which gaps will open later on. The vertical dotted red line at $x = -22.5 H_0$ shows the location (in the left-hand part of the disc) at which the imposed precession is zero. This is also the location at which the \textit{rate of change} of precession is greatest. The dashed horizontal black line in the surface density plot shows the initial surface density.}
\label{FIGURE_Fiducial_OrbitAveragedRadialProfilesOrb0Orb1}
\end{figure*}

\begin{figure*}
\centering
\includegraphics[scale=0.43]{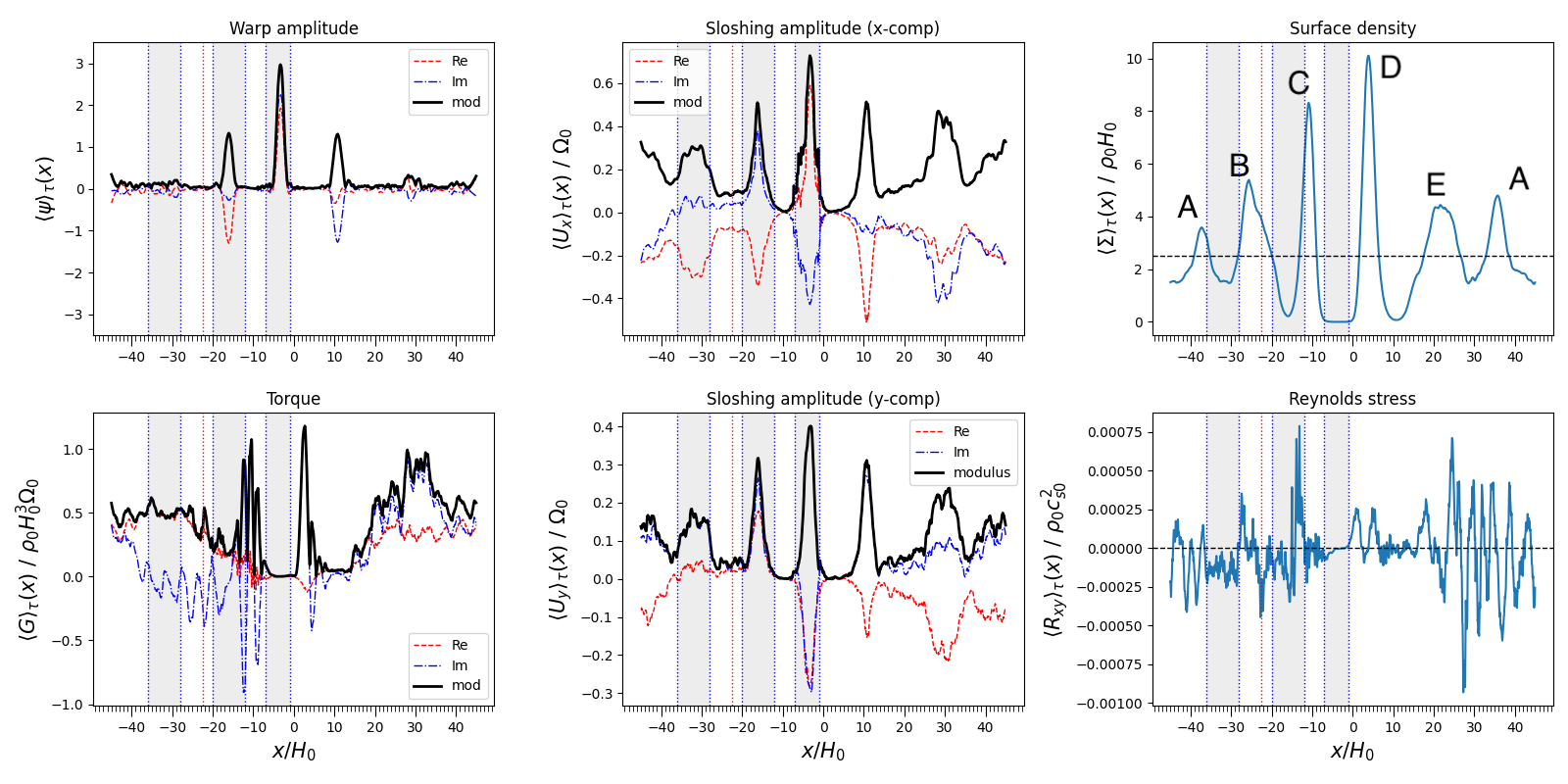}
\caption{Same as Fig.~\ref{FIGURE_Fiducial_OrbitAveragedRadialProfilesOrb0Orb1}, but the radial profiles have been orbit-averaged over orbit 99--100. At this point the disc has broken into five rings (cf. bottom right-hand panel of Fig.~\ref{FIGURE_FiducialSimulationFlowField}). Two, compact  Gaussian-shaped tori, C and D, straddle the centre of the domain and are separated by a deep gap. Further out, we have two wider rings, B and D, which are in the process of rejoining with the outermost ring (A).}
\label{FIGURE_Fiducial_OrbitAveragedRadialProfilesOrb99Orb100}
\end{figure*}

On longer time-scales, the envelope of the vertical KE decreases with time: by the end of the simulation at orbit 100 it is around $4 \times 10^{-2}\,c_{\text{s}0}^2\rho_0$, compared to its peak value of $E_{\text{kin},z} \sim 10\,c_{\text{s}0}^2\rho_0$ during the first orbit of the simulation. The radial KE decreases in a similar manner after the sloshing motions saturate at around orbit 15. The only source of dissipation is from shocks.\footnote{We show in Appendix \ref{APPENDIX_TiltDissipationStudy} that the numerical damping rate of the tilt at a resolution of $32/H_0$ is very small (around $\sim10^{-6}\Omega_0$).} These appear during the first few orbits of the simulation as standing shocks from converging horizontal sloshing motions: see the thin vertical feature in the density field around $x = 0$ in the top-right hand panel of Fig.~\ref{FIGURE_FiducialSimulationFlowField}. Later on we also observe the formation of \textit{oblique} shocks (see, for example, the thin oblique streams of gas between the rings B and C, and D and E, in the bottom middle panel of Fig.~\ref{FIGURE_FiducialSimulationFlowField}). These shocks do not form directly from converging sloshing motions, but rather due to interactions between neighbouring broken disc segments with different tilts (see also the discussion of interacting disc segments in Section \ref{RESULTS_GAUSSIANRINGSTUDY}). By orbit 80, the disc has broken into relatively well-separated rings, and both energies settle into a quasi-steady state for the remainder of the simulation.

We also inspected the time-evolution of the Reynolds stresses (not shown). The volume-averaged $xy$-component of the stress $\langle R_{xy} \rangle$, which is associated with radial angular momentum transport, oscillates about zero on orbital time-scales. Like with the kinetic energies, we observe a beating in the envelope of the stress: the amplitude of the oscillations increases early on, driven primarily by growth in the horizontal sloshing motions as the imposed precession warps the disc. The envelope of $\langle R_{xy}\rangle$ reaches a peak of around $\langle R_{xy} \rangle \sim 0.2\,c_{\text{s}0}^2 \rho_0$ around orbit 12, before decaying gradually, reaching $0.025\,c_{\text{s}0}^2 \rho_0$ around orbit 30. The envelope of the stress decreases monotonically after orbit 40, which is around the time the disc starts to break again, and the amplitude of the stress is negligible after orbit 60. 

These results naively suggest the dynamics associated with disc breaking are not effective at driving accretion by means of AM transport. However, this interpretation comes with two important caveats. In a global context, accretion can also be driven by mixing of the misaligned components of angular momentum (see $\nu_2$ term in Equation 2.2 of \cite{pringle1992simple}). Furthermore, our simulations are 2D and therefore axisymmetric. It is possible that in a fully 3D simulation a turbulent cascade or wave turbulence could emerge that is more three-dimensional, and which could, in principle, be more effective at extracting energy from the, shear resulting in positive $R_{xy}$.

Finally, the other off-diagonal stress components ($\langle R_{xz} \rangle$ and $\langle R_{yz} \rangle$; not shown) exhibit similar behaviour to $\langle R_{xy} \rangle$): short oscillations on orbital periods, and a decaying envelope over the duration of the simulation. However, unlike $R_{xy}$, which always appears to oscillate about 0, the off-diagonal components exhibit periods (of around 15 orbits, i.e, comparable to the bending wave time-scale) over which they maintain an oscillatory, but non-zero stress which can reach fairly large amplitudes (of $|R_{xz}| \sim 0.06\,c_{\text{s}0}^2\rho_0$ and $|R_{yz}| \sim 0.01\,c_{\text{s}0}^2\rho_0$, respectively). Their magnitudes are subdominant compared to $\langle R_{xy} \rangle$ during the first 60 orbits of the simulation, but become comparable to (and eventually larger than) $\langle R_{xy}\rangle$ once the disc breaks into rings.

\begin{figure}
    \centering
    \includegraphics[width=1\linewidth]{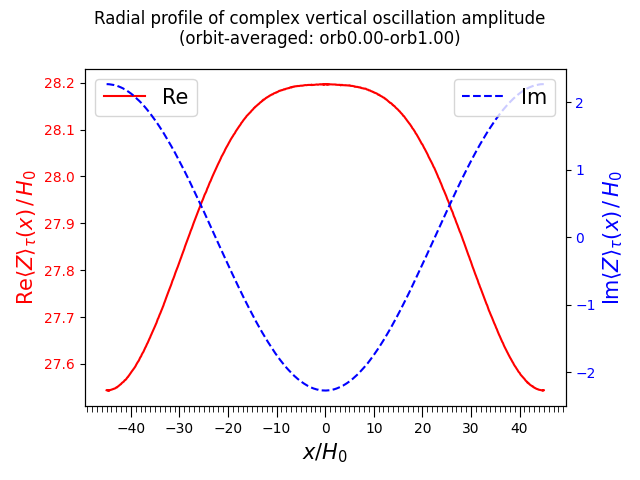}\\
    \includegraphics[width=1\linewidth]{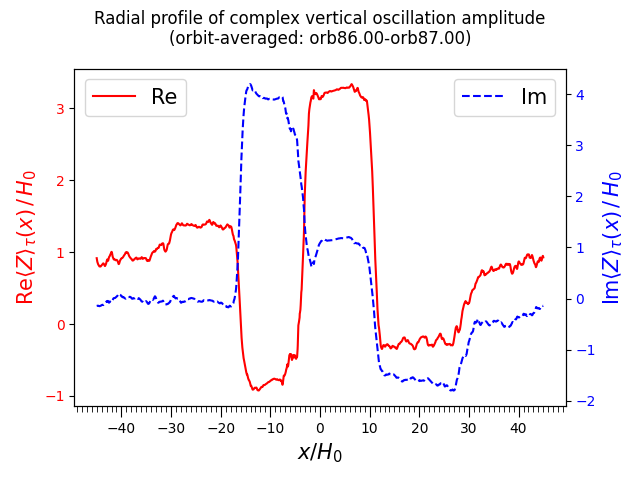}
    \caption{Orbit-averaged radial profiles of the real (red) and imaginary (blue) components of the complex tilt $Z$ (see equation \ref{EQUATION_ComplexTilt}) from the fiducial simulation (Precession02: $z_0 = 28H_0, a = 0.05$). Profiles are shown over the first orbit of the simulation (top panel), and over orbit 86--87, by which time the disc has broken into five rings.}
\label{FIGURE_FiducialSimulationComplexTiltProfile}
\end{figure}

\subsubsection{Orbit-averaged radial profiles}
\label{RESULTS_FiducialSimulation_OrbitAveragedRadialProfiles}
Finally we turn to the evolution of orbit-averaged quantities (see Section \ref{METHODS_Diagnostics} for definitions). In Fig.~\ref{FIGURE_Fiducial_OrbitAveragedRadialProfilesOrb0Orb1} we show radial profiles averaged over the first orbit of the simulation. Even within the first orbit we see growth in the warp amplitude $\langle \psi \rangle_\tau$ (top-left-hand panel), which develops two peaks at $x \sim \pm 22.5H_0$. This is precisely where the rate of change of precession is strongest. The warp drives horizontal sloshing motions which in turn mediate a torque, hence the corresponding double-peaked structures in $\langle U_x \rangle_\tau$, $\langle U_y \rangle_\tau$, and $\langle G \rangle_\tau$, respectively. The $xy$-component of the Reynolds stress (bottom right-hand panel), on the other hand, has developed two small and negative peaks at the same locations ($\langle R_{xy} \rangle_\tau \sim -5\times10^{-4}$). The surface density (top-right-hand panel) remains unchanged where the disc is most warped, but has been redistributed elsewhere, developing a wide gap of around $\langle \Sigma \rangle_\tau \sim 2.34\,\rho_0 H_0$ around $x/H_0 = 0$, a decrease of around $7\%$ compared to its initial (and uniform in $x$) value of  $\sim 2.51\,\rho_0 H_0$. Note that during the first orbit the warp is small ($\langle \psi \rangle_\tau \sim 0.1$), and the radial sloshing motions are not yet large enough to converge into shocks. 

\begin{figure}
    \centering
    \includegraphics[width=1
    \linewidth]{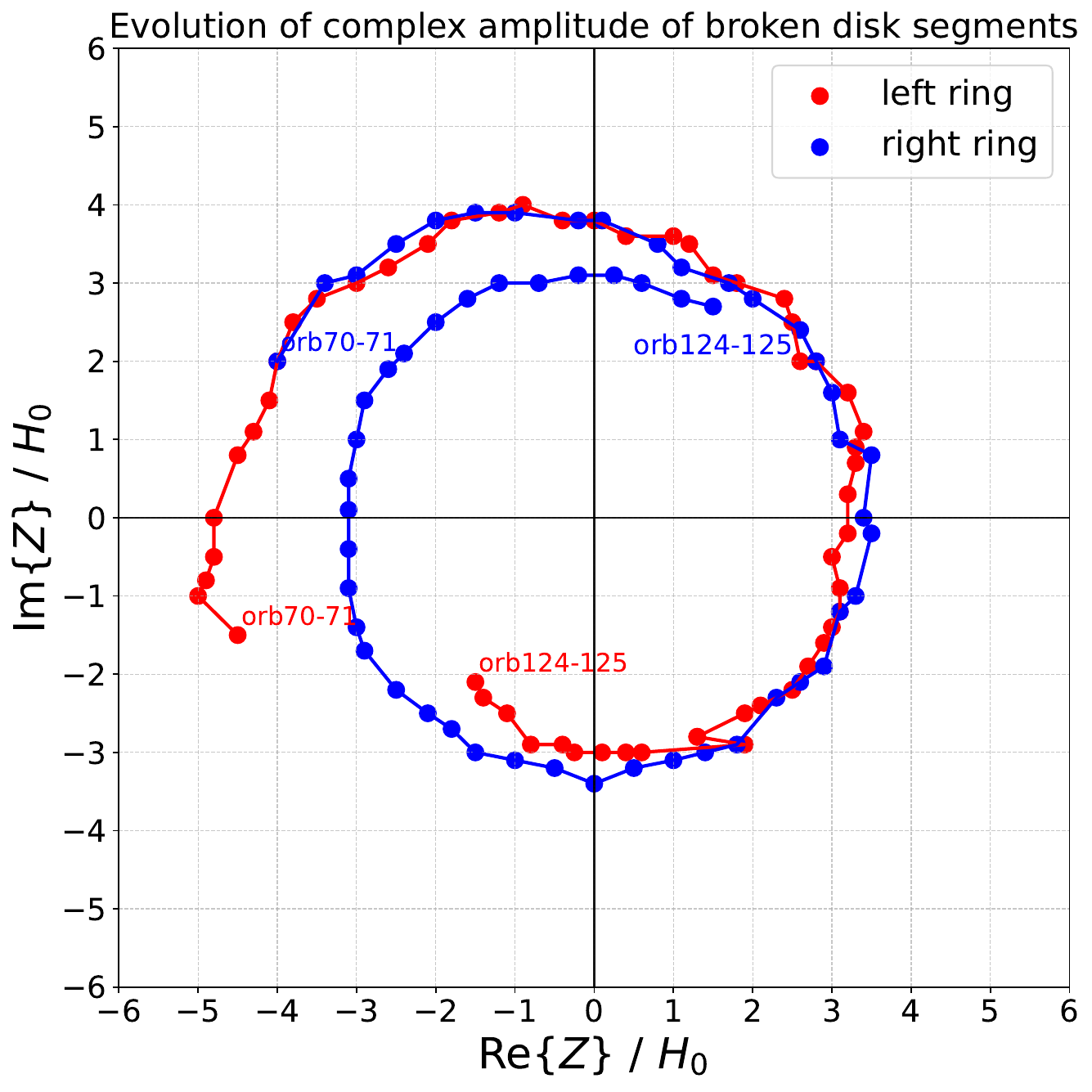}
    \caption{Orbit-averaged tilt $Z$ in the complex plane for the broken disc segments C ("left-ring"; red) and D ("right-ring"; blue) from the fiducial simulation (see Fig.~\ref{FIGURE_FiducialSimulationFlowField}). Each data point is separated by a time-interval of one orbit. By orbit 125 the two rings are nearly out-of-phase, indicating that they are precessing independently.}
    \label{FIGURE_ComplexTiltComplexPlaneFiducialSim}
\end{figure}

The warp amplitude continues to grow after the first orbit (not shown), and its peaks reach as high as $\langle \psi \rangle_{\tau} \sim 9$ by orbit 14. However, the location of the peaks shift compared to first orbit: by orbit 12 the peaks in the warp amplitude are concentrated around $x/H_0 \sim 35$, which is precisely where the disc first breaks. Accompanying this increase in the warp, we observe an increase in the amplitude of the horizontal sloshing motions, which peak around $\langle U_x \rangle_\tau \sim 3 \Omega_0$ and $\langle U_y \rangle_\tau \sim  \Omega_0$ at orbit 12. However, the peaks in the sloshing motions are broader than the peaks in the warp amplitude, and they are centred around $|x|/H_0 \sim 20$, which is close to where the rate of change of precession is greatest. The surface density has developed five gaps at this point located at $x/H_0 \sim -35, -15, -2, 15, 35$ of depth around $\langle \Sigma \rangle_{\tau} \sim 1.6, 1.9, 2, 1.9, 1.6\,\rho_0 H_0$, respectively, indicating that the breaking process has started. Note that, starting around orbit 2.8 (by which time the radial sloshing motion has grown to $\langle U_x \rangle_\tau \sim 1.5 \Omega_0$), we observe the formation of a shock at $x = 0$, which is precisely where the sloshing motions converge. This is forms a spike in the orbit-averaged surface density at $x = 0$ (not shown).

Finally, in Fig.~\ref{FIGURE_Fiducial_OrbitAveragedRadialProfilesOrb99Orb100} we show the orbit-averaged radial profiles at the \textit{end} of the simulation (orbit $99$--$100$). As can be seen in the flow field (bottom-middle and bottom-right-hand panel of Fig.~\ref{FIGURE_FiducialSimulationFlowField}), the disc has broken into five rings by this point (the left- and right-hand most rings are the same ring due to the radial periodic boundary conditions). Two tilted broken disc segments (tori) with a Gaussian density structure in the radial and vertical direction straddle the centre of the domain, one centred around $x \sim -11H_0$ (ring C), the other around $x \sim 4H_0$ (ring D). These are separated by very deep gaps, as can be seen in the surface density profile where $\langle \Sigma \rangle_{\tau} \sim 0.01 \rho_0 H_0$ (top-right-hand panel of Fig.~\ref{FIGURE_Fiducial_OrbitAveragedRadialProfilesOrb99Orb100}). The warp amplitude is close to zero in the rings, indicating that they are essentially flat, and peaks in the deep gap between them at $\langle \psi \rangle_{\tau} \sim 3$. The sloshing motions also peak in the gaps, with $\langle U_x \rangle_\tau \sim 0.7 \Omega_0$. We observe similar peaks in the warp amplitude and sloshing motions in the deep gaps separating these compact broken disc segments from their wider, neighbouring rings (rings A and B, and E and A in Fig.~\ref{FIGURE_FiducialSimulationFlowField}). These neighbouring rings are in the process of rejoining, which is reflected in the corresponding shallow gaps in surface density between them, and in the lack of a matching peak in the warp amplitude at $|x| \sim 35H_0$.

\subsubsection{Evolution of complex tilt}
\label{RESULTS_FiducialSimulation_EvolutionOfComplexTilt}
In Fig.~\ref{FIGURE_FiducialSimulationComplexTiltProfile} we show orbit-averaged radial profiles of the real (in red) and imaginary (in blue) components of the \textit{complex tilt} (equation \ref{EQUATION_ComplexTilt}). Over the first orbit (top panel) the radial profile of the real part of the tilt increases smoothly from the boundaries, where $\text{Re}\{Z\} \sim 27.5H_0$, to $x/H_0 = 0$ where it is around $x \sim 28.2$; the latter is very close to the initial tilt of the disc ($z_0 = 28H_0)$. The variation with radius is due to the imposed precession, which is distorting the disc in the first orbit, as can be seen in the flow field (top middle panel of Fig.~\ref{FIGURE_FiducialSimulationFlowField}). The imaginary component of the complex tilt is out of phase with the real part. It is smaller than the real part, but non-zero, ranging from $\text{Im}\{Z\} \sim 2\,H_0$ near the radial boundaries to $\text{Im}\{Z\} \sim -2\,H_0$ at $x/H_0=0$, i.e.~it is greatest in magnitude precisely where the imposed precession rate is largest (see top panel of Fig.~\ref{FIGURE_ComplexTiltAndWarpAnalytical}). Over the next ten orbits the real part of the tilt is damped as the sloshing motions converge into a standing shock at $x/H_0 = 0$, whereas the imaginary part grows and becomes comparable to $\text{Re}\{Z\}$ around orbit 10. This shows that the imposed precession is inducing a twist in the disc.

At later times, i.e.~once the disc has broken into rings, we observe discontinuities in both $\text{Re}\{Z\}$ and $\text{Im}\{Z\}$. See the bottom panel of Fig.~\ref{FIGURE_FiducialSimulationComplexTiltProfile}, which shows profiles of both quantities orbit-averaged over orbit 86--87. This is particularly evident in the two central broken disc segments that we described earlier (rings C and D in the bottom-middle panel of  Fig.~\ref{FIGURE_FiducialSimulationFlowField}). The ring to the left of $x/H_0 = 0$ has $\{\text{Re}Z, \text{Im}Z\} \sim \{-1, 4\}H_0$, while the right has $\{\text{Re}Z, \text{Im}Z\} \sim \{3, 1\}H_0$. 

As the rings oscillate about $z/H_0 = 0$, both components of the tilt evolve. To show this, we plot the complex pair $\{\text{Re}Z, \text{Im}Z\}$ in Fig.~\ref{FIGURE_ComplexTiltComplexPlaneFiducialSim} for the left-hand ring (i.e.~the C ring) in red, and for the right-hand ring (i.e.~the D ring) in blue over a time interval of about 50 orbits. Several interesting features emerge from this figure. First, the rings start out close to one another in phase (orbit 70--71), but end up almost completely out of phase by orbit 124--125. When differential precession is too strong, the disc breaks so that broken pieces can precess independently, which is precisely what we see in the figure. The C and D rings are thus dynamically disconnected from a precession point of view. However, repeated interactions of these rings with their immediate neighbours (i.e.~with the B and E rings, respectively, in Fig.~\ref{FIGURE_FiducialSimulationFlowField}) allows for a decay in their relative tilt, which is why the complex tilt of each ring traces a spiral in the complex plane. 

The left (right)-hand ring is centred around $x \sim -11H_0$ ($x \sim 4H_0$). The precession frequency at this location is $|\omega_\text{n}| = (a/2)\cos{kx} \sim 0.018\Omega_0$ ($\sim 0.024\Omega_0$). The precession period is then $1/\omega_\text{n}\sim 56$ ($\sim42$) orbits. From Fig.~\ref{FIGURE_ComplexTiltComplexPlaneFiducialSim} we see that the phase of the left- (right-) ring comes back to its starting position after 56 (45) orbits, so this is in excellent agreement with what we expect. Note that the left-hand ring (red curve in Fig.~\ref{FIGURE_ComplexTiltComplexPlaneFiducialSim}) is positioned where the precession frequency is slightly smaller (i.e.~the precession period is slightly longer) so it should take longer to return to its starting position in the complex $Z$ plane than the right-hand ring (blue curve), which is exactly what we see in the figure.

\section{Parameter study}
\label{RESULTS_PARAMETERSTUDY}
In the previous section we found that a disc initialized with a large initial tilt ($z_0 = 28H_0$) and imposed precession amplitude ($a = 0.05$) breaks into rings. We now examine how varying the initial tilt (Section \ref{RESULTS_ParameterStudy_InitialTilt}), the precession amplitude (Section \ref{RESULTS_ParameterStudy_PrecessionAmplitude}), or both (Section \ref{RESULTS_ParameterStudy_SmallTiltSmallPrecession}) affects this outcome. In Section \ref{RESULTS_ParameterStudy_ViscosityStudy} we discuss simulations in which we included an explicit viscosity in order to probe the diffusive regime of warp evolution.

\begin{figure}
    \centering
    \includegraphics[width=1\linewidth]{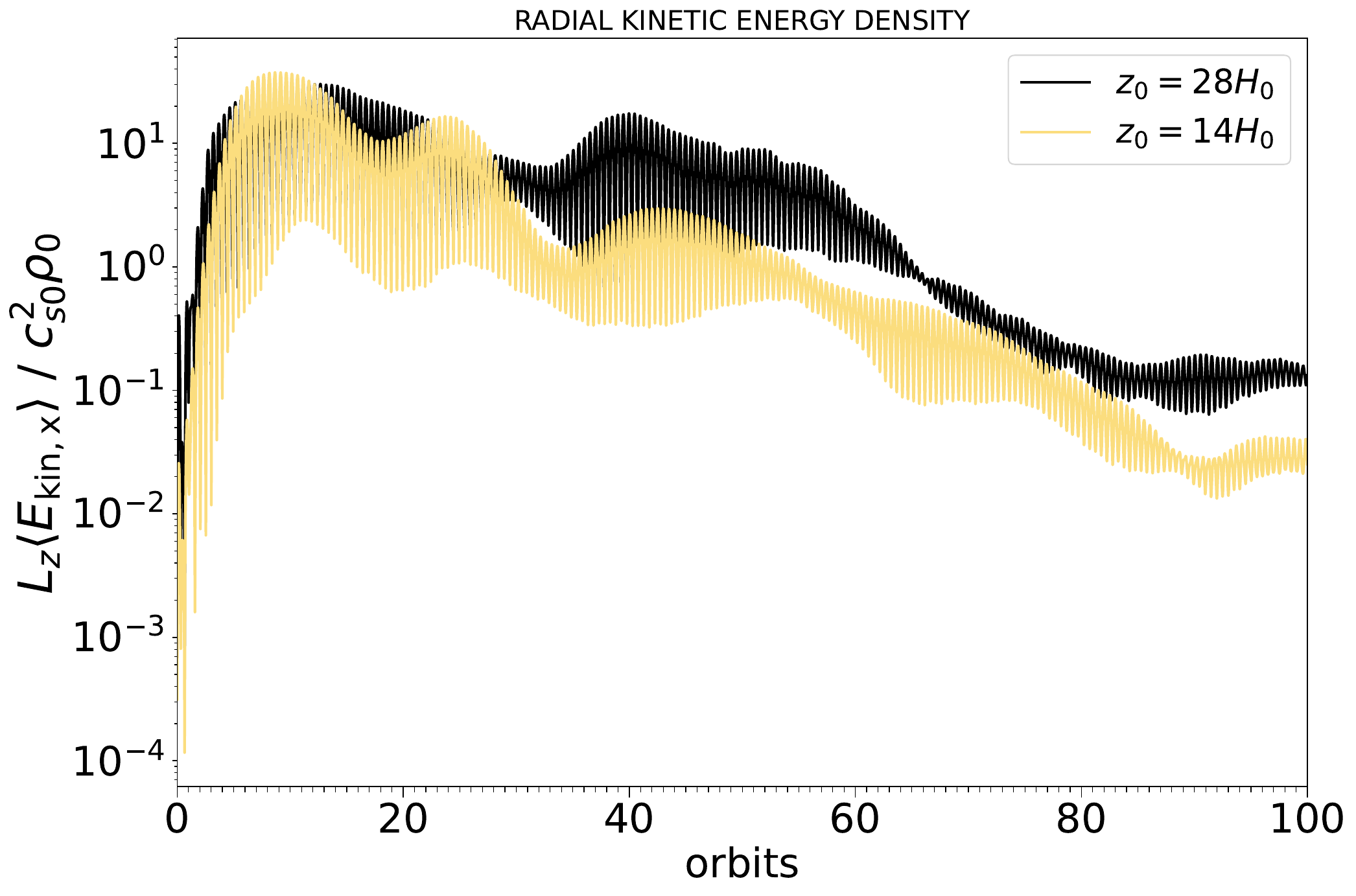}
    \includegraphics[width=1\linewidth]{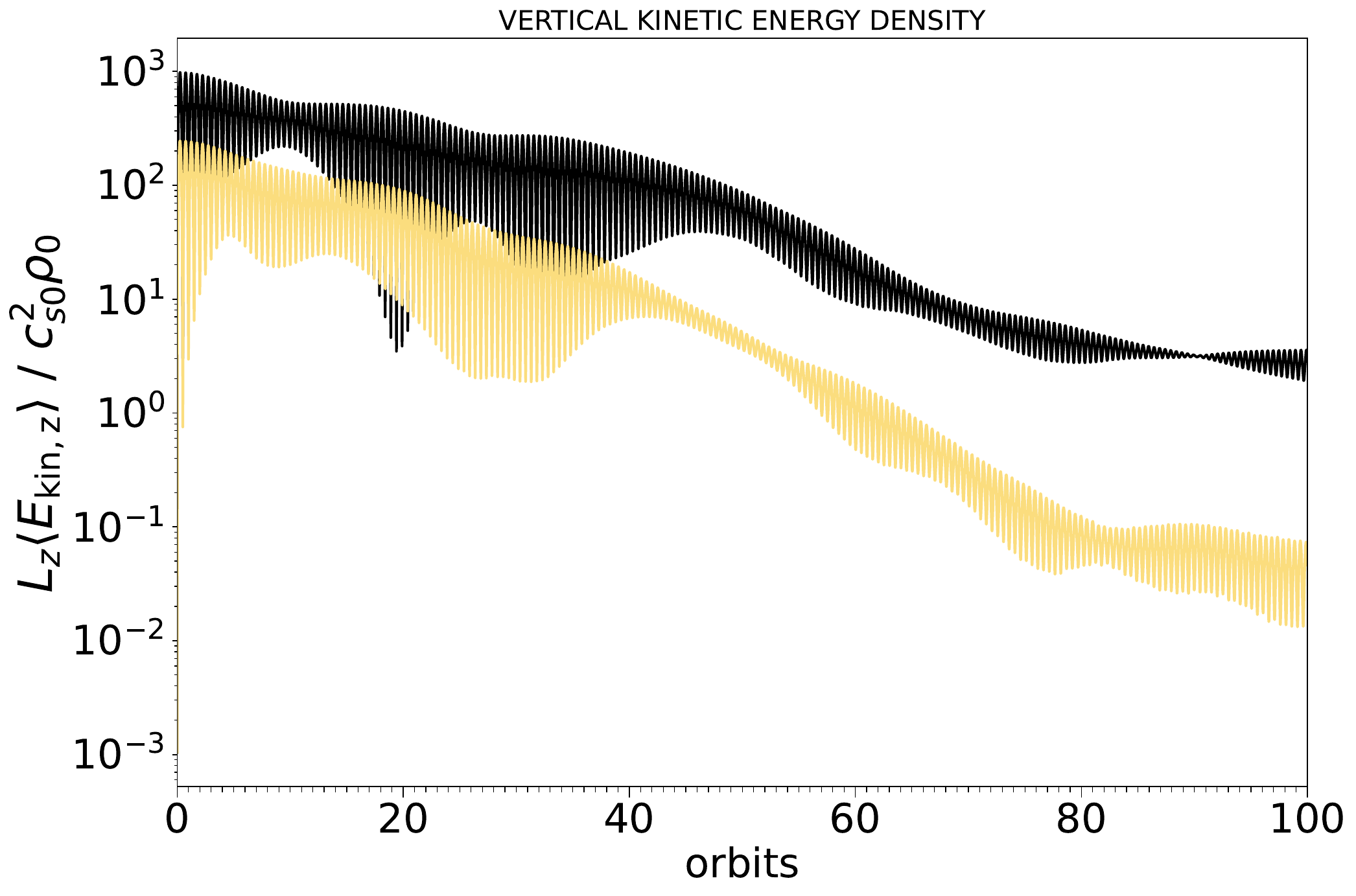}
    \caption{Initial disc tilt study: comparison of time-series of volume-averaged radial (top) and vertical (bottom) kinetic energy densities between two simulations with the same imposed precession amplitude ($a=0.05$), but different initial disc tilts: $z_0 = 28H_0$ (black curves; simulation Precession02) and $z_0 = 14H_0$ (yellow curves; simulation Precession03). The two simulations were run in boxes of vertical size $L_z = 68H_0$ and $L_z = 48H_0$, respectively, so we have multiplied the volume-averages by their respective vertical box sizes to facilitate easier comparison between them.}
\end{figure}

\subsection{Initial disc tilt study}
\label{RESULTS_ParameterStudy_InitialTilt}
In this section we compare the consequences of changing the initial disc tilt $z_0$, keeping the precession amplitude $a$ the same at $a = 0.05$. We compare our fiducial simulation (Precession02), which was initialized with $z_0 = 28H_0$, with Precession03, which was initialized with half that tilt ($z_0 = 14H_0$). We used a shorter box for the smaller tilt simulation ($L_z = 48H_0$ compared to $L_z = 68H_0$); otherwise the simulations were run with the same numerical set-up. The reader who is short on time can jump to the one-paragraph summary at the end of this subsection.

After initialization, the imposed precession warps the disc in both cases, although the maximum warp amplitude is significantly smaller in the simulation initialized with the smaller tilt ($\langle \psi \rangle_{\tau} \sim 2$ in the $z_0 = 14H_0$ run,  compared to $\langle \psi \rangle_{\tau} \sim 9$ in the $z_0 = 28H_0$ run). This is to be expected, since the disc with the larger tilt experiences a stronger differential precession because the vertical gravitational acceleration scales with the vertical coordinate [$g_z = -\Omega_0^2 z(1+a\cos kx)$]. Note that in the linear theory of Section~3 the warp is proportional to the tilt, whereas here the maximum warp does not scale linearly with the initial tilt, so there is evidence of nonlinearity in these simulations. 

\begin{figure*}
    \centering
    \includegraphics[scale=0.5]{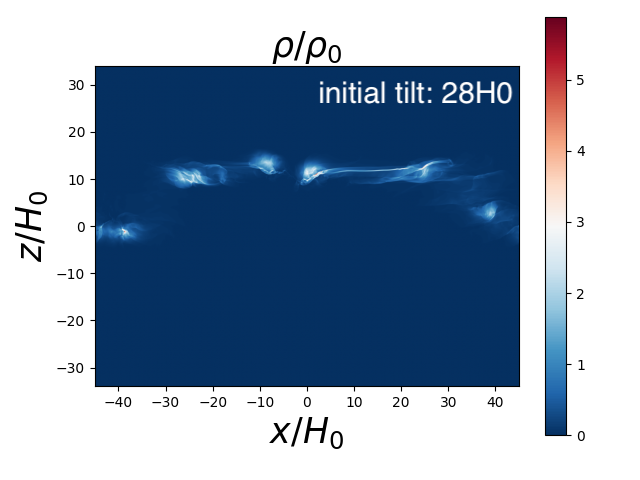}
    \includegraphics[scale=0.5]{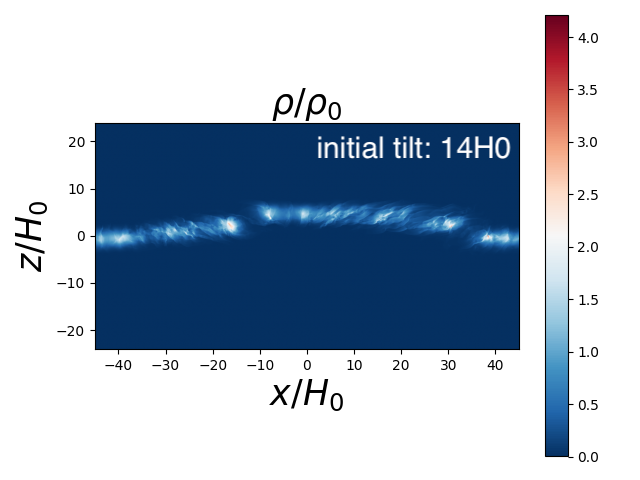}
    \caption{Initial disc tilt study: comparison of flow field (density in the $xz$-plane) at similar times (around orbit 40) in two simulations with the same precession amplitude ($a = 0.05$), but with different initial disc tilts, $z_0 = 28H_0$ (left; simulation Precession02) and $z_0 = 14H_0$ (right; simulation Precession03). The disc initialized with the smaller tilt breaks into fewer rings.}
    \label{FIGURE_InitialDiskTiltStudyFlowField}
\end{figure*}

The smaller warp in the $z_0 = 14H_0$ run (and the resultant smaller horizontal sloshing motions) result in the disc becoming parametrically unstable between $|x|\sim 10$--$30H_0$ (not shown). In the larger tilt simulation the sloshing motions are so large that the PI modes are sheared out (we examine the PI in greater detail in Section \ref{RESULTS_ParameterStudy_SmallTiltSmallPrecession}). Otherwise the simulations share many of the same features: for example, we see the sloshing motions converge into shocks even at the lower tilt, although it takes longer for the sloshing motions to grow sufficiently in amplitude for this to occur (5 orbits at $z_0 = 14H_0$, compared to 2 orbits at $z_0 = 28H_0$). 

A key result is that the disc breaks into multiple rings after around 40 orbits in both simulations, but we find fewer breaks and larger rings in the run initialized with a smaller initial tilt. See Fig.~\ref{FIGURE_InitialDiskTiltStudyFlowField} which compares the flow field around orbit 40 between the two simulations. The large initial tilt run is in the process of breaking into five rings, of which the two straddling $x/H_0 = 0$ are compact `Gaussian-type' tori. In contrast, the run initialized with the smaller tilt breaks into just three rings: a flat ring spanning the radial boundaries, a warped left-hand ring centred around $x \sim -25H_0$ of width $\sim 15H_0$, and a much wider, warped right-hand ring of width $\sim 30H_0$ which is centred around $x = 5H_0$. Although fewer breaks develop in the run with the small tilt, those that do occur do so at roughly the same locations as in the run with the larger tilt: $x/H_0 \sim -35, 15, 35$. The gaps between the rings are typically more shallow in the $z_0 = 14H_0$ run: at orbit 40 we measure gap depths of $\langle \Sigma \rangle_{\tau} \sim 1.2, 0.6, 1.2$ at $x/H_0 \sim -35, -15, 35$. In the $z_0 = 28H_0$ run the gap depths at the same time are $\langle \Sigma \rangle_{\tau} \sim 0.8, 1, 0.5, 1.6$ at $x/H_0 \sim -35, -15, -5, 35$.

The shallower gaps in the smaller initial tilt run leads to considerably more interaction between the broken disc segments. This ultimately serves to damp the tilt, which eventually begins to fill in the gaps (we investigate this process of disc rejoining in more detail in \ref{RESULTS_GAUSSIANRINGSTUDY}). By orbit 100, nearly all of the tilt has been damped in the $z_0 = 14H_0$ run, leaving three disc segments separated by shallow gaps. This is in contrast to the $z_0 = 28H_0$ run, where the disc at the end of the simulation consist of left- and right-hand rings with a small tilt ($\lesssim H_0$), and two compact Gaussian tori straddling the centre of the domain which have maintained a considerable relative tilt (of several $H_0$), and which are separated by deep gaps. We do not observe the development of any such Gaussian-ring-like structures in the simulation with smaller tilt.

In Fig.~\ref{FIGURE_InitialDiskTiltStudyFlowField} we compared time-series of volume-averaged radial (top) and vertical (bottom) kinetic energy densities between the two simulations ($z_0 = 28H_0$ in black, $z_0 = 14H_0$ in yellow). The saturated radial kinetic energy is comparable in magnitude between the two runs and we observe a clear beating in the radial KE (with period of around 15 orbits), which is related to the propagation of an underlying bending wave through the disc. The beating is actually more regular in the run with the smaller tilt, probably because breaking occurs later in this run and so the bending wave is able to develop more fully. By orbit 40 the disc has broken into multiple rings in both simulations as described above, and this is associated with a greater dissipation rate due to the formation of oblique shocks in the gaps between disc segments where supersonic flow with different tilts intersects. Because the gaps in the $z_0 = 14H_0$ run are shallower than in the $z_0 = 28H_0$ run, there are more such interactions between broken disc segments in this run, and this actually drives a larger rate of tilt damping, as can be seen by comparing the slopes in the vertical KE between orbit 40 and orbit 80 (bottom panel).

In conclusion, reducing the initial disc tilt from $z_0 = 28H_0$ to $z_0 = 14H_0$, but keeping the precession amplitude the same at $a = 0.05$, results in the disc first becoming parametrically unstable where the induced warp is greatest. The disc does break over several tens of orbits, but into fewer segments (three rather than five); the resultant disc segments are wider in the $z_0 = 14H_0$ run, and we do not observe the formation of compact Gaussian-type tori. The gaps between the rings are shallower, too, and there is therefore considerably more interaction between the rings in the form of oblique shocks. This leads to greater damping of the tilt in the $z_0 = 14H_0$ run.

\subsection{Precession amplitude study}
\label{RESULTS_ParameterStudy_PrecessionAmplitude}
Next we investigate the effect of decreasing the \textit{precession amplitude} $(a)$, keeping the initial disc tilt $(z_0)$ the same. We compare two simulations, both initialized with a large initial tilt of $z = 28H_0$: Precession02 (our fiducial simulation) was run with a larger precession amplitude of $a = 0.05$, while simulation Precession01 was run with $a=0.01$. The reader who is short on time can jump to the one-paragraph summary at the end of this subsection. 

\begin{figure}
    \centering
    \includegraphics[width=1\linewidth]{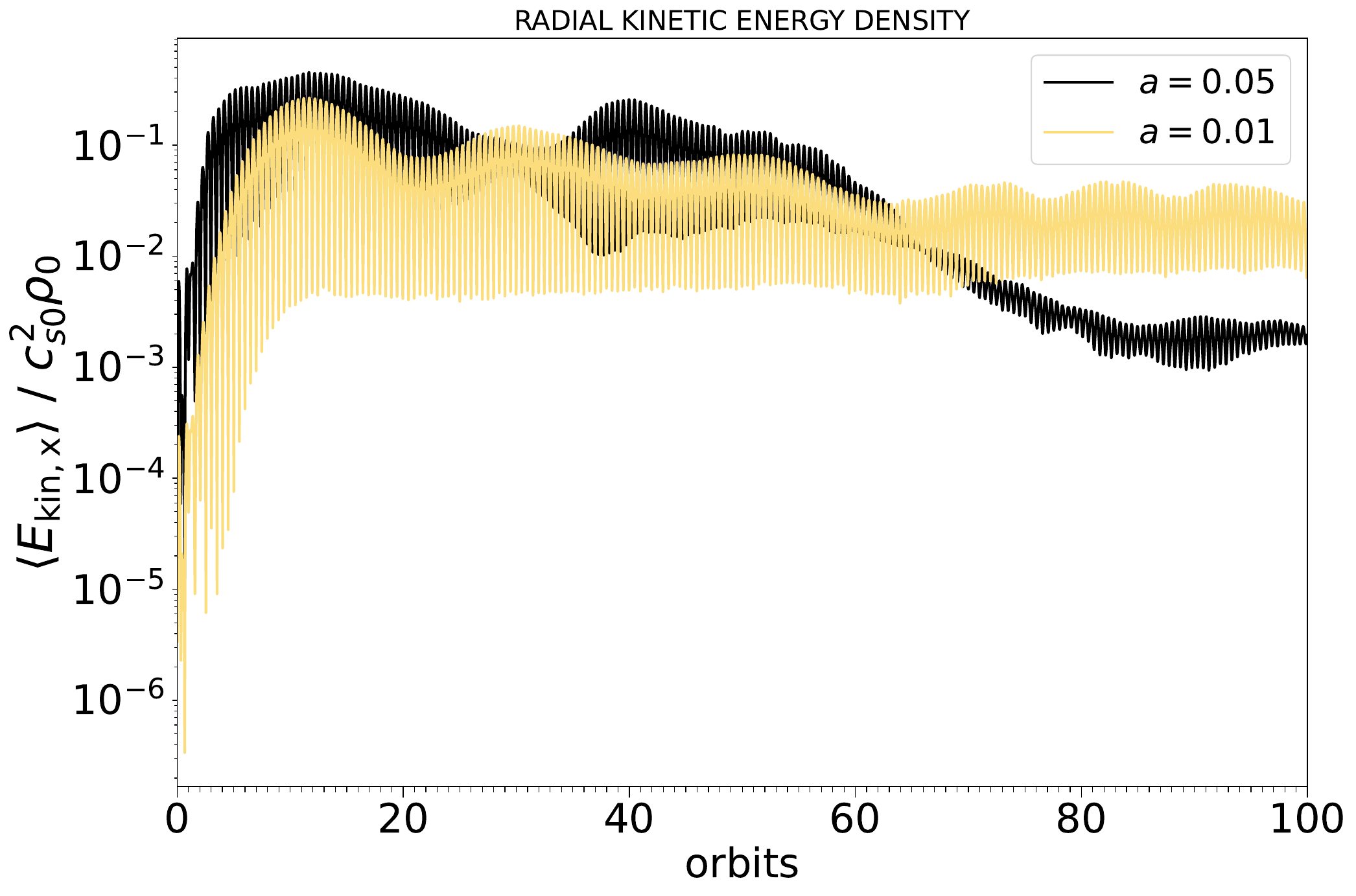}
    \includegraphics[width=1\linewidth]{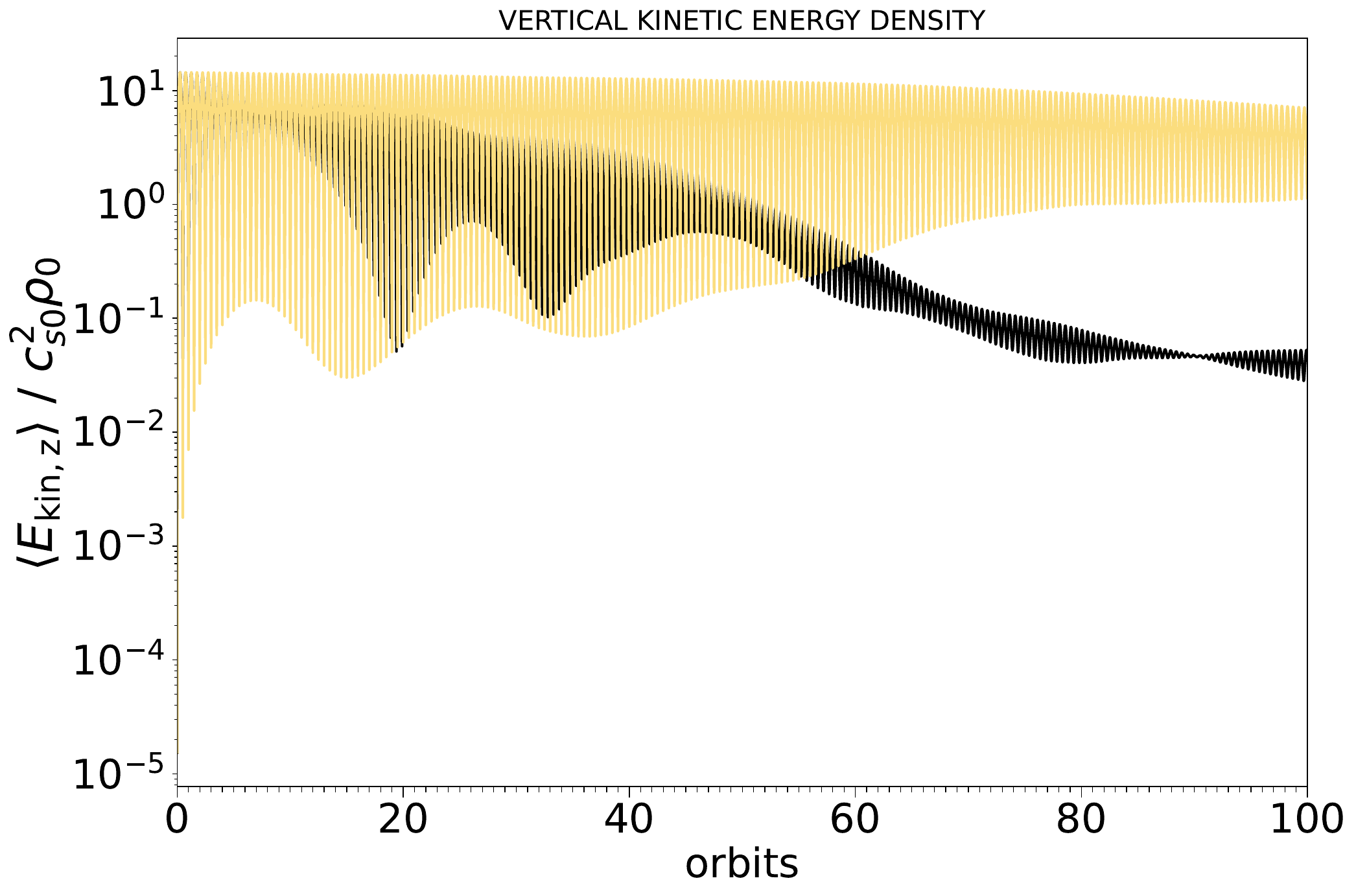}
    \caption{Precession amplitude study: comparison of time-series of volume-averaged radial (top) and vertical (bottom) kinetic energy densities between two simulations with the same initial disc tilt ($z_0 = 28H_0$), but different precession amplitudes, $a = 0.05$ (black curves; run Precession02) and $a = 0.01$ (yellow curves; Precession01).}
    \label{FIGURE_PrecessionAmplitudeStudy}
\end{figure}

Over the first fifty orbits or so, the two runs display markedly different behaviour. The disc subject to stronger precession develops a significantly larger warp within the first two orbits of initialization. The warp amplitude in the run with stronger precession peaks at $\langle \psi \rangle_{\tau} \sim 9$ around orbit 14; by contrast it is much noisier and barely reaches $\sim 0.5$ in the run with weaker precession. Associated with the larger warp in the $a = 0.05$ run are rapidly growing sloshing motions that converge into shocks at $x = 0$. This results in more rapid damping of the tilt (and hence the vertical kinetic energy density), as can be seen by comparing the black ($a = 0.05$) and yellow curves ($a = 0.01$) in the bottom panel of Fig.~\ref{FIGURE_PrecessionAmplitudeStudy}. The disc in the $a = 0.01$ instead becomes parametrically unstable where the warp amplitude is largest ($10 \lesssim |x|/H_0\lesssim 30$). The parametric instability (PI)  eventually spreads throughout the entire the disc. The growth and saturation of the PI in this run is similar to what we found in one of our free warp simulations (see simulation FreeWarp19 in \cite{held2026disc}, which was initialized with $\psi_\text{max} \sim 0.14$).

By orbit 50 the discs in the two runs appear to have evolved into completely different states. By this point, in the $a = 0.05$ run the disc has broken into five tilted rings, as described earlier. By contrast, the disc in the $a = 0.01$ run remains continuous, and is instead characterized by small-scale wave action (with characteristic features of size $\sim H_0$), representing the saturated state of the parametric instability. However, between orbit 50 and 60, we begin to observe new features in the density flow field in the $a = 0.01$ run (not shown). Over-densities appear at certain locations in the disc, giving it a `knotted' structure. Closer inspection reveals the emergence of gaps between these `proto-rings'. There is significant material between the gaps at this stage, which results in the formation, twice per orbit, of oblique shocks in the gaps as the tilted rings either side of the gaps move past one another. Despite embarking on a different evolutionary path over the first 50 orbits of the simulation compared to the $a = 0.05$ run, by orbit 70, the disc in the $a = 0.01$ run, too, has broken into five rings rings: a flat, left- (right-) ring spanning the radially periodic boundaries which extends from $|x|/H_0 \sim 32$--$45$, a central flat ring which extends from $x/H_0 \sim -10$ to $x/H_0 \sim 12$, and three, compact ($\sim H_0$), Gaussian-type tori centred around $x/H_0 \sim -28, 20, 25$. 

\begin{figure*}
    \centering
     \includegraphics[scale=0.5]{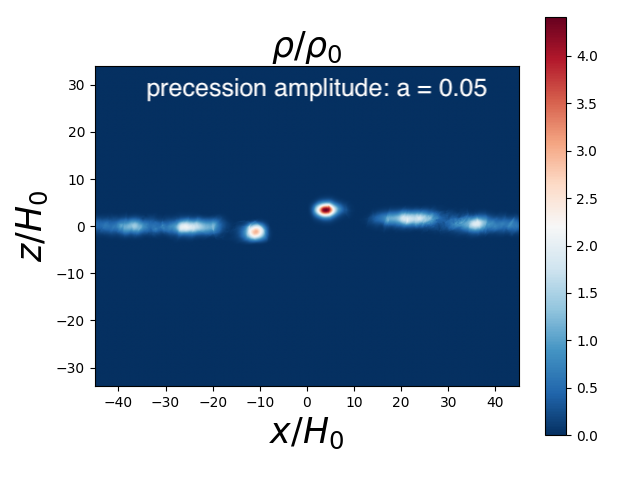}
    \includegraphics[scale=0.5]{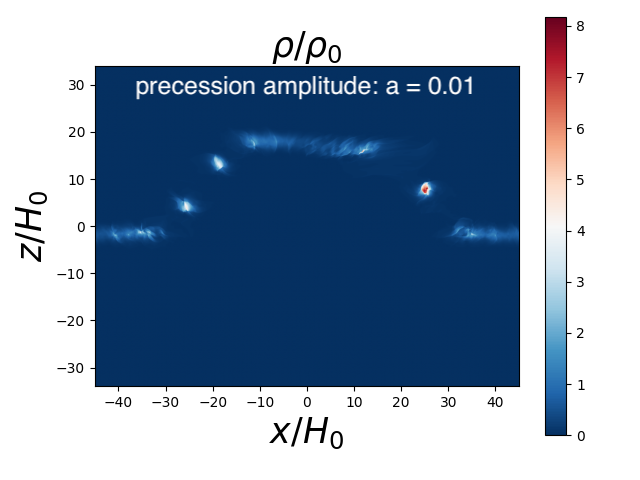}
s    \caption{Precession amplitude study: comparison of flow field (density in the $xz$-plane) at similar times (orbit 95) in two simulations run with the same initial disc tilt ($z_0 = 28H_0$), but with different precession amplitudes, $a = 0.05$ (left; simulation Precession02) and $a = 0.01$ (right; simulation Precession01)). The disc breaks into five rings in both cases, but the dissipation of the tilt is greater in the run with the larger precession amplitude, which causes the rings around $x \sim \pm30H_0$ to align and eventually rejoin.} 
    \label{FIGURE_PrecessionAmplitudeStudyFlowField}
\end{figure*}

The onset of breaking in the small-precession-amplitude run around orbit 50 is associated with a slight increase in the dissipation rate, as can be seen in time series of the vertical kinetic energy (yellow curve, bottom panel of Fig.~\ref{FIGURE_PrecessionAmplitudeStudy}). However, the rate of change is significantly smaller than in the large-precession-amplitude run. As a result, the broken disc segments in the small-precession-amplitude run have significantly larger tilts by the end of the simulation (orbit 100) than in the corresponding large-precession-amplitude run. Fig.~\ref{FIGURE_PrecessionAmplitudeStudyFlowField}  shows a comparison of the flow field near the end of the two simulations (orbit 95).

In conclusion, we have studied the consequences of decreasing the precession amplitude by a factor of five (from $a=0.05$ to $a=0.01$), keeping the initial disc tilt the same ($z_0 = 28H_0$). The discs in the two runs initially undergo demonstrably different evolution: in the $a = 0.01$ run the disc develops a gentle warp and becomes parametrically unstable, while in the $a = 0.05$ the disc develops a large warp, as well as large amplitude sloshing motions which converge into shocks. By orbit 50 the disc in the large-precession-amplitude run has already broken into five rings, while the disc in the small-precession-amplitude run remains continuous and characterized by small-scale wave action resembling the saturated state of the parametric instability. The disc in the small-precession-amplitude run eventually does break into five rings. The locations of the break are at $x/H_0 \sim -30, -25, -15, 25$  compared to $x/H_0 \sim -15, 5, 10$ in the $a = 0.05$ run. The rate of dissipation of the tilt following breaking is also significantly smaller in the small-precession-amplitude run: all the disc segments in that run retain significant tilt by the end of the simulation, whereas in the corresponding large-precession-amplitude run only the two central tori (rings C and D) retain an appreciable tilt.

\subsection{Parametrically unstable discs}
\label{RESULTS_ParameterStudy_SmallTiltSmallPrecession}
So far all the simulations that we have discussed exhibited disc breaking. In this section we discuss two simulations in which the disc is in a region of the tilt - precession amplitude space ($z_0, a$) in which it develops parametric instability (PI) but does remains continuous over the course of the simulation. 

\subsubsection{Parametric instability in a free warp}
\label{RESULTS_ParameterStudy_FreeWarpSimWithPI}
We first present a new analysis of an inviscid (wavelike-regime) \textit{free warp} simulation (FreeWarp19) first presented in \cite{held2026disc}. The disc was initialized with a small initial warp amplitude ($A = 2H_0, \psi_{\text{max}} \sim 0.14$) and run for 50 orbits. The resolution was $64/H_0$. The disc became parametrically unstable, and the PI was the dominant dissipation mechanism. 

Although we did not include an imposed precession in this simulation, we nevertheless observe a marked growth in the orbit-averaged complex tilt $\langle \text{Im}\{Z\} \rangle _{\tau}$. This develops a sinusoidal profile over the first orbit with amplitude $\langle \text{Im}\{Z\} \rangle _{\tau} \sim 0.008H_0$. The amplitude increases over the course of the simulation, and reaches $\sim 0.2H_0$ at orbit 28. This is in fact, comparable to the tilt at this stage, which peaks at around $\langle \text{Re}\{Z\} \rangle_{\tau} \sim 0.4H_0$ at the same time. Evidently the free warp becomes significantly twisted in the presence of parametric instability, in a way that cannot be explained by imposed differential precession (as there is none) or by numerical errors (see Appendix~\ref{APPENDIX_TiltDissipationStudy}). Presumably, then, the internal torques associated with the PI alter the phase relationships in the ideal Keplerian bending wave, as well as damping that wave.

\subsubsection{Parametric instability in a precessing disc}
\label{RESULTS_ParameterStudy_PrecessionDiskWithPI}
Next we turn to a simulation (Precession04) in which we did include a weak imposed precession. The disc in this run was initialized with a tilt of $z_0 = 14H_0$ and a precession amplitude of $a = 0.01$.

\begin{figure}
\centering
\includegraphics[scale=0.64]{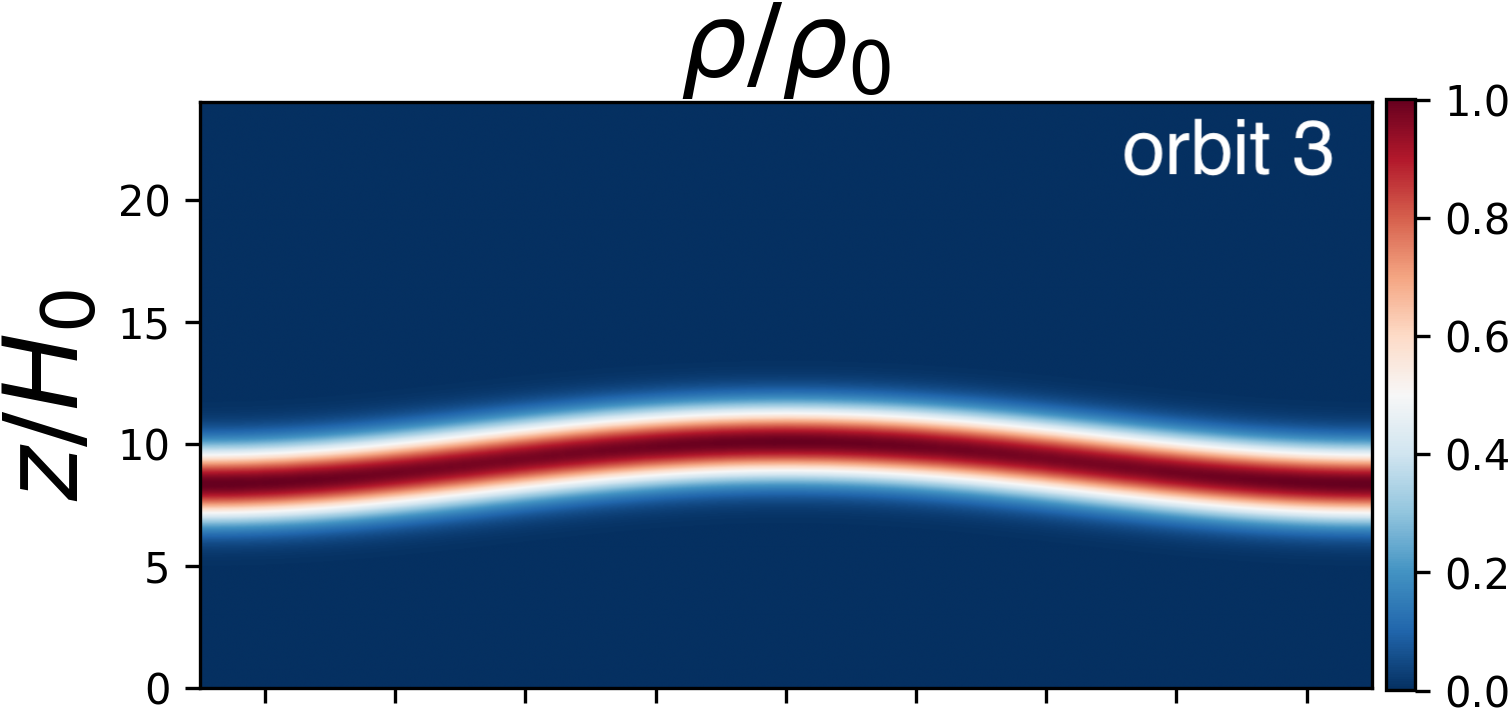}
\includegraphics[scale=0.64]{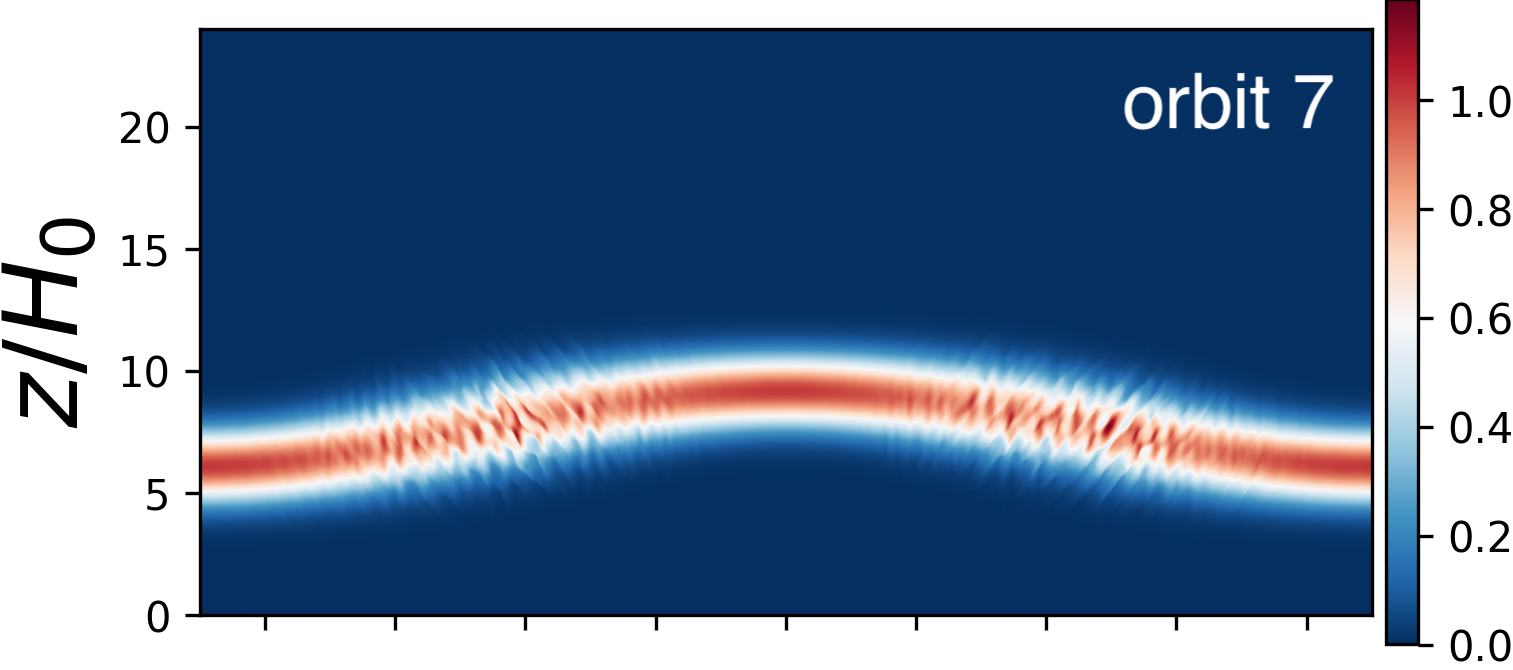}
\includegraphics[scale=0.64]{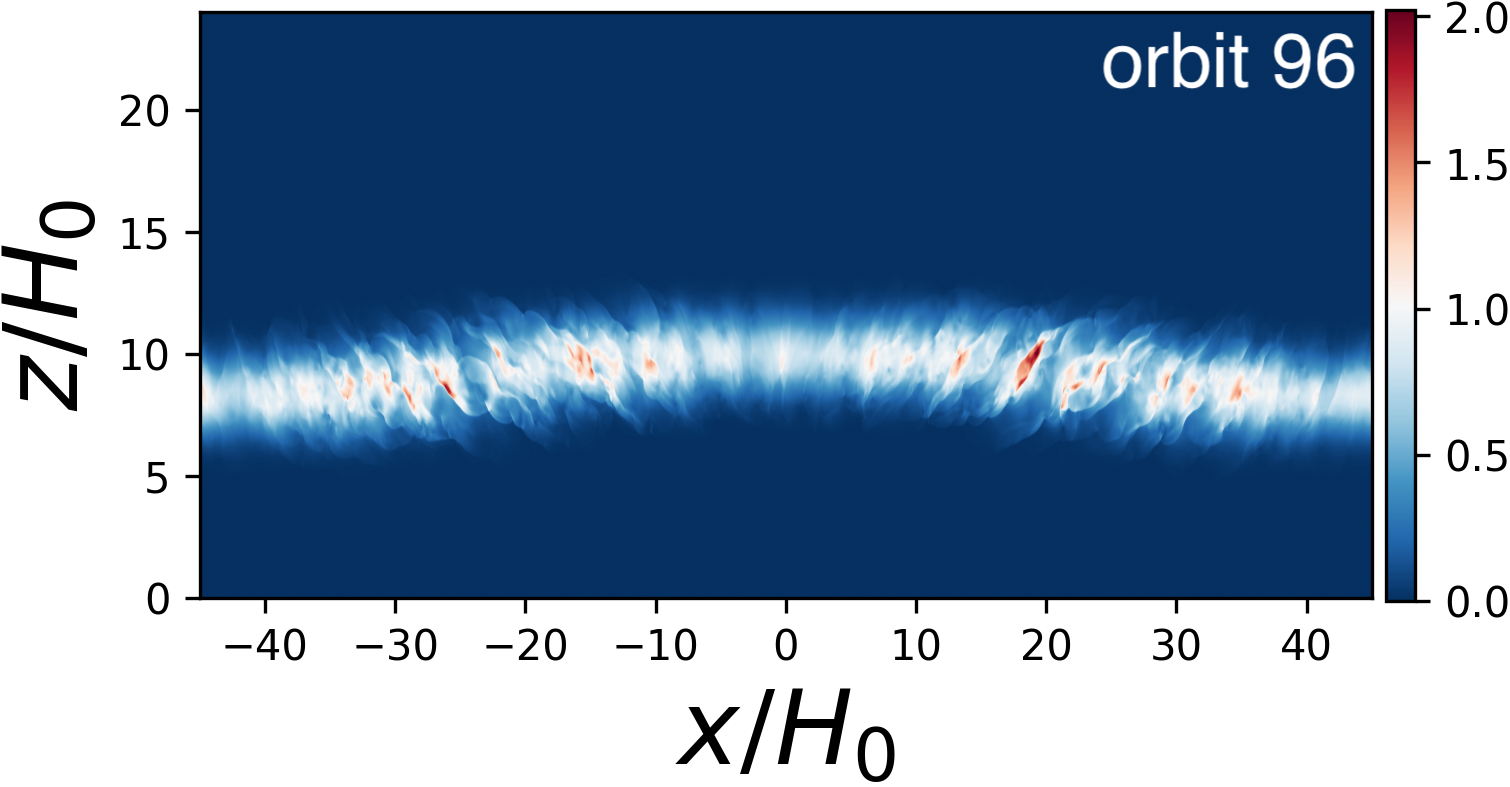}
\caption{Density snapshots in the $xz$-plane from a run (Precession04) with a smaller initial tilt ($z_0=14H_0$) and precession amplitude ($a = 0.01$) than those used in our fiducial simulation (compare to Fig.~\ref{FIGURE_FiducialSimulationFlowField}). The disc develops a small warp ($\psi_{\text{max}} \sim 0.2$) due to the imposed precession. Rather than breaking into rings, it becomes parametrically unstable.}
\label{FIGURE_SmallTiltAndPrecessionSimulation}
\end{figure}

In Fig.~\ref{FIGURE_SmallTiltAndPrecessionSimulation} we show density snapshots from the simulation in the $xz$-plane. The disc develops a gentle warp after the first few orbits on account of the imposed precession (top panel). We do not observe any shocks, indicating that the horizontal sloshing motions induced by the warp are small in amplitude. We confirmed this by inspecting orbit-averaged radial profiles (not shown): the warp amplitude grows over the first 8 orbits and peaks at around $\langle \psi \rangle_{\tau} \sim 0.2$, and the radial sloshing amplitude peaks around $\langle U_x \rangle_{\tau} \sim \Omega_0$. These values are comparable to those which we measured in the small-warp-amplitude ($\psi_\text{max} \sim 0.14$) free warp simulation (FreeWarp19) in \cite{held2026disc}. The resultant dynamics in our simulation here are indeed similar to those of that free warp simulation. The disc becomes parametrically unstable around orbit 7 (middle panel). The PI first develops where the warp amplitude is greatest (around $x \sim \pm 22H_0$, which is precisely where the rate of change of imposed precession is greatest. The disc does not break over the course of the simulation (100 orbits). Instead the PI spreads from its initial location throughout the disc, and by the end of the simulation (bottom panel) the entire disc is characterized by small-scale wave action due to the PI.

In Fig.~\ref{FIGURE_SmallTiltAndPrecessionTimeSeriesEkinxEkinz} we show the corresponding time-series of radial (black dash-dotted line) and vertical (blue solid line) kinetic energy (KE) density. These are considerably less complicated than in our fiducial simulation run with $a = 0.05$ and $z_0 = 28H_0$ (the reader may wish to compare to Fig.~\ref{FIGURE_FiducialSimulationTimeSeriesEkinxEkinz}). The radial KE is nearly always several orders of magnitude smaller than the vertical KE, as expected since the warp is too small for the horizontal sloshing motions to grow to large amplitudes. The envelope of the vertical kinetic energy decreases very gradually over the duration of the simulation. We do not observe any shocks in the simulation, and so the dominant source of dissipation is the parametric instability, which takes energy from the large scale flow and puts it into small scales. 

The volume-averaged turbulent stress $\langle R_{xy} \rangle$ (not shown) oscillates about zero twice per orbit, and has an envelope that peaks at $\langle R_{xy} \rangle \sim 0.035\,c_{\text{s}0}^2\rho_0$ around orbit 10 (which is when the PI saturates). It then beats with successively smaller peaks for the remainder of the simulation and is $\sim 0.015\,c_{\text{s}0}^2\rho_0$ around orbit 100. Thus the PI derives energy from the deformation (i.e.~from the warp) rather than from the background Keplerian shear, and is therefore not a good driver of angular momentum transport (but see caveats in the second-to-last paragraph of Section \ref{RESULTS_FiducialSimulation_TimeSeries}). Note that \cite{1994ryunonlinearevolution} carried out 2D local simulations of the PI in tidally distorted discs and drew similar conclusions about the efficacy of the PI with regards to driving accretion.

\subsection{Viscosity study: wavelike vs diffusive regimes}
\label{RESULTS_ParameterStudy_ViscosityStudy}
So far we have only considered inviscid simulations in which the disc was in the \textit{wavelike} regime of warp propagation. Now we include an explicit $\alpha$-viscosity in order to probe the \textit{diffusive} regime. For our set-up, the appropriate criterion for the diffusive regime is given by $\alpha > kH_0$ as explained in \cite{held2026disc}. For a radial wavelength of $k = 2\pi/L_x$ and a radial box size of $L_x = 90H_0$, this works out to be $\alpha > 0.07$. Therefore, to ensure that our runs are securely in the diffusive regime, we employ an explicit viscosity of $\alpha = 0.1$ in the simulations described in this section.\footnote{The warp damping rate is given by $\alpha\Omega_0/2$, which corresponds to a viscous damping period of $1/(2\pi)\times 2/(\alpha \Omega_0) \sim 3$ orbits.} We present a suite of simulations in the diffusive regime with different initial tilts ($z_0 \in \{28,38\}H_0$) and precession amplitudes ($a \in \{0.01, 0.05, 0.1\}$).

\subsubsection{Wavelike vs. diffusive regime: small precession amplitude}
We start by comparing the wavelike and diffusive regimes in discs with a small precession amplitude ($a = 0.01$) and a large initial tilt ($z = 28H_0)$, i.e.~simulation Precession01 (inviscid) vs. Precession08 ($\alpha = 0.1$).

The evolution of the disc in the diffusive regime is fairly simple. It does not become parametrically unstable, nor does it break over the duration of the simulation; instead it remains smooth, laminar, and continuous throughout the entire duration of the simulation. The disc develops a gentle warp over the first thirty orbits, with the orbit-averaged warp amplitude peaking at $\langle \psi \rangle_{\tau} \sim 0.7$. The orbit-averaged sloshing motions similarly grow from zero to a maximum of order $\sim \Omega_0$ over the same time-span. Both the warp amplitude and the sloshing motions peak where the rate of change of precession is strongest, i.e.~at $x/H_0 = -22.5, 22.5$. 

\begin{figure}
    \centering
    \includegraphics[scale=0.24]{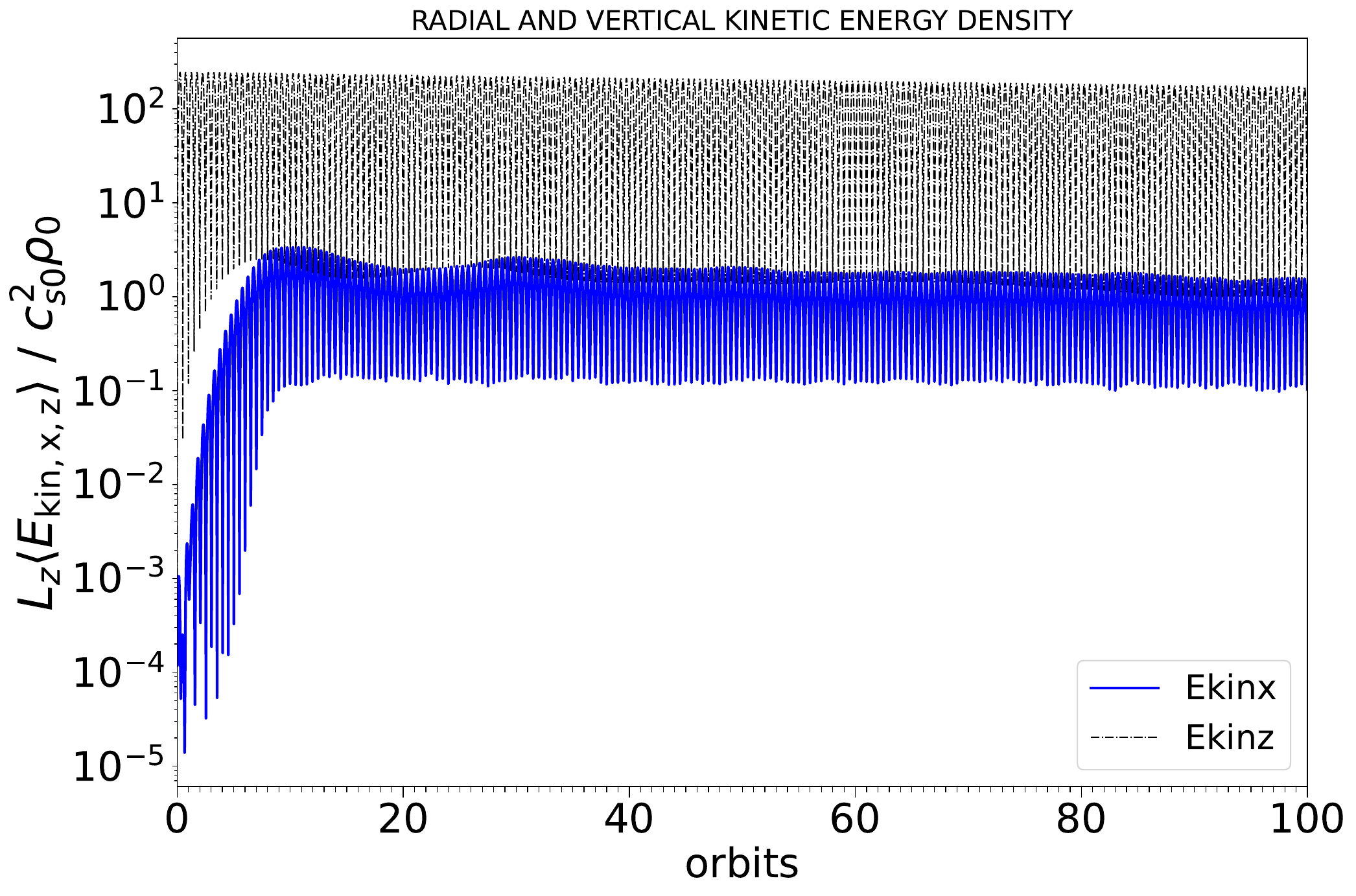}
    \caption{Time-evolution of volume-averaged radial (blue solid curve) and vertical kinetic energy density (black dot-dashed curve) from a run (Precession04) with smaller initial disc tilt ($z_0 = 14H_0$) and precession amplitude ($a = 0.01$) than those employed in our fiducial simulation (see Fig.~\ref{FIGURE_FiducialSimulationTimeSeriesEkinxEkinz}). This simulation was run in a box of smaller  vertical size ($L_z = 48H_0$) than our fiducial simulation was, so we have multiplied the volume-averages by the vertical box size to facilitate easier comparison between the two.}
\label{FIGURE_SmallTiltAndPrecessionTimeSeriesEkinxEkinz}
\end{figure}

The time-series of vertical and radial kinetic energy (not shown) bear greater similarity to those in the wavelike run Precession04 with $z_0 = 14H_0$ and $a = 0.01$ (see Fig.~\ref{FIGURE_SmallTiltAndPrecessionTimeSeriesEkinxEkinz}) than they do to Precession01 (yellow curves in Fig.~\ref{FIGURE_PrecessionAmplitudeStudy}), even though the latter simulation employed the same initial tilt ($z_0 = 28H_0$) and precession amplitude ($a = 0.01$) as the diffusive regime run. The radial KE saturates at $\langle E_\text{kin} \rangle \sim 5\times10^{-5}c_{\text{s}0}^2 \rho_0$. We do not observe beating in the envelope of the radial KE in the diffusive regime run, however. This is not surprising, given that this beating is caused by a bending wave propagating through the disc. In the diffusive regime the warp propagates in a diffusive manner, rather than as a wave, which explains the lack of a beat observed in this simulation.

Most of the energy is in the tilt rather than in the sloshing motions. The tilt (as measured by measured by $\text{Re}\{Z\}$; not shown) develops a double-peaked profile over the first ten orbits, with the peaks located near the locations where the change of precession is strongest ($x/H_0 \sim \pm 22$). The imaginary part of the complex tilt also develops a sinusoidal profile, with positive peaks at the radial boundaries and a negative peak at $x = 0$. The peaks and the trough grow over the first 35 orbits of the simulation, reaching extrema of $\text{Im}\{Z\} \sim 10$ at $x = \pm 45$ and $\sim -7.5$ at $x = 0$ at orbit 35, and decrease thereafter. In fact, the radial profiles for $\text{Re}\{Z\}$ and $\text{Im}\{Z\}$ resemble those shown in red and blue, respectively, in our theoretical plots of $Z=A\exp[\ii\omega_\text{n}(x)t]$ (see second-to-last panel in Fig.~\ref{FIGURE_ComplexTiltAndWarpAnalytical}). The theoretical result did not take into account pressure effects, and can be interpreted as the limit in which each fluid column oscillates independently of the others in the vertical direction. Given that the sloshing motions are small in the diffusive run, the coupling of different columns is likely also weak, so it is not surprising that the tilt profile is similar to the theoretical result. 

An interesting question is what is damping the tilt in the diffusive run: over the duration of the simulation (100 orbits) the tilt decreases from $\text{Re}\{Z\} \sim 28H_0$ at initialization to $\sim 20H_0$ by orbit 100. This decrease in the tilt by around $30\%$ mirrors the gradual change in tilt that we observe in the vertical kinetic energy. In the wavelike regime, the damping of the tilt was dominated first by parametric instability, and later by oblique shocks that formed within the gaps between broken disc segments. However, we do not observe any of these features in the diffusive run.\footnote{The sloshing motions in the diffusive run do converge into shocks at $x/H_0 = 0$ in the disc \textit{atmosphere}, i.e.~several ($\sim 10H_0$) scale-heights above and below the local disc-midplane. However, the density is very low in these regions, so the energy loss due to these atmospheric shocks is likely also low.} Aside from numerical dissipation, which is very small at the resolution of $32/H_0$ which we employ here (see Appendix \ref{APPENDIX_TiltDissipationStudy}), the only source of dissipation is the explicit viscosity itself. As in the linear theory of Section 3, viscous damping of the sloshing motions causes the tilt to decay, although we note that the disc is bouncing twice per orbit in the simulation (because we started with a large initial tilt of $z = 28H_0$), so the disc is in the nonlinear regime even though it is laminar.

\subsubsection{Wavelike vs. diffusive regime: large precession amplitude}
Next we compare the wavelike and diffusive regime in discs with a larger precession amplitude $a = 0.05$, keeping the initial disc tilt the same at $z_0 = 28H_0$, i.e.~Precession02 (inviscid) vs. Precession11 ($\alpha = 0.1$) regimes.

\begin{figure*}
\centering
\includegraphics[scale=0.43]{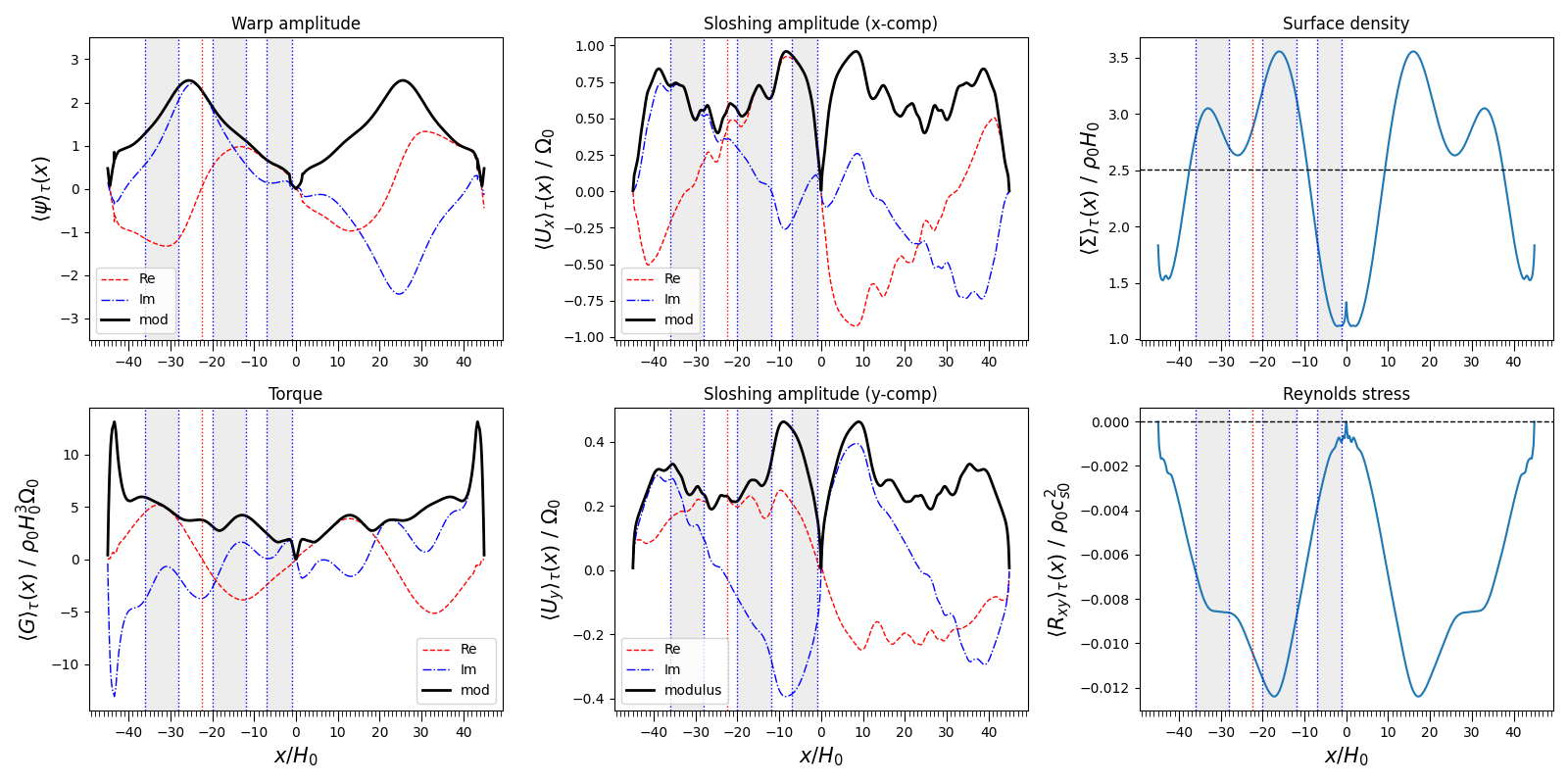}
\caption{Same as Fig.~\ref{FIGURE_Fiducial_OrbitAveragedRadialProfilesOrb0Orb1} but taken from a simulation (Precession11) in the \textit{diffusive} regime with $\alpha = 0.1$. The radial profiles have been orbit-averaged over orbit 16--17. The shaded regions show the approximation locations of the breaks (gaps) in the wavelike fiducial simulation (Fig.~\ref{FIGURE_Fiducial_OrbitAveragedRadialProfilesOrb0Orb1}).}
\label{FIGURE_ParameterStudy_DiffusiveRegimeOrbitAveragedProfilesOrb16to17}
\end{figure*}

\begin{figure*}
\centering
\includegraphics[scale=0.25]{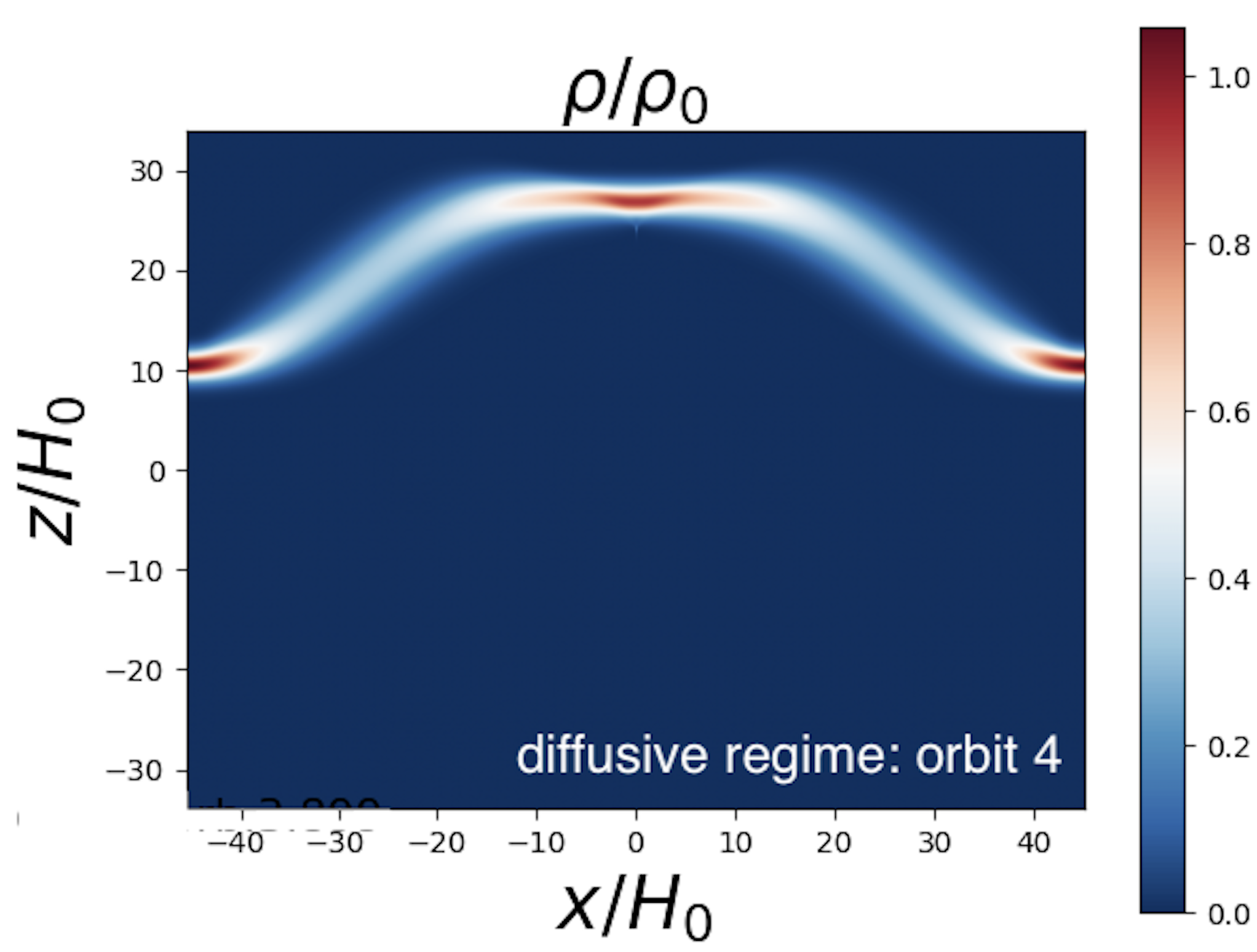}
\includegraphics[scale=0.25]{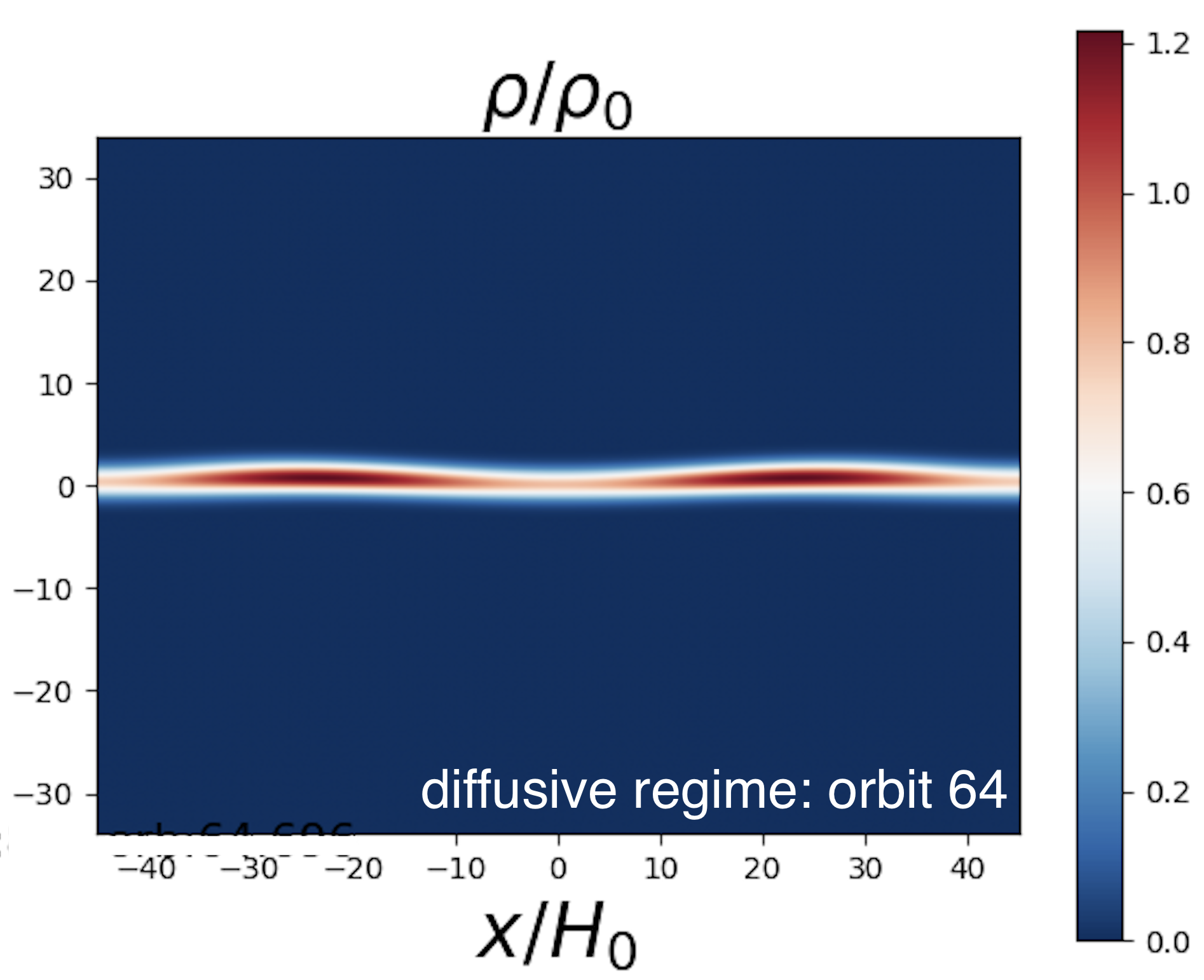}
\caption{Density snapshots taken at orbit 4 (left) and orbit 64 (right) showing the evolution of a disc under the action of an imposed precession in \textit{diffusive} regime with $\alpha = 0.1$ (Precession11). The simulation was run with an initial tilt of $z_0 = 28H_0$ and a precession amplitude of $a = 0.05$. The reader may wish to compare to Fig.~\ref{FIGURE_FiducialSimulationFlowField}, which shows the corresponding evolution of the disc in the wavelike regime.}
\label{FIGURE_DiffusiveRegimeFlowField}
\end{figure*}

At early times we observe similar dynamics in both the wavelike and diffusive runs. In the diffusive regime the disc, too, develops a large warp after initialization (see left-hand panel of Fig.~\ref{FIGURE_DiffusiveRegimeFlowField}). This is reflected in the orbit-averaged warp amplitude, which grows over the first 14 orbits of the simulation, though it saturates at a significantly smaller value of $\langle \psi_{\text{max}} \rangle_{\tau} \sim 2.5$ (see top-right hand panel of Fig.~\ref{FIGURE_ParameterStudy_DiffusiveRegimeOrbitAveragedProfilesOrb16to17}), compared to $\langle \psi_{\text{max}} \rangle_{\tau} \sim 9$ in the wavelike regime. After orbit 14, the warp amplitude decreases as the tilt is damped. The sloshing motions (middle panels) grow after initialization, too, and peak at $\langle U_x \rangle_{\tau} \sim \Omega_0$ at orbit 3. Despite the large explicit viscosity, the sloshing motions converge into shocks at $x/H_0 \sim 0$ around this time (one such shock is just visible below the disc at $x/H_0 = 0$ in the left-hand panel of Fig.~\ref{FIGURE_DiffusiveRegimeFlowField}), but the flow field is considerably more laminar than in the inviscid run (compare to top-right panel of Fig.~\ref{FIGURE_FiducialSimulationFlowField}). 

Even at a precession amplitude of $a = 0.05$, the disc in the diffusive regime does not break. The damping of the tilt is significantly greater in this run compared to its lower precession amplitude $a = 0.01$ counterpart (Precession08), which was discussed in the previous subsection. Note that even in the linear regime (i.e.~when there are no shocks and bouncing/compression of the disc is weak), when the only source of damping is viscosity, the tilt will damp more quickly when the precession amplitude is larger simply because there is a greater warp, and thus more energy in the sloshing motions, which are damped by the explicit viscosity. By the end of the simulation both the tilt and the warp have been almost completely erased, and the disc has effectively settled into a flat state aligned with the shearing box orbital plane at $z/H_0 = 0$ (see right-hand panel of Fig.~\ref{FIGURE_DiffusiveRegimeFlowField}). 

Interestingly, we observe significant structure in the orbit-averaged radial profile of the surface density (shown for orbit 16-17 in the top right-hand panel of Fig.~\ref{FIGURE_ParameterStudy_DiffusiveRegimeOrbitAveragedProfilesOrb16to17}). Its shape after the first orbit is very similar to that in the inviscid simulation (bottom right-hand panel of Fig.~\ref{FIGURE_Fiducial_OrbitAveragedRadialProfilesOrb0Orb1}), consisting of a sinusoidal profile with peaks near the radial boundaries and a broad trough at $x/H_0 \sim 0$. Converging sloshing motions at $x/H_0 \sim 0$ produce a narrow spike in the surface density there, although the spike is smaller in the diffusive run (it reaches a peak of $\langle \Sigma \rangle_{\tau} \sim 2.4\,\rho_0H_0$ by orbit 5, compared to $\sim 5.4\,\rho_0H_0$ at orbit 8 in the wavelike regime). Over the first 25 orbits of the diffusive regime run these spikes are smoothed out by viscous diffusion. The \textit{trough} in the surface density centred at $x/H_0 \sim 0$, on the other hand, actually deepens over this time, reaching as low as $\langle \Sigma \rangle_{\tau} \sim \,\rho_0H_0$. Nevertheless, we do not observe any sign of breaking in the density flow field, which appears smooth at this location, if under-dense compared to its initial value ($\Sigma (t=0) \sim 2.51\,\rho_0H_0$). Typically, locations where the disc actually breaks are associated with both a trough in the surface density \textit{and} a corresponding peak in the warp amplitude (e.g.~compare top-left and top-right-hand panels in Fig.~\ref{FIGURE_Fiducial_OrbitAveragedRadialProfilesOrb99Orb100}). In the diffusive regime run, on the other hand, the trough in surface density at $x/H_0 = 0$ is not accompanied by an increase in the warp amplitude at the same location. 

We note that the dominant dissipation mechanisms differ in the inviscid (wavelike) and viscous (diffusive) simulations. In the former, dissipation is dominated by shocks that form where the sloshing motions converge (at $x/H_0 \sim 0$). In the diffusive regime runs, on the other hand,  the imposed viscous diffusion can act along the entire length of the sloshing motions. Thus we expect the corresponding surface density evolution to differ as well.

In order to attempt to push the disc to break in the diffusive regime, we have run a simulation (Precession19) with an even larger initial tilt ($z_0 = 38H_0$), keeping the precession amplitude the same at $a = 0.05$. The warp amplitude (not shown) develops a significantly more narrow peak than what we found in the diffusive $z = 28H_0$ run and saturates at a slightly larger value of $\langle \psi \rangle_{\tau} \sim 3$. However, even with the larger tilt, the disc does not break, and the evolution is similar to that described in the preceding paragraphs. Finally, we have also run a simulation with an initial tilt of $z_0 = 28H_0$ and a very large precession amplitude of $a = 0.1$, although it is hard to motivate astrophysically how one would get precession this large. The results are similar to the run with $a = 0.05$ and $z_0 = 38H_0$.

\section{Study of interactions between broken disc segments}
\label{RESULTS_GAUSSIANRINGSTUDY}

A common feature of disc breaking that we have found is that the broken disc segments typically \textit{interact} with each other. This is the case both for the simulations with an imposed precession, which we discussed earlier in this paper, as well as in simulations of free warps discussed in \cite{held2026disc}. These broken disc segment interactions occur in the form of an oblique shock in the gap between the rings (see bottom-middle panel of Fig.~\ref{FIGURE_FiducialSimulationFlowField}), where supersonic gas streams with different tilts intersect. In the shearing box reference frame, these interactions occur twice per orbit. (The equivalent in global geometry would be the two azimuthal locations at which the planes of the mutually tilted rings intersect, i.e.~along the line of nodes of the pair of rings.) These interactions typically lead to the development of radially propagating waves/shocks and other small scale structure in the rings. Such disc segment interactions constitute a way for the different rings to remain in communication with each other, leading to the transfer of tilt (and possibly mass) between them. The the net effect is to significantly damp the tilt of the rings provided they are sufficiently close together.  Note that `streamers' linking broken disc segments have also been observed observed in global simulations \citep{liska2021disc, kaaz2023nozzle}.
 
So far we have considered runs in which we started with an initially continuous disc which subsequently broke into rings.  In this section we instead begin with two broken rings and monitor their subsequent interactions. We initialize simulations of two rings with different initial tilts ($z_0$) and an initial radial separation $L$. The rings are initialized with a Gaussian density structure in the radial and vertical direction (see equations \ref{EQUATION_GaussianRingsDensityProfile1}--\ref{EQUATION_GaussianRingsDensityProfile2}), mirroring the structure of some of the rings observed in our fiducial simulation (Section \ref{RESULTS_FIDUCIALSIMULATION}). The left-hand ring is always initialized without any tilt, while the right hand ring is given a tilt by displacing it from the shearing-box orbital plane at $z/H_0 = 0$. Motivated by the typical separations observed in previous simulations, we study three different initial separations between the ring centres of $L \in \{10, 15, 20\}H_0$. For each separation we investigated two different initial tilts for the right-hand ring: a smaller tilt of $z_0 = H_0$, and a larger tilt of $z_0 = 5H_0$. The simulation parameters are summarized in Table \ref{TABLE_GaussianRingStudy}. Note that we do not include an imposed precession in these simulations.

\begin{figure}
    \centering
    \includegraphics[width=1\linewidth]{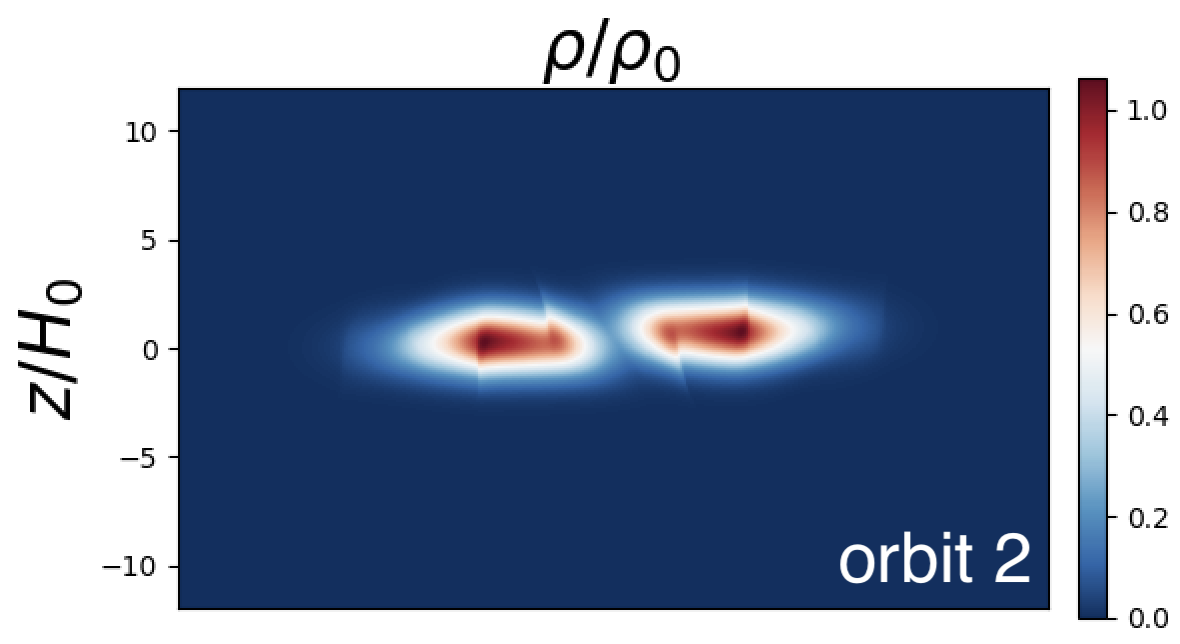}
    \includegraphics[width=1\linewidth]{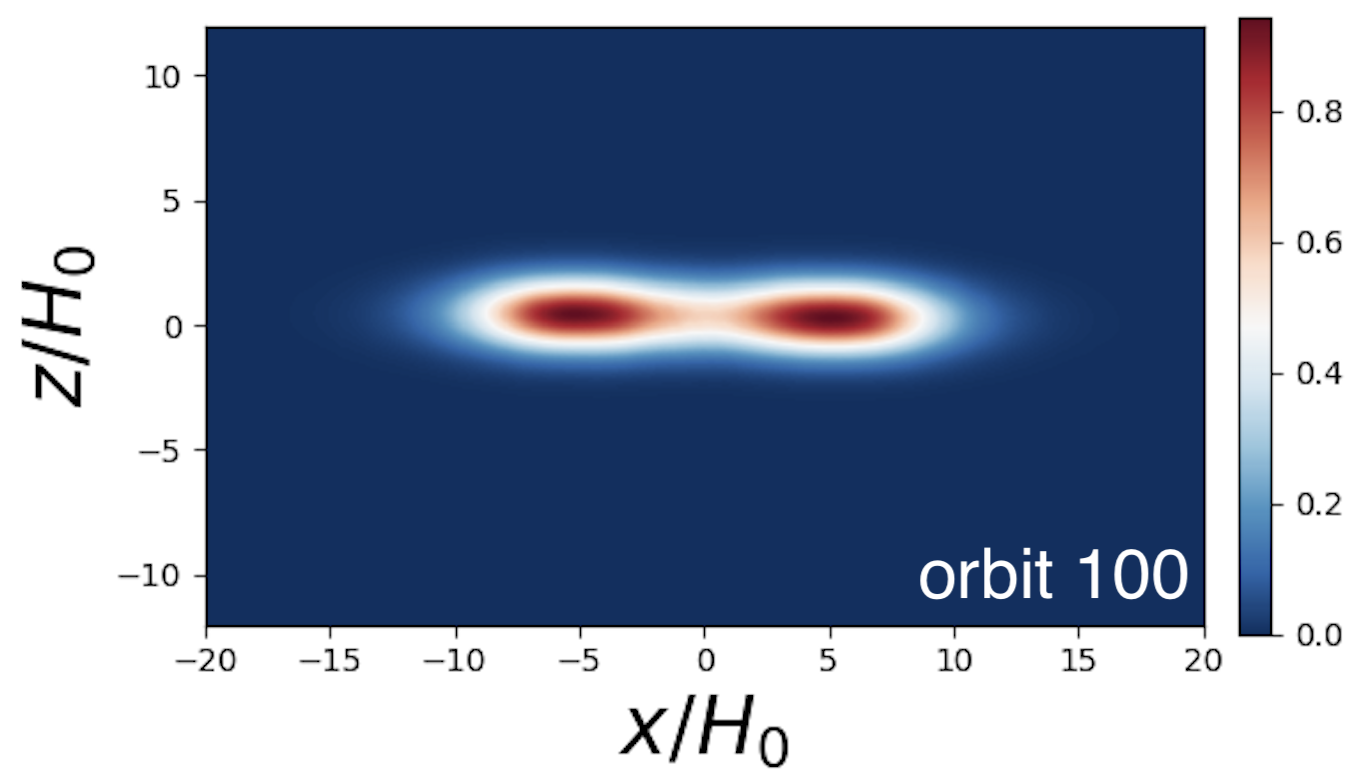}
    \caption{Density in $xz$-plane in a simulation of two broken disc segments (simulation GaussianRing01). The initial radial separation of the rings is $L=10H_0$, while their initial relative tilt is $z_0 = H_0$. Top panel: as the rings move past each other an oblique shock forms twice per orbit where they intersect. Bottom panel: by the end of the simulation the rings have synchronized their tilts, and the gap between them has been partially filled, indicative of disc `rejoining'. }
    \label{FIGURE_GaussianRing01FlowField}
\end{figure}

\subsection{Effect of increasing ring separation}
\label{RESULTS_GaussianRingStudy_EffectofIncreasingRingSeparation}
Let us first consider a run (GaussianRing01) in which we start with two rings whose centres are separated by $10H_0$ in the radial direction. We also initialize the rings with a small relative initial tilt of $z_0 = H_0$. An oblique shock forms at the interface between the two rings ($x/H_0=0$) as soon as the simulations start. See top panel of Fig.~\ref{FIGURE_GaussianRing01FlowField}, which shows the density in the $xz$-plane shortly after initialization. This occurs every time the rings come within a vertical distance of $H_0$ of each other (twice per orbit). These shocks subsequently propagate through the two rings. The shocks typically have a Mach number of around $M = \sqrt{\rho_2/\rho_1} \sim 1.1$, where $\rho_2/\rho_1$ is the compression ratio (i.e.~the density ratio across the shock).

Repeated interactions between the two rings excite a tilt in the left-hand ring within the first five orbits, and by orbit 15, the rings are oscillating in sync, indicating that they have adopted the same tilt. Oblique shocks are no longer excited in the gap between them, and the flow field appears considerably more laminar than it did shortly after initialization. As the tilts of the two rings synchronize we also observe that they appear to be \textit{rejoining} (bottom panel of Fig.~\ref{FIGURE_GaussianRing01FlowField}). We also observe this gradual filling of the gap in the orbit-averaged surface density (not shown): at initialization $\langle \Sigma \rangle_{\tau} \sim \,\rho_0H_0$ in the gap; by the end of the simulation it has increased to $\sim 1.5\,\rho_0H_0$.  It appears that the rings are thus joining up, similar to what we observed between the A and B (and E and A) rings in our fiducial simulation (see bottom-right panel of Fig.~\ref{FIGURE_FiducialSimulationFlowField}).

\begin{figure}
    \centering
    \includegraphics[scale=0.24]{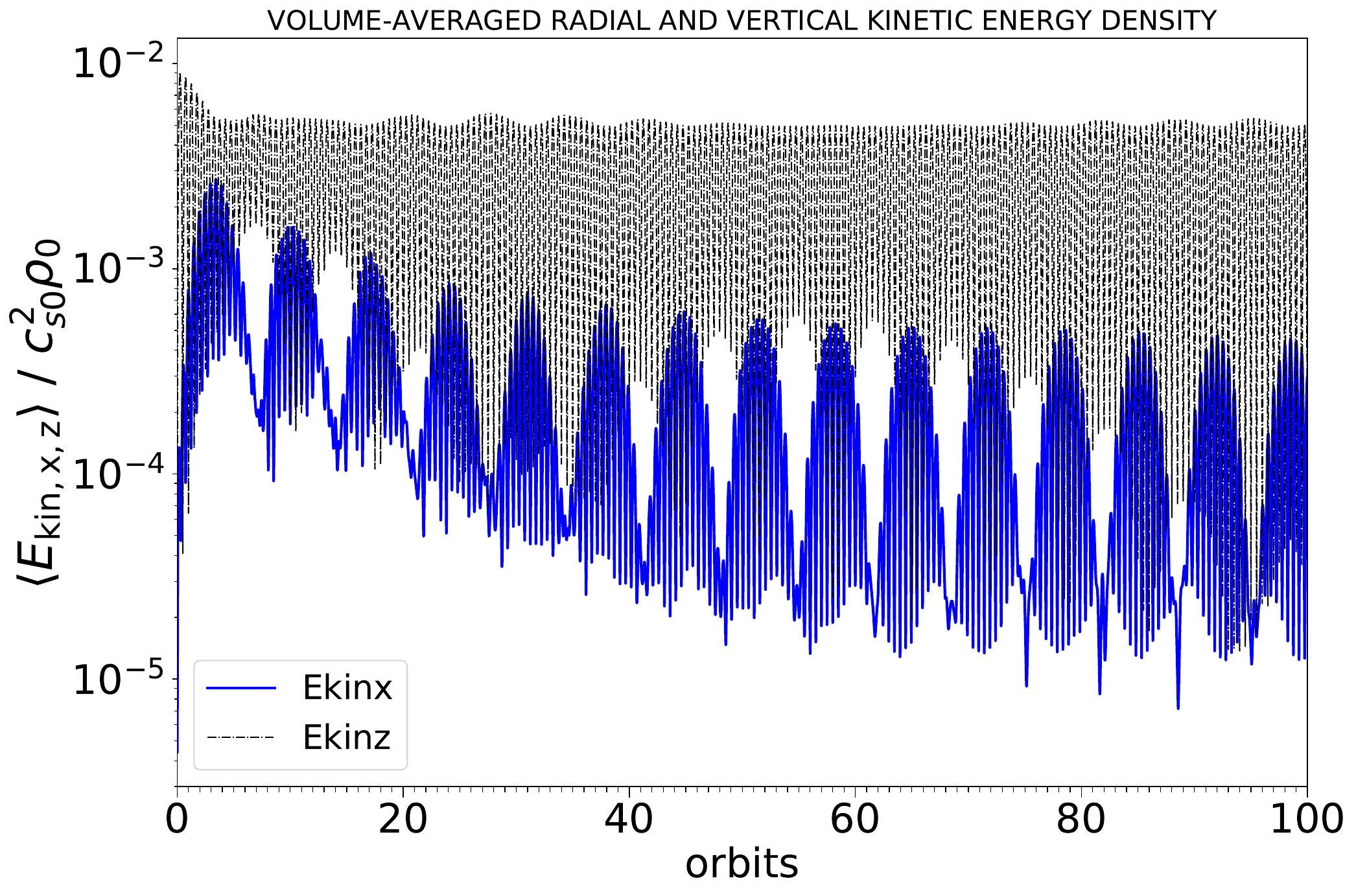}
    \caption{Time-evolution of volume-averaged radial (blue solid curve) and vertical kinetic energy density (black dot-dashed curve) taken from a simulation of two tilted broken disc segments (GaussianRing01). The initial ring separation was $10H_0$, and the right-hand ring was initialized with a tilt of $z_0 = H_0$. }
\label{FIGURE_GaussianRing01TimeSeriesEkinxEkinz}
\end{figure}

The behaviour mentioned in the flow field is mirrored in the volume-averaged vertical and radial kinetic energy (KE) densities (Fig.~\ref{FIGURE_GaussianRing01TimeSeriesEkinxEkinz}), which both oscillate twice per orbit on short time-scales, while their envelopes exhibit variations on longer time-scales. The vertical KE decreases rapidly within the first four orbits, dropping from its first peak of around $\langle E_\text{kin,z} \rangle \sim \,9\times 10^{-3} \rho_0 c_{\text{s}0}^2$ to $\sim 5.5\times 10^{-3}$. During the same time-interval the radial KE increases from its initial value of $\langle E_\text{kin,x} \rangle \sim 5\times10^{-6} \rho_0 c_{\text{s}0}^2$ to $\sim 3\times 10^{-3} \rho_0 c_{\text{s}0}^2$. From orbit 4 onward,  the envelope of the vertical KE remains effectively unchanged for the remainder of the simulation, indicating that the damping of the tilt is very small once the rings have synchronized their tilts and oblique shocks can therefore no longer form in the gap between them. The envelope of the radial KE exhibits a distinct beating with a period of $\sim 7$ orbits. This is only slightly larger than (half) the 
period of a bending wave with radial wavelength $k=2\pi/L_x$ equal to the radial box size $L_x = 40H_0$, given by $(1/2)\times 1/[(1/2)c_{\text{s}0}k] \sim 6.3$ orbits.\footnote{A bending wave with period $T$ will result in a beat in the radial kinetic \textit{energy} with period $(1/2)T$.}  We conclude that the beat is caused by a underlying bending wave propagating through the domain.\footnote{Although we have a highly non-uniform density in these simulations, there is still material far away from the ring centres that a long-wavelength bending wave can propagate in, and bending wave theory predicts that the wave should propagate independently of the surface density. This does, however, raise the question of whether a bending wave reflects at the edge of a disc, and where this edge actually occurs.}

\begin{figure}
    \centering
    \includegraphics[width=1\linewidth]{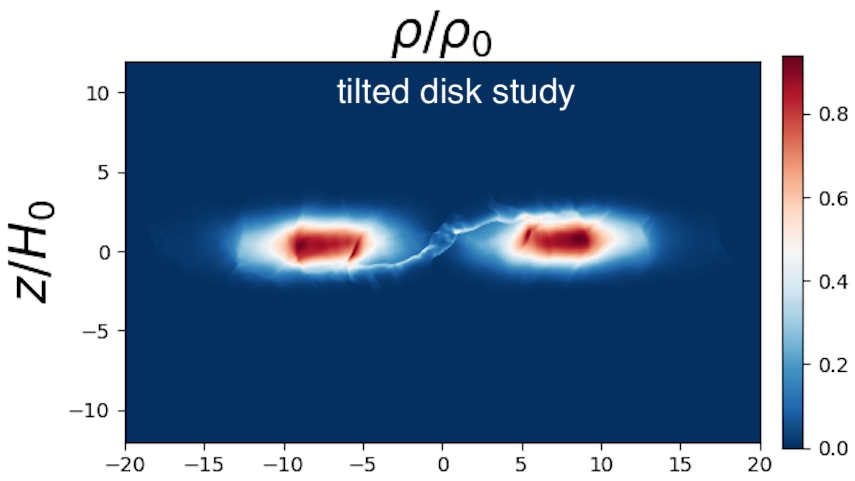}
   \includegraphics[width=1.07\linewidth]{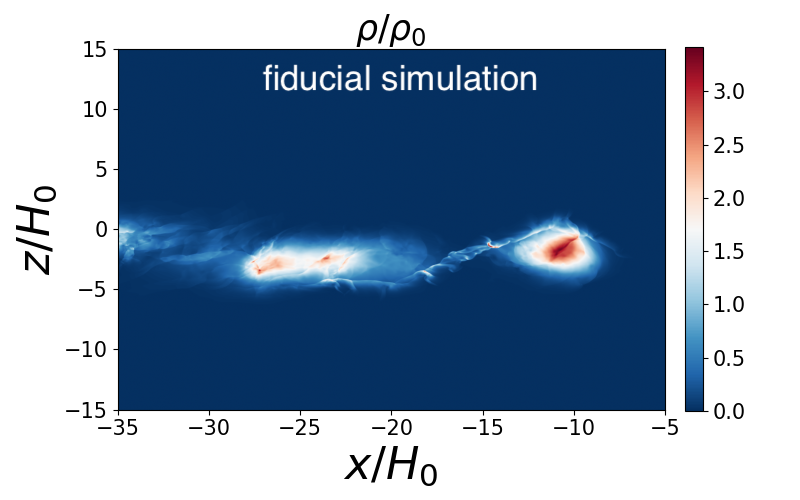}
    \caption{Comparison of interactions between broken disc segments. Top panel: density snapshot (orbit 9) from a simulation (GaussianRing06) in which we started with two broken disc segments separated by $L=15H_0$ with a relative tilt of $z_0 = 5H_0$. As we move around an orbit, the right hand ring oscillates about $z = 0$ twice per orbit, and an oblique shock forms in the gap between the rings. Bottom panel: snapshots taken at orbit 68 from our fiducial simulation (Precession02; see  Section \ref{RESULTS_FIDUCIALSIMULATION}.). The imposed precession naturally caused the disc to break into rings, and the flow field bears many similarities to the top panel.}
    \label{FIGURE_GaussianRing06FlowField}
\end{figure}

Next, we briefly comment on what happens as we increase the radial distance $L$ between the two rings (keeping their initial relative tilt the same at $z_0 = H_0$). When $L=15H_0$ (simulation GaussianRing04), we observe a similar outcome as when the separation was $10H_0$. An oblique shock forms in the gap between the rings twice per orbit as they move past each other. These shocks propagate through the disc over the first 30 orbits of the simulation, although the shocks are weaker than in the $L=10H_0$ run with typical Mach numbers of $M = \sqrt{\rho_2/\rho_1} \sim  1.05$. The maximum value reached by the volume-averaged radial kinetic energy density (not shown) is an order of magnitude smaller in this run than in the aforementioned run in which the rings were initially separated by $10H_0$, and the beating in the radial KE is less prominent, too, especially after 20 orbits. The envelope of the vertical KE also decrease more gradually than it does in the $L=10H_0$ run ($\langle E_{\text{kin}z}\rangle$ drops from a maximum of around $10^{-2}\,c_{\text{s}0}^2\rho_0$ within the first orbit to around $6\times10^{-3}\,c_{\text{s}0}^2\rho_0$ by orbit 26), and exhibits a more irregular beat with a longer period (between 26 and 29 orbits). Over the duration of the simulations the rings gradually synchronize their tilts, although this takes longer than in the simulation where the rings were initialized closer together in the radial direction. We again observe a gradual filling of the gap between the rings: the minimum orbit-averaged surface density in the gap increases from its initial value of  $\langle \Sigma \rangle_{\tau} \sim 0.1\,\rho_0H_0$ to $\sim 0.5\,\rho_0H_0$ over the duration the simulation (100 orbits).

Finally, in the simulation in which the ring centres have the largest radial separation of $L=20H_0$ (GaussianRing07), we do not observe any clear signs of oblique shocks forming between them in the 2D density flow field. We do observe what could best be described as waves (rather than shocks) propagating through both rings, indicating that there is still some interaction between them even at the largest separation. As in the previous cases, this interaction gradually excites a small tilt in the left-hand ring, although the rings do not have time to synchronize their tilts over the duration of the simulation. 

\subsection{Effect of increasing ring tilt}
\label{RESULTS_GaussianRingStudy_EffectofIncreasingRingTilt}
Next we investigate the effect of increasing the initial \textit{tilt} $z_0$ of the right-hand ring. We first describe a run (GaussianRing06) in which the initial relative tilt of the rings is $z_0 = 5H_0$, and the radial separation between the ring centres is $L=15H_0$. This most closely matches the relative tilt and separation observed between the broken disc segments in our fiducial simulation with precession, in particular the B and C, and D and E rings (see Fig.~\ref{FIGURE_FiducialSimulationFlowField}).\footnote{Our fiducial simulation includes an imposed precession, whereas our simulation of interacting rings does not. However, the precession acts on long time-scales ($\sim 56$ orbits at the location of the C ring), whereas a lot of the other dynamics that are relevant to breaking appear to take place on shorter time-scales, so the comparison should be valid.}

A strong oblique shock forms between the two rings almost immediately after initialization (see top panel of Fig.~\ref{FIGURE_GaussianRing06FlowField}). Depending on the snapshot, we measure Mach numbers $M = \sqrt{\rho_2 / \rho_1}$ between $M \sim 1.2$ and $1.9$. In the bottom panel, we show a snapshot taken from our fiducial simulation (Section \ref{RESULTS_FIDUCIALSIMULATION}) around orbit 68, by which time the disc has broken into several rings. The flow field appears remarkably similar in the two cases, suggesting our simple two-ring set-up is capturing similar interactions between the rings to those that emerged naturally from disc breaking in our fiducial simulation.

After about 15 orbits these periodic interactions  have excited a tilt in the initially untilted left-hand ring. Due to shock damping, the vertical KE (not shown) drops from a maximum of around $\langle E_\text{kin,z} \rangle \sim 2.5\times 10^{-1}\,c_{\text{s}0}^2\rho_0$ to around $ 10^{-1}\,c_{\text{s}0}^2\rho_0$ over 40 orbits. By orbit 50 the two rings are essentially oscillating in tandem, the flow field appears considerably smoother, and the envelope of the vertical KE remains steady for the remainder of the simulation. We also observe a beating in the radial KE (not shown), but it is much less pronounced than in the run with the smaller initial tilt and radial separation between the rings shown in Fig.~\ref{FIGURE_GaussianRing01TimeSeriesEkinxEkinz}. Finally, we again observe a gradual filling of the gap between the rings: the minimum orbit-averaged surface density in the gap increases from its initial value of  $\langle \Sigma \rangle_{\tau} \sim 0.25\,\rho_0H_0$ to $\sim 0.5\,\rho_0H_0$ over the duration the simulation (100 orbits).

In summary, our simulations of tilted rings with radial separations between $10$--$20H_0$ have revealed repeated interactions between the rings in the form of waves and shocks. These interactions have the net effect of synchronizing the ring tilts by transferring the misaligned component of angular momentum between them, and (particularly when shocks are present) of damping the tilts. This occurs on time-scales of tens of orbits when the ring separations are $<15H_0$, while for separations of $20H_0$ or greater, the interactions are weaker, and synchronization of the tilts takes longer (likely $>100$ orbits). We do not observe any significant radial migration of either ring relative to its initial position. However, for small ($<15H_0$) ring separations we observe a noticeable filling of the gap between the two rings as they synchronize, indicating that interactions between broken ring segments cause the discs to rejoin. 

This behavior is similar to what we observe between the A and B, and the E and A rings, in our fiducial simulation with precession (see Fig.~\ref{FIGURE_FiducialSimulationFlowField}), and, to a lesser extent between the B and C, and the D and E rings. Conversely, the deep gap that forms between the C and D rings in that run indicates that the break is sustained. This implies that an active mechanism must be at work in that run to maintain the breaks against these tendencies towards synchronization and gap refilling.

\section{Discussion}
\label{RESULTS_DISCUSSION}

\subsection{Astrophysical applications}
\label{DISCUSSION_AstrophysicalApplications}
Differential precession in discs can be driven by both Newtonian and relativistic effects \citep{fragile2026tilted}. The former occurs in the presence of a potential that differs from that of a point mass and leads to equal and opposite apsidal precession. Relativistic precession, on the other hand, drives an apsidal precession that is greater than nodal precession precession, and they are both in the same direction.

In this paper we have taken nodal and apsidal precession to be equal and opposite (i.e.~we have taken $b = a$ in equations \ref{EQUATION_gxprecession}--\ref{EQUATION_gzprecession}). Our simulations are hydrodynamic, and most of them are in the wavelike regime for which $\alpha < H/r$, which corresponds to $\alpha < kH_0$ in the local formalism \citep{held2026disc}. Applications include the inner regions of tilted circumbinary protoplanetary discs, or the outer regions of circumstellar discs around a young star in the presence of a tilted external companion. 

\subsubsection{Astrophysical applications: wavelike regime}
An interesting question is whether precession can be induced in a disc by means other than a potential that differs from that of a point mass, or relativistic effects. MWC 758 is a protoplanetary disc around a Herbig Ae star known for exhibiting spiral patterns in scattered light \citep{boehler2018complex}. Recently \cite{winter2025exoalma} have inferred a small warp (of amplitude $\psi \sim 0.1$) and twist in the disc from residual velocity maps. However, no misaligned companion star (or planet) has been detected in this object, so it is unclear what is causing the disc to develop a twist. In \cite{held2026disc} we carried out a simulation (FreeWarp19) of a \textit{free} warp (i.e.~without an imposed precession) in the wavelike regime with a small initial maximum warp amplitude of $\psi \sim 0.14$, which happens to be in the same regime as MWC 758. We found that the disc in our simulation became parametrically unstable. We have reanalysed data from this simulation, and found that the disc actually does develop a small twist even though we have not included any precession in that run (see Section \ref{RESULTS_ParameterStudy_FreeWarpSimWithPI}). Thus we expect that a disc with a small warp amplitude ($\psi \sim 0.1$) will become parametrically unstable, and that the PI might be able to induce a small twist (or warp) in the disc.

\subsubsection{Astrophysical applications: diffusive regime}
To reach the diffusive regime ($\alpha > H/r$, which corresponds to $\alpha > kH_0$ in the local formalism) we have included an effective viscosity in select simulations by means of an alpha viscosity with $\alpha = 0.1$. The astrophysical application is to hot discs that are sufficiently ionized that they are unstable to the magnetorotational instability. Applications include dwarf novae discs around white dwarfs in binaries (at least in their hot, bright, outburst phases). In these systems, apsidal and nodal precession are induced in the outer parts of the disc by the external stellar companion. Another application is to circumbinary discs around black holes (or neutron stars), because the inner disc does not get very close to the central binary, so Newtonian precession will be more important there than relativistic precession, or to the outer regions of circumstellar discs around a black hole (or neutron star) in a binaries with a tilted external companion.\footnote{If the disc is being fed via Roche lobe overflow, it will tend to form in the binary orbital plane, but it can become tilted out of the plane by an instability such as the radiation-driven warping instability \citep{pringle1996self, wijerspringle1999warped}.} The nodal precession of such tilted discs has been used to explain superorbital variability in several XRBs \citep[e.g.][]{ogilviedubus2011precessing}.

Finally, although we have not considered it in this paper, our formalism allows for the case where apsidal precession is greater than nodal precession and in the same direction: this could be achieved by taking $a < 0, b > 0$ and $|a| > |b|$ in equations \ref{EQUATION_gxprecession}--\ref{EQUATION_gzprecession}. This then mimics the effects of relativistic precession. Applications include modelling the inner regions of discs around tilted black holes.

\subsection{Comparison with disc breaking models}
\label{DISCUSSION_ComparisontoTheory}
We already summarized various disc breaking models in section~6 of \cite{held2026disc}. Here, we briefly reiterate those arguments that rely explicitly on precession (in the wavelike regime) or on viscosity (in the diffusive regime).

\subsubsection{Comparison to theory: wavelike regime}
A popular argument for disc breaking by means of differential precession in the \textit{wavelike} regime involves comparing the time it takes for a bending wave to propagate across the disc to the (differential) precession time-scale \citep{papaloizou1995dynamicsoftilted, lubow2018linear, rabago2024warps}, though we caution that this is a linear concept. When the bending-wave timescale is shorter, the disc can communicate effectively and maintain coherent precession, similar to a rigid body, with only a small warp in comparison to its tilt. When the precession timescale is shorter, the warp is large in comparison to the tilt. Since breaking is an outcome of nonlinear warp dynamics, it is more likely to occur when the ratio of the bending-wave timescale to the precession timescale is larger and the disc behaves less rigidly. Nevertheless, it is still true that the tilt must exceed a threshold in order for the warp to be sufficiently nonlinear to break.

Based on the linear theory of tilted discs subject to differential precession which we developed in Section \ref{THEORY}, the bending-wave angular
frequency, in the absence of viscous damping, is given by $\omega = (c_{\text{s}0}/2)k$ (see equation \ref{EQUATION_BendingWaveFrequency}). The corresponding bending-wave period in our wavelike runs should then be $1/[(c_{\text{s}0}/2)k] \approx 28.6$ orbits. (We confirmed this by measuring a beat of about half that, i.e.~14.3 orbits in the kinetic energy time series.) In our fiducial simulation, we set the precession amplitude to $a = 0.05$. The corresponding precession angular frequency is given by $|\omega_\text{n}/\Omega_0| = (a/2) \cos(kx)$ (see top panel of Fig.~\ref{FIGURE_ComplexTiltAndWarpAnalytical}). At $x/H_0 = 0$, which is the location of the the right-hand edge of the gap that forms between the C and D rings (see bottom-middle panel of Fig.~\ref{FIGURE_FiducialSimulationFlowField}), the precession angular frequency is $\omega_\text{n} = 0.025\Omega_0$, which corresponds to a precession period of $2\pi/\omega_\text{n} = 40$ orbits. 

Thus the precession period is actually slightly longer than the bending-wave period in our fiducial simulation, indicating that the disc behaves semi-rigidly. Our linear theory provides an estimate for when it is expected to break down because of nonlinearity (in the form of significant compression, or `bouncing', of the disc; see Section \ref{METHODS_DiscussOfNonlinearEffects}). In the wavelike regime, linear theory should be valid provided that $a^2|z_0/H_0|^2\lesssim2kH_0$ , where $z_0$ is the initial disc tilt and $k = 2\pi/L_x$ is the wavelength of the bending wave. For the radial box size of $L_x = 90H_0$ which we employ in all our simulations, the RHS of the criterion is then $2kH_0 \sim 0.14$. Our fiducial simulation (Precession02: $a = 0.05$, $z_0 = 28H_0$), which exhibited disc breaking, is thus firmly in the non-linear regime, whereas our simulation Precession04 ($a = 0.01, z_0 = 14H_0$), in which the disc became parametrically unstable but did not break, is in the linear regime.

\subsubsection{Comparison to theory: diffusive regime}
\cite{dougan2018instability} developed a theory of disc breaking in the diffusive regime involving an instability of the warp amplitude. We investigated a number of initial disc tilts ($z_0 \in \{28, 38\}H_0$) and precession amplitudes ($a \in \{0.05, 0.1\}$) in the diffusive regime (see Section \ref{RESULTS_ParameterStudy_ViscosityStudy}). As explained in Section 5.2 of \cite{held2024instability}, the usual criterion for diffusive warp propagation ($\alpha > H/r$) is modified in the local formalism to $\alpha > kH_0$. We included an explicit alpha viscosity of $\alpha = 0.1$ in all our diffusive regime runs. For this $\alpha$ viscosity the stability boundary predicted by \cite{dougan2018instability} occurs at $\psi_{\text{crit}} \sim 2.7$ (see their fig. 5).

We observed growth in the warp amplitude after initialization, and a distinct sharpening of the peak in $\psi$ as it grows, which is indicative of the anti-diffusion of the warp predicted by \cite{dougan2018instability}. However, the disc did not develop a break in any of our diffusive regime runs. At ($a=0.05, z_0 = 28H_0$) the warp amplitude peaked around $\psi_\text{max} \sim 2.5$, which fell just below the critical value. At even larger tilts ($z_0 = 38H_0$) or precession amplitudes ($a = 0.1$), $\psi_\text{max}$ did reach as high as $ \sim 3.5$, however, the disc exceeded the critical value of 2.7 only in a narrow annulus of width around $2H_0$, and only for a short amount of time (around 13 orbits). 

We note that for $\alpha = 0.1$ and warp amplitudes between $\psi = 2.7$ and 3.5, the instability predicted by \cite{dougan2018instability} actually has a very small growth growth rate (see their fig. 3). Thus, even though the warp amplitude exceeds the critical value needed for instability in some of our diffusive regime runs, it does so only for a short time and in a limited region of space, so probably the instability does not have time to manifest. For the warp wavelength of $\lambda = 90H_0$ used in our simulations it is difficult to probe smaller viscosities while still remaining in the diffusive regime. Therefore, future work of disc breaking in the diffusive regime using our formalism should investigate different wavelengths.

\section{Conclusions}
\label{CONCLUSIONS}
Accretion discs undergo differential precession in the presence of a non-spherical potential (such as a circumbinary disc), or due to relativistic effects (such as a tilted disc surrounding a spinning black hole). Global simulations have found that such tilted, precessing discs can break or tear into rings, though the breaking region is often poorly resolved, and the disc is in a non-steady state due to the short time interval over which the simulations are run. 

To overcome these limitations, we have developed a formalism for modelling differential precession in local (shearing-box) simulations (see Fig.~\ref{FIGURE_TiltedDiskShearingBoxView}). The formalism requires only minor modifications to the standard shearing-box equations. This enabled us to  carry out the first local simulations of disc breaking (or tearing) and parametric instability under the effects an imposed differential precession. This work extends our previous study of disc breaking and parametric instability in a freely evolving warp \citep{held2026disc}.

Our key result is that, for a sufficiently large initial tilt ($z_0$) and precession amplitude ($a$), a tilted disc in the wavelike regime becomes warped and twisted, and ultimately breaks into five rings over a time-span of around 60 orbits. We observe novel behaviour compared to our earlier work on the evolution of free (unforced) warps: when precession drives the warp, the disc breaks into a greater number of rings, and the morphology of the rings is more variable. The disc breaks into both compact rings (of size $\sim H_0$), which more closely resemble tori rather than flat discs, and which tend to be separated by deep gaps allowing them to precess independently of one another, as well broader, flatter rings (of radial size $\sim 10H_0$) which tend to be separated by shallower gaps. We observe significant interactions between disc segments in the form of oblique shocks that form in the gaps between them. These shocks damp the relative tilt between the rings, ultimately causing the gaps between them to refill as they align -- a clear example of disc rejoining.

We have carried out a detailed parameter study of different initial disc tilts ($z_0 \in \{14, 28\}H_0)$ and precession amplitudes $a \in \{0.01, 0.05\}$. When both the tilt and the precession are sufficiently small ($z_0 = 14H_0, a = 0.01$), the disc becomes parametrically unstable and remains continuous instead of breaking. We have extended the linear theory of bending waves first presented in \cite{held2026disc} to include the effects of precession as well as viscosity. Finally, by including an effective viscosity of $\alpha = 0.1$, we have also investigated warp evolution in the \textit{diffusive} regime, which is relevant to hot, fully ionized discs around tilted black holes. At the tilts ($z_0 \lesssim 38H_0$) and precession amplitudes ($a = 0.05$) which we investigated, the disc in the diffusive regime does not break.

Finally, to investigate the impact of numerical dissipation on tilted discs, we have carried out a suite of simulations (without any imposed precession) at different resolutions ($8,16$ and $32/H_0$) and initial disc tilts ($z_0 \in \{10,28\}H_0$). At low resolutions ($8/H_0$), which are comparable to the resolution in current global simulations, we find that numerical dissipation both significantly damps the disc tilt and induces an artificial rigid precession in the disc. These effects decrease as a power law with resolution (see Fig.~\ref{FIGURE_TiltedDiskResolutionStudy}).

For simplicity, we have assumed that apsidal and nodal precession are equal and opposite in this work, which is the case when the disc is tilted with respect to the symmetry plane of a non-spherical potential, such as in a circumbinary disc or the outer radii of a disc in a binary with a tilted external companion. However, our formalism can easily be adapted to study the case where apsidal precession is greater than nodal precession (and in the same direction). This would be applicable to relativistic precession in a tilted disc around a black hole. Furthermore, all of our simulations were hydrodynamic, isothermal, and where either inviscid, or included a constant alpha viscosity -- in reality, precessing, tilted discs around black holes are highly ionized, hot, and unstable to the magnetorotational instability (MRI), which results in a turbulent viscosity. The local formalism that we have presented in this paper for modelling tilted, differentially precessing accretion discs brings high-resolution, long-duration MHD simulations of disc breaking and tearing well within the capabilities of current computational resources. We will investigate this regime in future work.

\section*{Acknowledgments}
We thank Carolin N. Kimmig and Daniel Price for useful discussions. This research was funded by STFC through grant ST/X001113/1. Simulations were run on the CSD3 cluster at the University of Cambridge through an STFC DiRAC computing grant, and on the Swirles cluster at DAMTP, University of Cambridge.

\section*{Data availability}
The data underlying this article will be shared on a reasonable request to the corresponding author.


\bibliographystyle{mnras}
\bibliography{BrokenDisksBib} 


\appendix

\section{Tilt dissipation study}
\label{APPENDIX_TiltDissipationStudy}

Global grid-based simulations of tilted discs have found that insufficient numerical resolution can induce an artificial precession in the disc. See, for example, Fig. 8 of \cite{kimmig2024warped}.\footnote{Numerical dissipation can arise due to (i) \textit{misalignment} of the flow with the grid (as is the case with tilted discs in global simulations), and (ii) advection error, which can occur even when the flow is \textit{aligned} with the grid. In our local set-up a tilted disc is actually aligned with the Cartesian grid, but oscillates relative to that grid, which will result in an advection error.} Here we present a suite of inviscid simulations of tilted discs to determine the impact of numerical dissipation on the complex tilt (equation \ref{EQUATION_ComplexTilt}). We remind the reader that $|Z|$ corresponds to the tilt of the disc, while $\arg Z$ corresponds to its precession.

\begin{figure}
    \centering
    \includegraphics[width=1\linewidth]{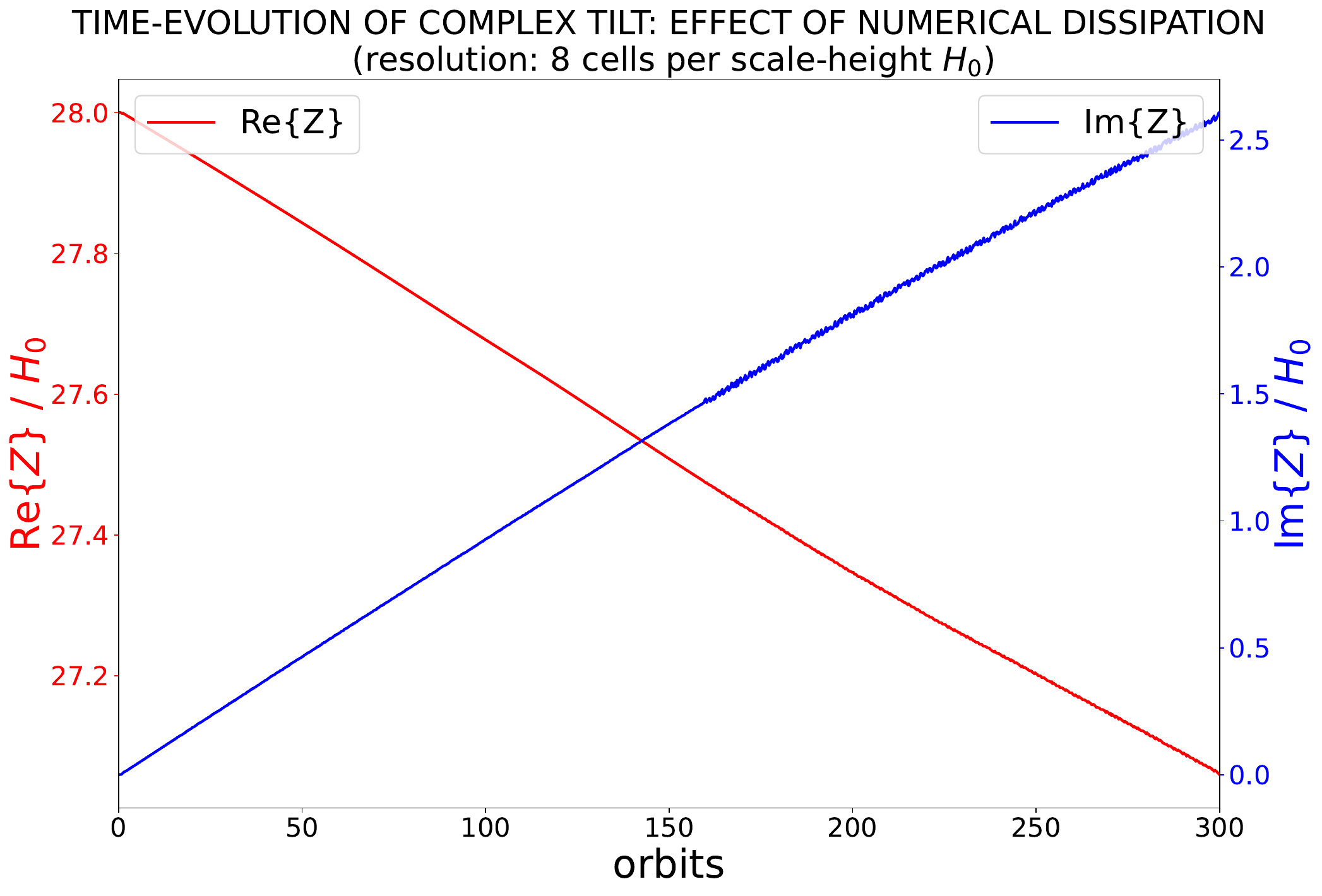}
    \caption{Tilted disc study: effect of numerical dissipation on the complex tilt $Z$ (see equation \ref{EQUATION_ComplexTilt}) at a resolution of ($8/H_0$). The initial condition consisted of a tilted disc ($z_0 = 28H_0$) without any imposed precession. The left-hand axis (red) shows the evolution of the real part of the complex tilt $Z$ (which measures the tilt of the disc); the right-hand axis (blue) shows $\text{Im}\{Z\}$ (which measures precession of the disc). The results show that numerical dissipation leads to damping of amplitude of the tilt, and a linear increase the phase of the complex tilt (which can be interpreted as a rigid precession of the disc). We observe similar trends at higher resolutions ($16/H_0$ and $32/H_0$), but the rates are significantly smaller: see Fig.~\ref{FIGURE_TiltedDiskResolutionStudy}.}
\label{FIGURE_TiltedDiskTimeSeriesReZImZRes8}
\end{figure}

\begin{figure}
    \centering
    \includegraphics[width=1\linewidth]{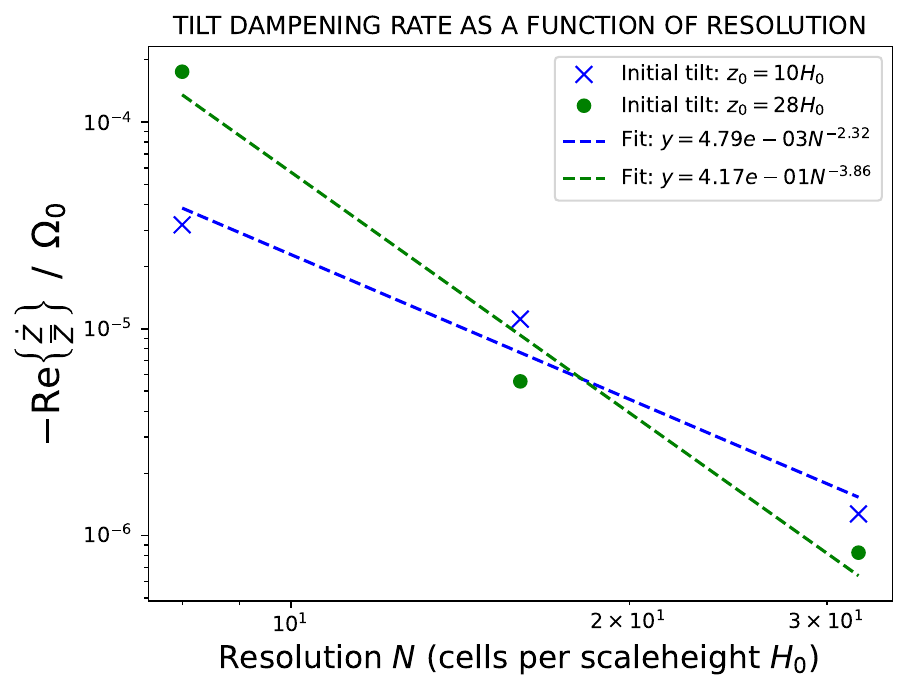}
    \includegraphics[width=1\linewidth]{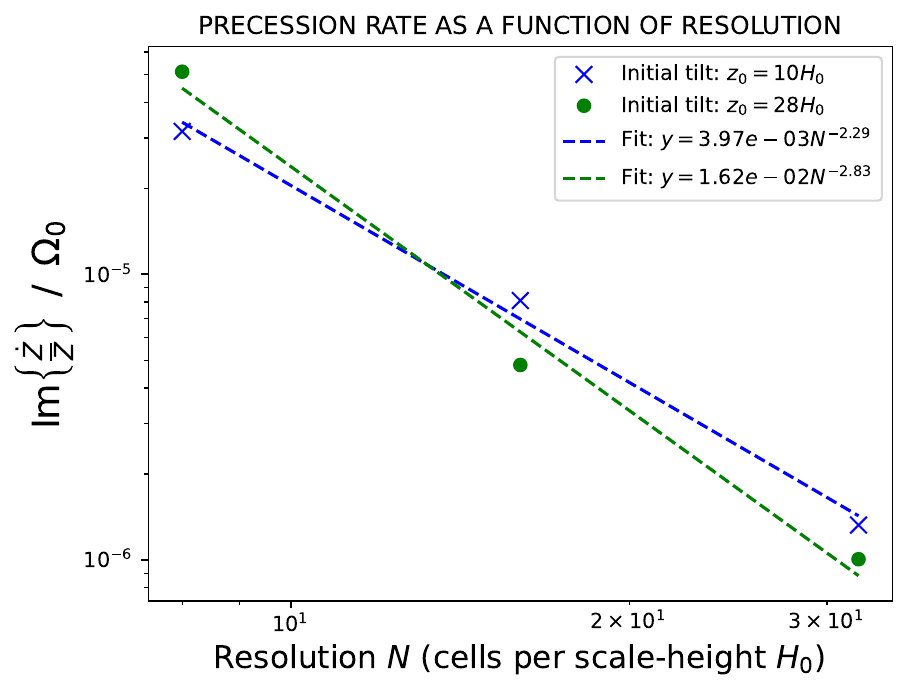}
    \caption{Effect of numerical resolution on disc tilt ($\text{Re}\{Z\}$) and precession ($\text{Im}\{Z\}$). \textit{Top panel}: rate of change of real part of tilt $-\text{Re}\{\dot{Z}/Z\}$ as a function of resolution $N$ (cells per scale-height $H_0$). \textit{Bottom panel}: rate of change of imaginary part of tilt $\text{Im}\{\dot{Z}/Z\}$. The results are plotted for two initial disc tilts: $z_0 = 10H_0$ (blue crosses) and $z_0 = 28H_0$ (green circles), and for three resolutions: $8/H_0, 16/H_0, 32/H_0$. (The precession rate for the $z_0 = 28H_0$, res $16/H_0$ simulation was negative, so we have plotted its absolute value.)}
\label{FIGURE_TiltedDiskResolutionStudy}
\end{figure}

The initial condition in all runs discussed in this section consists of a flat disc with a uniform tilt (equation \ref{EQUATION_InitialDensityProfileTiltedDiskSims}). We \textit{do not} include an imposed precession in these runs. We consider two sets of simulations: in one set the disc is initialized with a smaller initial tilt ($z_0 = 10H_0$), while in the other set the disc has a larger tilt ($z_0 = 28H_0$). The latter tilt is the same as that used in our fiducial simulation (Section \ref{RESULTS_FIDUCIALSIMULATION}). For a given initial tilt, we repeat the simulation at three different resolutions: $8/H_0$, $16/H_0$, $32/H_0$.\footnote{The smaller tilt runs runs were carried out in a box of vertical extent $L_z = 48H_0$, while the larger tilt runs employed taller boxes ($L_z = 68H_0$).} The runs are listed in Table \ref{TABLE_TiltedDiskStudyNOPrecession}. All runs employed a density floor of $\rho_{\text{floor}} = 10^{-14}\rho_0$. We perturbed the velocity components with random noise at initialization to seed any instabilities that might be present. All simulations were run for 300 orbits.

\subsection{Very low resolution induces an unphysical warp in the disc}
After initialization we observe the disc oscillating uniformly about $z/H_0 = 0$ in all cases, as expected. We focus first on the simulation TiltedDisk04: this has the lowest resolution ($8/H_0$) and largest initial tilt ($z_0 = 28H_0$). We observe a short wavelength ($\lambda \sim 4H_0$) low amplitude ($0.01 \lesssim \langle \psi_{\text{max}} \rangle_{\tau} \lesssim 0.03$) corrugation develop in the disc around orbit 90. The corrugation first arises between $10 \lesssim x/H_0 \lesssim 30$, and spreads throughout the entire radial extent of the disc within 10-20 orbits. In the lowest-resolution run, we also observe exponential growth in the radial kinetic energy (not shown) between around orbit 80 and orbit 120: $\langle E_{\text{kin},x}$ grows from about $10^{-6}\,c_{\text{s}0}^2 \rho_0$ to around $10^{-4}\,c_{\text{s}0}^2 \rho_0$ during this interval. Thereafter the radial KE continues to grow, but more slowly, and saturates around $4\times10^{-4}\,c_{\text{s}0}^2 \rho_0$). 

We surmise that the aforementioned corrugation is due to an artificial coupling between the tilt mode and two counter-propagating inertial waves. In the absence of any numerical dissipation this coupling should be zero.\footnote{This is not to be confused with the non-zero coupling between a \textit{bouncing} mode and two counter-propagating inertial waves which we found in our study of vertical oscillations in distorted discs \citep{held2024instability}; the latter is a physical effect, and leads to the development of a warp in the disc.} However, the large numerical dissipation in the $8/H_0$ run likely gives rise to some non-zero coupling coefficient between the tilt mode and inertial waves in the disc, which should increase with tilt. 

In a run with the same resolution ($8/H_0$), but a smaller initial tilt ($z_0 = 10H_0$), we observe the same phenomenon, but the corrugation sets in at later times (around orbit 200). Finally, we do not observe such a corrugation in the higher resolutions runs ($16/H_0$ and $32/H_0$).

\subsection{Effect of numerical dissipation on tilt and precession}
We have also found that numerical dissipation both damps the tilt and induces a rigid precession in the disc. In Fig.~\ref{FIGURE_TiltedDiskTimeSeriesReZImZRes8} we plot the time-evolution of the complex tilt from our lowest resolution run ($8/H_0$) with a large initial tilt ($z_0 = 28H_0$).\footnote{Here we are plotting the \textit{instantaneous} and radially-averaged value of $Z$, rather than the orbit-averaged radial profile defined in equation \ref{EQUATION_ComplexTilt}.} The real part of the complex tilt is shown in red, while the imaginary part is shown in blue. We observe a clear decrease in $\text{Re}\{Z\}$ (indicating that the tilt is decaying), whereas we observe an increase in $\text{Im}\{Z\}$ (indicating that the disc is precessing). We have also inspected orbit-averaged radial profiles of $\text{Im}\{Z\}$ (not shown) and found that these are noisy: this suggests that the artificial growth in $\text{Im}\{Z\}$ corresponds to a \textit{rigid} precession rather than a \textit{differential} one (i.e.~a \textit{twist}).

In the runs initialized with a smaller tilt ($z_0 = 10H_0$), the tilt at the end of the simulation (orbit 300) is around $\text{Re}\{Z\} /H_0\sim 9.4, 9.8, 9.98$ at resolutions of $8/H_0, 16/H_0$ and $32/H_0$, respectively. In the runs initialized with a larger initial tilt ($z_0 = 28H_0$) we measure  $\text{Re}\{Z\}/H_0 \sim 27, 27.7, 27.95$, respectively. The precession at the end of the simulations is $\text{Im}\{Z\}/H_0 \sim 0.55, 0.16, 0.03$ in the $z_0 = 10H_0$ runs, and $\text{Im}\{Z\}/H_0 \sim 2.5, -0.25, 0.06$ in the $z_0 = 28H_0$ runs. (Note that the precession in the $z_0 = 28H_0$ at res $16/H_0$ was in the opposite direction.) The tilt decreases exponentially, whereas the increase in precession is linear. In Fig.~\ref{FIGURE_TiltedDiskResolutionStudy} we plot the rate of change of tilt $-\text{Re}\{\dot{Z}/Z\}$ and precession $-\text{Im}\{\dot{Z}/Z\}$ as function of resolution for the two initial tilts. From this figure it is clear that the numerical tilt dampening rate, and the linear increase in the phase with time, both decrease as a power law as we increase the resolution $N$: we measure $-\text{Re}\{\dot{Z}/Z\} \propto N^{-2.29}$ and $\text{Im}\{\dot{Z}/Z\} \propto N^{-2.83}$ in runs with initial disc tilts of $z_0=10H_0$, and $z_0 = 28H_0$, respectively.

In summary, we find that numerical dissipation alone can both damp the initial tilt in the disc, as well as cause the disc to begin precessing (even if we do not include an imposed precession). However, this numerically-induced damping of the tilt, and linear increase in precession, are very small at the resolution of $32/H_0$ employed in all other simulations in this paper (of the order $10^{-6}\Omega_0$). This gives us confidence that numerical dissipation is subdominant to physical dissipation (e.g.~due to shocks or parametric instability) in those runs. On the other hand, simulations run at resolutions of $\sim 8/H_0$ induce tilt damping / precession rates that are of order 10-100 times larger.

\section{Tables of Simulations}
\label{APPENDIX_TablesOfSimulations}

\begin{table*}
\begin{threeparttable}[b]
\centering
\caption{Precession amplitude ($a$) and initial disc tilt ($z_0$) study: simulations of tilted discs \textit{with} imposed precession. The fiducial simulation is Precession02. The columns indicate: box size $[L_x, L_y, L_z]$ (in units of the initial mid-plane scale-height $H_0$), resolution (in cells per $H_0$), viscosity, whether the simulation is in the diffusive ($\alpha > kH_0$) or
wavelike ($\alpha < kH_0$) regime, where $k=2\pi/L_x$ is the radial wavenumber of the warp, the dimensionless precession amplitude $a$, the initial tilt $z_0$ (in units of $H_0$), vertical boundary conditions (zBC), Riemann solver, interpolation scheme, and runtime (in orbits). The last column gives the final outcome of the simulation: \textit{breaking} (disc breaks into rings; it may also exhibit parametric instability at early times), \textit{instability} (the disc becomes parametrically unstable, but does not break), and \textit{damping} (the disc remains continuous and the tilt is ultimately damped).}
\label{TABLE_PrecessionStudy}
 	\begin{tabular}{lccccccccccr}
		\hline
		Run	& Box Size & Res & $\alpha$\tnote{1} & Regime & $a$ & tilt $(H_0)$ & zBC &  Riemann & Reconstr. & Runtime & Outcome \\ 
        \hline
        Precession10 & [90,0.04,68] & $32/H_0$ & inviscid & wavelike & 0.05 & 0 &  reflective & HLLC & parabolic & 100 & equilibrium  \\
        \hline
        Precession01 & [90,0.04,68] & $32/H_0$ & inviscid & wavelike & 0.01 & 28 &  reflective & HLLC & parabolic & 100 & breaking \\
        Precession02 & [90,0.04,68] & $32/H_0$ & inviscid & wavelike & 0.05 & 28 &  reflective & HLLC & parabolic & 125 & breaking  \\
        Precession03 & [90,0.04,48] & $32/H_0$ & inviscid & wavelike & 0.05 & 14 &  reflective & HLLC & parabolic & 100 & breaking  \\
        Precession04 & [90,0.04,48] & $32/H_0$ & inviscid & wavelike & 0.01 & 14 &  reflective & HLLC & parabolic & 100 & instability \\
        \hline
        Precession08 & [90,0.04,68] & $32/H_0$ & 0.1 & diffusive & 0.01 & 28 &  reflective & HLLC & parabolic & 100 & damping \\	
        Precession11 & [90,0.04,68] & $32/H_0$ & 0.1 & diffusive & 0.05 & 28 &  reflective & HLLC & parabolic & 100 & damping \\	
        Precession15 & [90,0.04,68] & $32/H_0$ & 0.1 & diffusive & 0.1 & 28 &  reflective & HLLC & parabolic & 100 & damping \\	
        Precession19 & [90,0.04,88] & $32/H_0$ & 0.1 & diffusive & 0.05 & 38 &  reflective & HLLC & parabolic & 100 & damping \\	
        \hline
        \end{tabular}
    \begin{tablenotes}
       \item [1] Technically this is the kinematic viscosity, rather than $\alpha$, and we have taken $\nu \sim \alpha c_{\text{s}0}H_0$ as is standard practice in the literature (see, e.g.~\cite{held2021magnetohydrodynamic}). The actual relationship for a Keplerian disc is $\nu = (2/3)\alpha c_{\text{s}0} H_0$ \citep{heldlatter2018, armitage2022lecture}, although the factor of $2/3$ is often dropped in the literature.
     \end{tablenotes}
\end{threeparttable}
\end{table*}

\begin{table*}
\centering
\caption{Tilt dissipation study: simulations of tilted discs \textit{without} imposed precession. The columns have the same meaning as in Table \ref{TABLE_PrecessionStudy}. Numerical dissipation damps the tilt and induces a rigid precession in the disc. The corresponding tilt decay rate ($-\text{Re}\{\dot{Z}/Z\}$) and the linear increase in the phase of the complex tilt ($\text{Im}\{\dot{Z}/Z\}$) are given by $\gamma$ and $\sigma$, respectively, in units of $\Omega_0$. (Both have been rounded to one significant figure.) See Fig.~\ref{FIGURE_TiltedDiskResolutionStudy}.}
\label{TABLE_TiltedDiskStudyNOPrecession}
 	\begin{tabular}{lcccccccccccr}
		\hline
		Run	& Box Size & Res & $\alpha$ & Regime & tilt $(H_0)$ & zBC &  Riemann & Reconstr. & Runtime & $\gamma/\Omega_0$ & $\sigma/\Omega_0$ \\ 
        \hline
        TiltedDisk02 & [90,0.04,48] & $8/H_0$ & inviscid & wavelike & 10 &  reflective & HLLC & linear & 300 & $3\times10^{-5}$ & $3\times10^{-5}$  \\
        TiltedDisk03 & [90,0.04,48] & $16/H_0$ & inviscid & wavelike & 10 &  reflective & HLLC & linear & 300 & $10^{-5}$ & $8\times10^{-6}$  \\
        TiltedDisk01 & [90,0.04,48] & $32/H_0$ & inviscid & wavelike & 10 &  reflective & HLLC & linear & 300 & $10^{-6}$ &  $10^{-6}$ \\
        \hline
        TiltedDisk04 & [90,0.04,68] & $8/H_0$ & inviscid & wavelike & 28 &  reflective & HLLC & linear & 300 & $10^{-4}$ & $5\times10^{-5}$ \\
        TiltedDisk05 & [90,0.04,68] & $16/H_0$ & inviscid & wavelike & 28 &  reflective & HLLC & linear & 300 & $6\times10^{-6}$ &  $-5\times10^{-6}$\\
        TiltedDisk06 & [90,0.04,68] & $32/H_0$ & inviscid & wavelike & 28 &  reflective & HLLC & linear & 300 & $8\times10^{-7}$ &  $10^{-6}$\\    
        \hline
	\end{tabular}
\end{table*}

\begin{table*}
\centering
\caption{Simulations of of the interaction between broken disc segments (without an imposed precession). \textit{Tilt} refers to the initial relative tilt between the two rings (in units of the scale-height $H_0$), while $L$ refers to the radial separation between the ring centres.}
\label{TABLE_GaussianRingStudy}
 	\begin{tabular}{lcccccccccr}
		\hline
		Run	& Box Size & Res & $\alpha$ & Regime & tilt $(H_0)$ & $L (H_0)$ & zBC &  Riemann & Reconstr. & Runtime \\ 
        \hline
        GaussianRing01 & [40,0.04,24] & $32/H_0$ & inviscid & wavelike & 1 & 10 & reflective & HLLC & linear & 100 \\
        GaussianRing03 & [40,0.04,24] & $32/H_0$ & inviscid & wavelike & 5 & 10 & reflective & HLLC & linear & 100 \\
        \hline
        GaussianRing04 & [40,0.04,24] & $32/H_0$ & inviscid & wavelike & 1 & 15 & reflective & HLLC & linear & 100 \\
        GaussianRing06 & [40,0.04,24] & $32/H_0$ & inviscid & wavelike & 5 & 15 & reflective & HLLC & linear & 100 \\
        \hline
        GaussianRing07 & [40,0.04,24] & $32/H_0$ & inviscid & wavelike & 1 & 20 & reflective & HLLC & linear & 100 \\
        GaussianRing09 & [40,0.04,24] & $32/H_0$ & inviscid & wavelike & 5 & 20 & reflective & HLLC & linear & 100 \\
        \hline

	\end{tabular}
\end{table*}

\bsp	
\label{lastpage}
\end{document}